%% file: main.tex
\documentclass[acmtog, authorversion=True]{acmart}
\pdfoutput=1
\usepackage{array}
\usepackage{booktabs} % For formal tables
\usepackage{multirow}  
\usepackage{enumerate}

\usepackage[ruled]{algorithm2e} % For algorithms

\SetAlFnt{\small}
\SetAlCapFnt{\small}
\SetAlCapNameFnt{\small}
\SetAlCapHSkip{0pt}

\setcopyright{acmcopyright}\acmJournal{TOG}
\acmYear{2020}\acmVolume{39}\acmNumber{6}\acmArticle{234}\acmMonth{12} \acmDOI{10.1145/3414685.3417812}
\input{macros.tex}

\begin{document}
% Title portion
\title
[ShapeAssembly: Learning to Generate Programs for 3D Shape Structure Synthesis]
{ShapeAssembly: Learning to Generate Programs for\\3D Shape Structure Synthesis}

\author{R. Kenny Jones}
\affiliation{%
    \institution{Brown University}
}

\author{Theresa Barton}
\affiliation{%
    \institution{Brown University}
}

\author{Xianghao Xu}
\affiliation{%
    \institution{Brown University}
}

\author{Kai Wang}
\affiliation{%
    \institution{Brown University}
}

\author{Ellen Jiang}
\affiliation{%
    \institution{Brown University}
}

\author{Paul Guerrero}
\affiliation{%
    \institution{Adobe Research}
}

\author{Niloy J. Mitra}
\affiliation{%
    \institution{University College London, Adobe Research}
}

\author{Daniel Ritchie}
\affiliation{%
    \institution{Brown University}
}

\begin{abstract}
\input{00-abstract.tex}

\end{abstract}





\begin{teaserfigure}
  \centering
  \includegraphics[width=\linewidth]{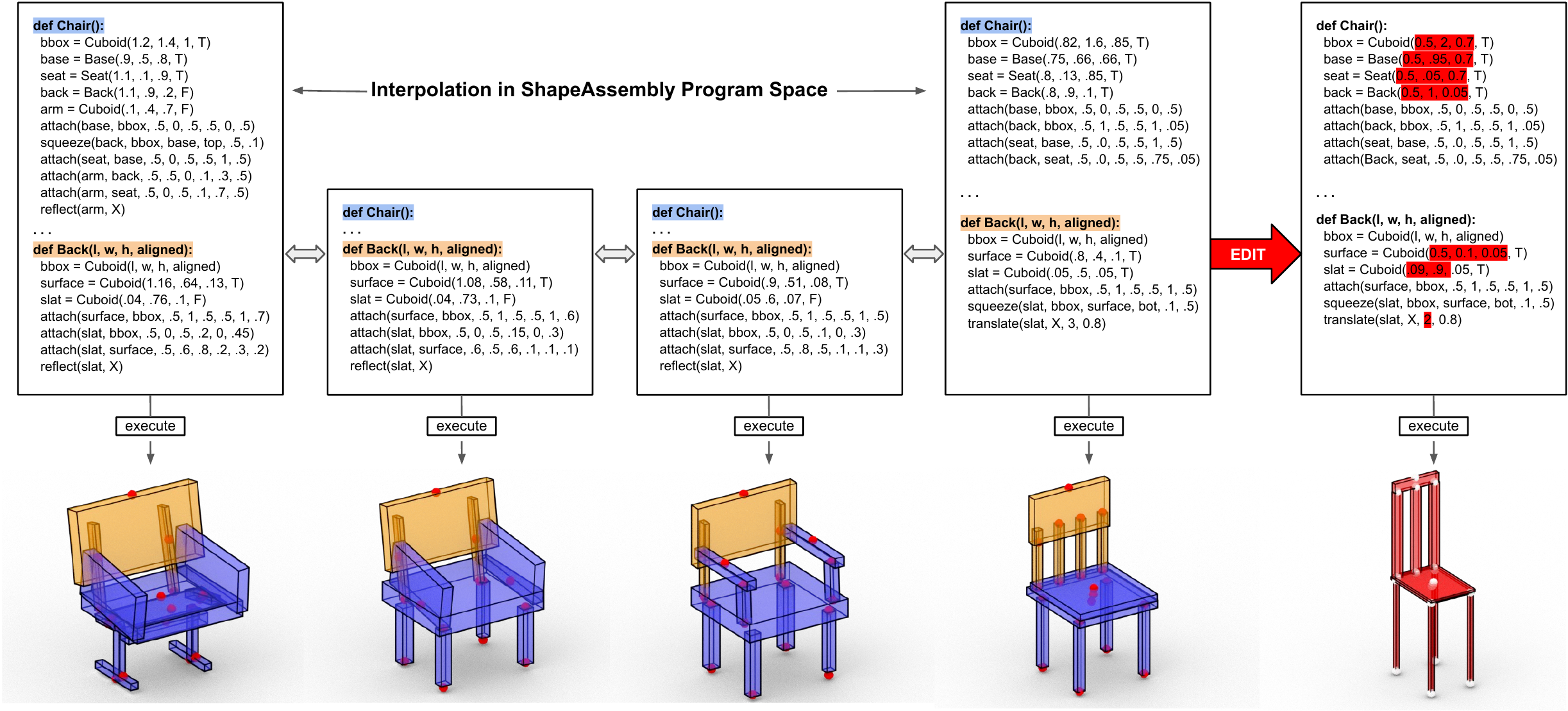}
\caption{
We present a deep generative model which learns to write novel programs in ~\dslname, a domain-specific language for modeling 3D shape structures. 
Executing a ~\dslname~ program produces a shape composed of a hierarchical connected assembly of part proxies cuboids. Our method develops a well-formed latent space that supports interpolations between programs. Above, we show one such interpolation, and also visualize the geometry these programs produce when executed. In the last column, we manually edit the continuous parameters of a generated program, in order to produce a variant geometric structure with new topology.}
\label{fig:teaser}
\end{teaserfigure}

\maketitle

\input{01-intro.tex}
\input{02-related.tex}
\input{03-approach.tex}
\input{04-method.tex}
\input{05-results.tex}

\input{06-conclusion.tex}
\input{07-acknowledgments.tex}

% Bibliography
\bibliographystyle{ACM-Reference-Format}
\bibliography{main}

\input{appendix.tex}

\end{document}

%% file: macros.tex
\definecolor{pink}{rgb}{1,0,1}
\newcommand{\revisiondelete}[1]{}

\newcommand{\reals}{\ensuremath{\mathbb{R}}}
\newcommand{\integers}{\ensuremath{\mathbb{Z}}}
\newcommand{\integersNonNeg}{\ensuremath{\integers^{+}}}

\newcommand{\denselist}{\itemsep 0pt\parsep=0pt\partopsep 0pt\vspace{-\topsep}}

\newcommand{\dslname}{\textsc{ShapeAssembly}}

%% file: 00-abstract.tex
Manually authoring 3D shapes is difficult and time consuming; generative models of 3D shapes offer compelling alternatives. Procedural representations are one such possibility: they offer high-quality and editable results but are difficult to author and often produce outputs with limited diversity. On the other extreme are deep generative models: given enough data, they can learn to generate any class of shape but their outputs have artifacts and the representation is not editable. 

In this paper, we take a step towards achieving the best of both worlds for novel 3D shape synthesis. First, we propose~\dslname, a domain-specific ``assembly-language'' for 3D shape structures.~\dslname~programs construct shape structures by declaring cuboid part proxies and attaching them to one another, in a hierarchical and symmetrical fashion.~\dslname~functions are parameterized with continuous free variables, so that one program structure is able to capture a family of related shapes. We show how to extract~\dslname~programs from existing shape structures in the PartNet dataset. Then, we train a deep generative model, a hierarchical sequence VAE, that learns to write novel~\dslname~programs. Our approach leverages the strengths of each representation: the program captures the subset of shape variability that is interpretable and editable, and the deep generative model captures variability and correlations across shape collections that is hard to express procedurally. 

We evaluate our approach by comparing the shapes output by our generated programs to those from other recent shape structure synthesis models. We find that our generated shapes are more plausible and physically-valid than those of other methods. Additionally, we assess the latent spaces of these models, and find that ours is better structured and produces smoother interpolations. As an application, we use our generative model and differentiable program interpreter to infer and fit shape programs to unstructured geometry, such as point clouds.

%% file: 01-intro.tex
\section{Introduction}
\label{sec:intro}

3D models of human-made objects are more in-demand than ever.
In addition to the traditional drivers of demand in computer graphics (visual effects, animation, games), new applications in artificial intelligence increasingly benefit from or even require high-quality 3D objects, such as producing synthetic training imagery for computer vision systems~\cite{PlayingForData,RenderingSUNCG,MetaSim} or training robots to perform tasks in virtual environments~\cite{AI2Thor,GibsonEnv,Habitat,BenKinematics}.
Despite the growing demand, the craft of 3D modeling largely remains as difficult and time-consuming as it has ever been.
The time and expertise required to create 3D content by hand will not scale to these demands.

One promising way out of this conundrum is the development of \emph{generative models} of 3D shapes, i.e. procedures which can be executed to generate novel shapes within some class~\cite{LSystemsBook,CGAShape,CityEngine}.
An ideal generative model would produce plausible output geometry, capture a wide range of shape variations, and use an interpretable representation which a user could subsequently manipulate and edit.
Unfortunately, no existing shape generative model achieves all of these properties.
On the one hand are \emph{procedural models}: structured computer programs which produce geometry when executed.
Procedural models can produce high-quality geometry, and their program-based representation makes them interpretable and editable to users with some programming background.
However, authoring a good procedural model from scratch is difficult (arguably at least as difficult as modeling an object by hand), and the amount of shape variation captured by a single procedural model is limited (e.g., it is difficult to write one program that can model all types of cars).
On the other hand are data-driven generative models, particularly \emph{deep generative models}: neural networks which learn how to generate 3D shapes from data~\cite{3DGAN,AtlasNet,PointSetGeneration,GRASS,StructureNet}.
Deep generative models capture variability with little human effort: given enough training data, they can in theory learn to generate any class of shape.
Since they lack the strict semantics of programs, however, their outputs often exhibit ``noise'' artifacts such as incomplete geometry and floating parts.
Additionally, the representations they learn are typically inscrutable to people, making them hard to edit or manipulate in predictable ways.

Our insight, in this work, is that these two approaches have complementary strengths: deep generative models are efficient to create and excel at broad-scale variability, and procedural models produce high-quality geometry by construction and better facilitate editing for fine-scale variability.
We take a first step toward achieving the best of both worlds by integrating these two approaches into a single pipeline: a deep generative model that \emph{learns to write programs}, which, when executed, themselves output 3D geometry.
We hypothesize that going through this intermediate program representation produces a generative model with a smoother latent space, whose outputs are more likely to be physically valid, compact, and editable.

As the motivating applications mentioned earlier demand 3D models of human-made objects, we focus on generating novel part-based shape structures in this paper.
We introduce~\dslname, an ``assembly language'' for 3D shape structures.
In~\dslname, shape structures are represented by hierarchical assemblies of connected parts, where leaf-level parts are approximated by a bounding cuboid (a similar representation as the ones used by PartNet~\cite{PartNet} and StructureNet~\cite{StructureNet}); these hierarchical cuboid structures can then be used to condition the generation of shape surface geometry in the form of e.g. point clouds.
A~\dslname~program constructs a shape by declaring cuboids, iteratively attaching them to one another, and specifying symmetric repetitions of connected cuboid assemblies.
The dimensions of these cuboids and the positions of these attachments are a program's parameters; manipulating them allows for exploring a family of related shapes.
Furthermore, our interpreter for executing~\dslname~programs is fully differentiable, meaning it is possible to compute gradients of a program's output geometry with respect to its continuous parameters. Figure~\ref{fig:teaser} shows some example hierarchical~\dslname~programs and the output shapes they produce.

While~\dslname~programs produce valid geometry under a range of parameter values, they do not exhibit \emph{structural} variability, and authoring them from scratch still takes time.
Thus, we train a neural network to write a variety of~\dslname~programs for us.
Using programs we extract from a shape dataset, we train a hierarchical sequence VAE which outputs hierarchical~\dslname~programs.
Each node in the hierarchy uses a recurrent language model to generate the program text at that level, and to decide which cuboids should be expanded into subroutine calls.
Furthermore, the well-defined semantics of~\dslname~allow us to identify semantically-invalid programs and modify the generator such that it never produces them.
The programs shown in Figure~\ref{fig:teaser} were written by our generative model, by decoding code vectors along a straight line in its latent space.
We show that this generative model indeed learns to generate plausible, novel shape programs that were never seen its the training set.

We evaluate our approach by comparing it to other recently-proposed generative models of 3D shape structure along several axes including plausibility, diversity, complexity, and physical validity.
We find that our generated shapes are both more plausible and more physically-valid than those of other methods.
Additionally, we assess the latent spaces of these models, and find that ours is better structured and produces smoother interpolations, both in terms of geometric and structural continuity.
As a bonus, we also show that~\dslname's decoder does a better job of fitting programs to unstructured point clouds while also maintaining physical validity, and that this performance difference is magnified by optimizing the program fit via our differentiable interpreter.

In summary, our contributions are:
\begin{enumerate}[(i)]
    \item The~\dslname~language and its differentiable interpreter, allowing the procedural specification of shape structures represented as connected part assemblies.
    \item A deep generative model for~\dslname~programs, coupling the ease-of-training and variability of neural networks with the precision and editability of procedural representations. 
\end{enumerate}

Code and data used for all of our experiments can be found at https://github.com/rkjones4/ShapeAssembly .

%% file: 02-related.tex
\begin{figure*}[t!]
  \centering
  \includegraphics[width=\linewidth]{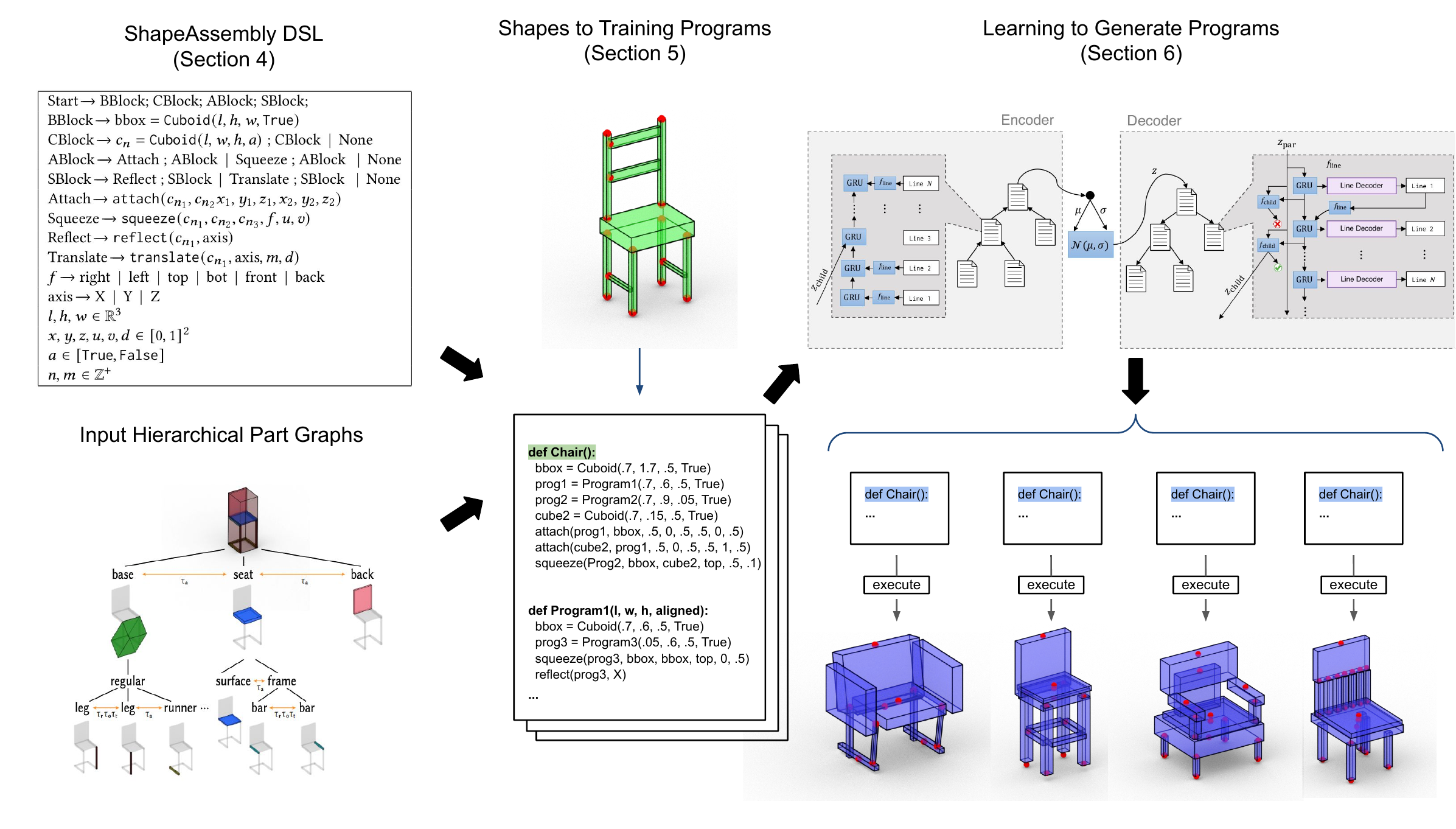}
\caption{
Our pipeline for generating 3D shape structure programs. We first define a DSL language for 3D shapes, ~\dslname~. Then, given a dataset of hierarchical part graphs, we extract~\dslname~programs from them. Finally, we use these programs as training data for a deep generative model. Our method learns to generate novel program instances that can be executed to produce complex and interesting 3D shape structures. }
\label{fig:pipeline}
\end{figure*}

\section{Related Work}
\label{sec:relatedWork}

\subsubsection*{Deep Generative Models of 3D Shapes}
Recent years have seen an explosion of activity in applying deep generative models to 3D shape generation.
Some of the earliest approaches generated shapes as 3D occupancy grids~\cite{3DShapeNets,3DGAN}; later work has explored generative representations of point clouds~\cite{PointSetGeneration}, 2D surface patches~\cite{AtlasNet}, and implicit surfaces~\cite{IMNet,DeepSDF,DeepLevelSets,BSPNet}.
Our approach is more closely related to generative models of \emph{part-based} shapes, wherein a complete object is synthesized by generating and assembling multiple subparts.
These include approaches for iteratively adding parts to partially-complete shapes~\cite{ComplementMe}, generating symmetry hierarchies~\cite{GRASS}, composing parts from two different shapes~\cite{SCORES}, and generating hierarchical connectivity graphs~\cite{StructureNet}.
Our method is different in that we do not aim to learn a generative model that outputs a single shape; rather, ours outputs a procedural program which then itself generates a related family of shapes.

\subsubsection*{Procedural and Inverse Procedural Modeling}
There is a rich history of methods for procedural modeling in computer graphics: especially noteworthy examples include its use in modeling plants~\cite{LSystemsBook} and urban environments~\cite{CityEngine,CGAShape}.
Most procedural modeling systems use some form of (context-free) grammar, i.e., a recursive string re-writing system (which may be interpreted as e.g., recursively splitting a spatial domain, in the case of shape grammars). 
Additionally, attachment-based grammars of part assemblies have been used to aid in the structural analysis of shapes~\cite{10.1145/2010324.1964980}.
Our procedural representation is fundamentally different: we use an imperative language which iteratively constructs shapes via declaring and then connecting parts represented as simple proxy geometry.
Also related to our work is the line of research on \emph{inverse} procedural modeling, i.e., inferring a procedural model from a set of examples~\cite{BayesianGrammarInduction,FacadeInduction,BayesianProgramMerging,ProcmodLearn,Nishida2016,Nishida2018}.
These methods all strive to infer an interpretable, stochastic program which generates multiple output shapes. 
In contrast, we represent shapes via \emph{deterministic} programs, and then we use a stochastic neural network to generate those programs.

\subsubsection*{Visual Program Induction}
Another related line of work to ours is \emph{visual program induction} (VPI): the practice of inferring a program which describes a single visual entity, such as a 3D shape.
We address a fundamentally different problem: training a generative model to generate \emph{novel} 3D shape programs from scratch.
We do use a VPI-like process as a subroutine, to convert every shape in a large dataset into training programs for our generative model.
Prior work in this area can be roughly divided into two categories: methods that assume that clean, segmented geometry is available and then use geometric heuristics to infer a program~\cite{InverseLSystems,InverseProceduralArchitecture}, and methods which use learning or optimization to operate directly on ``raw'' visual inputs such as images and occupancy grids~\cite{ellis2018learning,liu2019learning,sharma2018csgnet,InverseCSG,tian2019learning,zhou2019treeLSTM,zou20173d,ellis2019write}.
Our approach to extracting programs from shapes to formulate training data is more similar to the former.

One could consider solving our problem of novel shape program generation by first generating novel 3D \emph{shapes} with an existing shape generative model and then using a VPI-like system to infer a program describing that shape.
However, as we will later show, the programs produced by such a process are less clean and editable than ones generated by our model; furthermore, training to generate programs rather than shapes directly actually produces a better-structured latent space.\\

Our complete pipeline of using a neural network to generate a program and then using that program to generate the ultimate output is also related to work in visual question answering which uses neural networks to generate a ``query program'' for each question which then analyzes the input image and produces an answer~\cite{VQAPrograms}.
It is also related to work that tries to combine the advantages of neural guidance with symbolic search for performing inference over structured domains~\cite{lu2019neurally}.

Our work is the first to train a deep generative model to produce \emph{novel} shape programs from scratch, each of which outputs a parametric family of related 3D shapes.
This pipeline combines the advantages of neural and procedural shape modeling.

%% file: 03-approach.tex
\section{Approach}
\label{sec:approach}
Our approach (Figure~\ref{fig:pipeline}) is divided into the following stages:
\subsubsection*{Input}
Our pipeline takes as input a large dataset of \emph{hierarchical 3D part graphs}~\cite{PartNet,StructureNet}.
This is a shape representation in which each node represents a part in a shape consisting of an assembly of parts.
Nodes are connected via edges that denote physical part attachments.
They can also be connected via parent-child edges that denote hierarchy relationships (i.e., that one part is composed of several other smaller parts).
At the leaf level of this hierarchy, atomic parts are represented by cuboid proxy geometry (typically minimum-volume bounding boxes of more detailed part meshes).

\subsubsection*{Defining a DSL for connected, hierarchical shapes}
To represent shapes as programs, we introduce a domain-specific language (DSL).
Since our input shapes are characterized by graphs of parts, where graph edges denote physical part connections, we introduce a DSL based around declaring parts and then attaching them to one another.
We call this language~\dslname~(as in, an ``assembly language'' for shapes).
Section~\ref{sec:dsl} describes the language.

\subsubsection*{Creating a dataset of shape-program pairs}
Given the language described above, we present a method for finding programs that represent the shapes in our dataset.

In our procedure, we first extract the program content based on a combination of data cleaning and geometric analysis. Then, we create canonical programs through a series of ordering and filter steps. Section~\ref{sec:progextraction} describes this procedure in more detail.

\subsubsection*{Learning to generate programs}
Finally, we treat the programs extracted from each shape as training data for a generative model.
Section~\ref{sec:generation} describes our deep generative model's architecture, the procedure we use to train it, and how we sample from it to synthesize new programs, which when executed produce novel shape structures.

%% file: 04-method.tex
\section{An Assembly Language for Shapes}
\label{sec:dsl}

\begin{figure}[b!]
  \centering  
  \includegraphics[width=\linewidth]{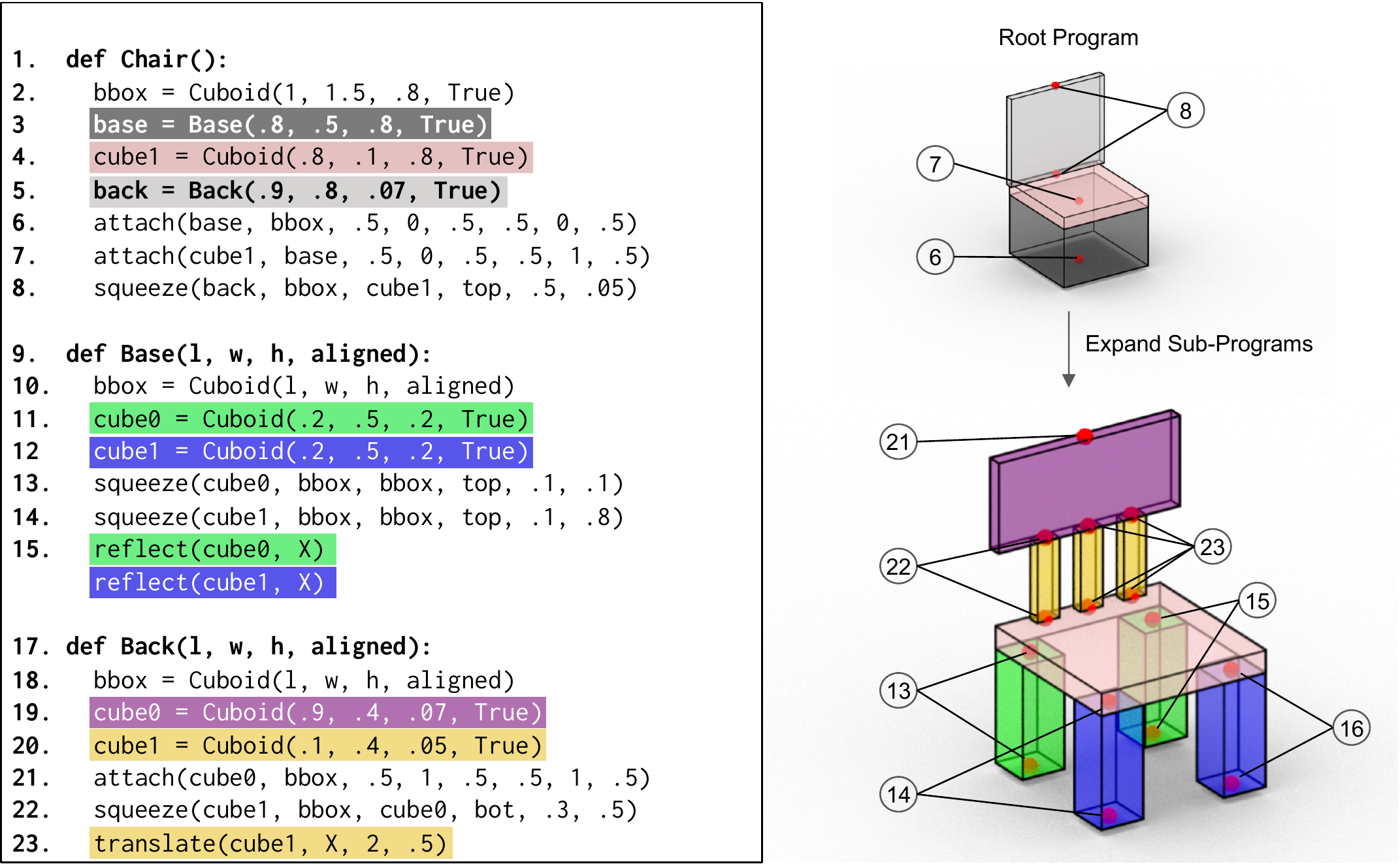}
\caption{
An example~\dslname~program and the shape that it generates.
Parts are colored according to the line of the program which instantiates them, and attachment points are numbered accordingly. In the top shape, we show the executed Chair program without hierarchy. In the bottom shape, we show the Chair program executed hierarchically with its sub-programs (Base and Back). For instance, the light grey back part is expanded into the purple back surface and gold slats.
}
\label{fig:dsl_examples}
\end{figure}

\begin{figure}[t!]
  \centering
  \includegraphics[width=\linewidth]{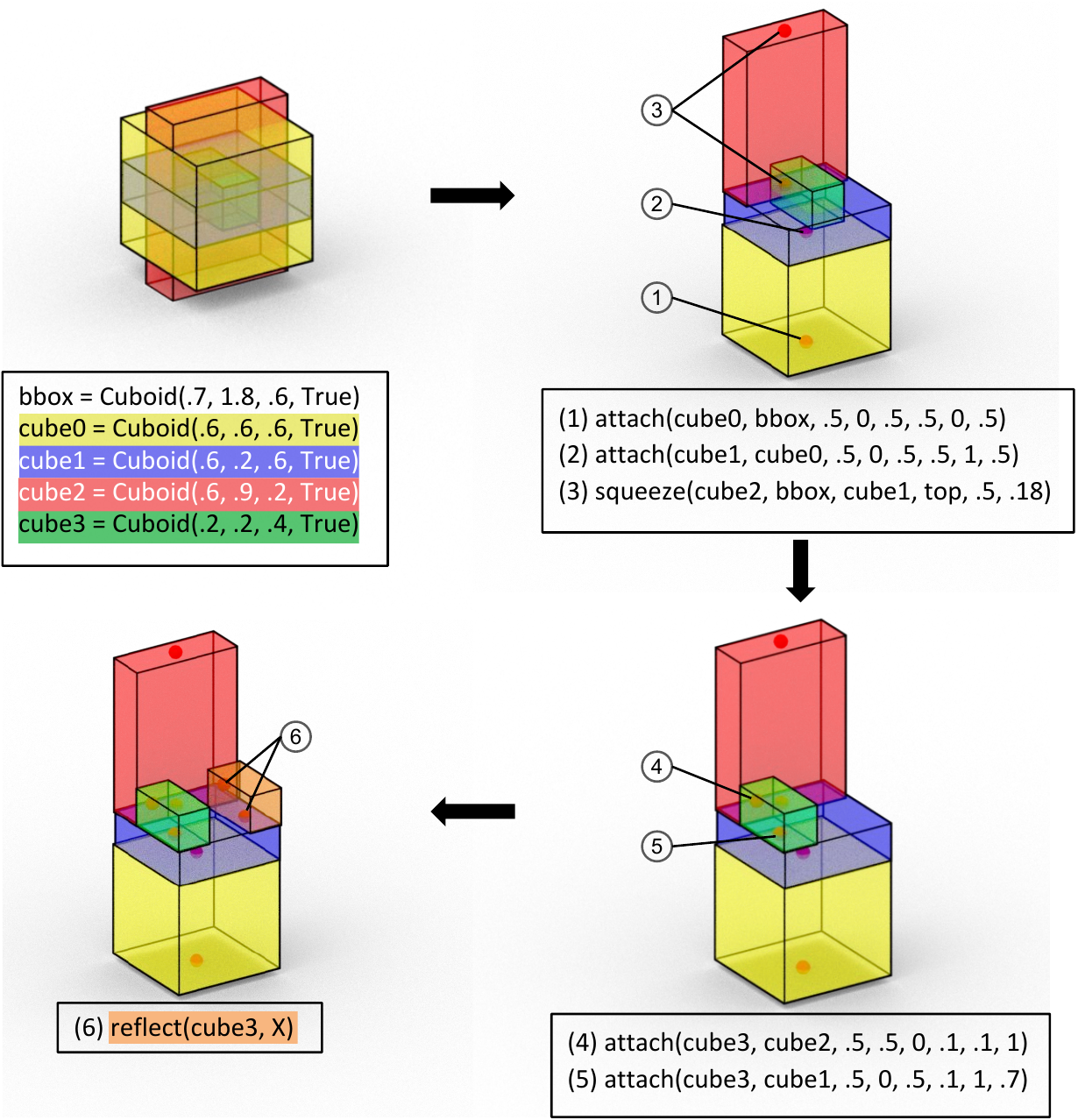}
\caption{
An illustration of how the~\dslname~interpreter incrementally constructs shapes by imperatively executing program commands. Cuboids are instantiated at the origin and are moved through attachment. Notice how the \texttt{reflect} command in line 6 acts as a macro function, creating a new cuboid and two new attachments.}
\label{fig:semantics}
\end{figure}

Our goal in this section is to define a domain-specific language for shapes which are specified as connected assemblies of parts. As we focus on the problem of shape structure synthesis, cuboids, serving as part proxy geometry, are the only data type in our language.
In Section~\ref{sec:results}, we show how to use other existing techniques to convert these proxies into surface geometry.

The primary operation in the language is attaching these cuboids together.
Attachment turns out to be a very powerful and flexible operation. 
In fact, our language does not include any operations for explicitly positioning or orienting cuboids: all of this is accomplished via attachment operations. 
Additionally, the language includes higher-level \emph{macros} that capture more complex spatial relationships, such as symmetry. 
At execution time, each macro is expanded into a series of cuboid declarations and attachment operations.

We call this DSL~\dslname, because it is an ``assembly language for shapes'': a low-level language for creating shapes, in which shapes are created by assembling parts. 
Table~\ref{tab:dsl} shows the grammar for~\dslname, and 
Figure~\ref{fig:dsl_examples} shows an annotated hierarchical program along with its executed 3D shape. 

A~\dslname~program consists of four main blocks:
\begin{itemize}
\denselist
    \item \textbf{BBlock:} Declares a non-visible bounding volume of the overall shape. This bounding volume is treated as a physical entity to which other parts can be connected.
    
    \item \textbf{CBlock:} Declares all the cuboid part proxies that will be used by the remainder of the program. The \texttt{Cuboid} command takes in $l,w,h$ parameters that control the starting dimensions of the part, and an aligned flag $a$ that specifies if the part has the same orientation as its bounding volume.  
    
    \item \textbf{ABlock:} Connects cuboids by iteratively attaching them to one another. The \texttt{attach} command takes in two cuboids, $c_{n1}$, $c_{n2}$, and attaches the point $(x_1, y_1, z_1)$ in the local coordinate frame of $c_{n1}$ with the point $(x_2, y_2, z_2)$ in the local coordinate frame of $c_{n2}$. The \texttt{squeeze} macro expands into two \texttt{attach} statements, such that $c_{n1}$ is placed in-between $c_{n2}$ and $c_{n3}$ along the specified face $f$ at the face-coordinate position $(u,v)$.
     
    \item \textbf{SBlock:} Generates symmetry groups by instantiating additional \texttt{Cuboid} and \text{attach} commands. The \texttt{reflect} macro reflects cuboid $c_{n}$ over axis $axis$ of the bounding volume. The \texttt{translate} macro creates a translational symmetry group starting at $c_n$ with $m$ additional members along axis $a$ of the bounding volume that ends distance $d$ away.
    
\denselist
\end{itemize}

\begin{table}[b!]
\small
\begin{tabular}{|l|}
\hline
Start  $\xrightarrow{}$  BBlock; CBlock; ABlock; SBlock; \\
BBlock $\xrightarrow{}$ $\text{bbox} = \texttt{Cuboid}(l, h, w, \texttt{True})$  \\
CBlock $\xrightarrow{}$ $c_n = \texttt{Cuboid}(l, w, h, a)$ ;  CBlock\: |\: None \\
ABlock $\xrightarrow{}$ Attach ; ABlock\: |\:  Squeeze ; ABlock \: |\: None \\
SBlock $\xrightarrow{}$ Reflect ; SBlock\: |\:  Translate ; SBlock \: |\: None \\
Attach $\xrightarrow{}$ $\texttt{attach}(c_{n_1}, c_{n_2}, x_1, y_1, z_1, x_2, y_2, z_2)$ \\
Squeeze $\xrightarrow{}$ $\texttt{squeeze}(c_{n_1}, c_{n_2}, c_{n_3}, f, u, v)$ \\
Reflect $\xrightarrow{}$ $\texttt{reflect}(c_{n}, \text{axis})$ \\
Translate $\xrightarrow{}$ $\texttt{translate}(c_{n}, \text{axis}, m, d)$ \\

$f \xrightarrow{}$ right\: |\: left \:|\: top\: |\: bot\: |\: front\: |\: back \\
$\text{axis} \xrightarrow{}$ X\: |\: Y \:|\: Z\: \\
$l, h, w \in \reals^+$ \\
$x, y, z, u, v, d \in [0,1]^2$ \\
$a \in [\texttt{True}, \texttt{False}]$ \\
$n, m \in \integersNonNeg$ \\
\hline
\end{tabular}
\caption{
The grammar for~\dslname, our low-level domain-specific ``assembly language'' for shape structure.
A program consists of \texttt{Cuboid} statements which instantiate new geometry and \texttt{attach} statements which connect these geometries together at specified points on their surfaces. Macro functions (\texttt{reflect}, \texttt{translate}, \texttt{squeeze}) form complex spatial relationships by expanding into multiple \texttt{Cuboid} and  \texttt{attach} statements.}
\label{tab:dsl}
\end{table}

\begin{figure*}[t!]
        \centering
        \includegraphics[width=\linewidth]{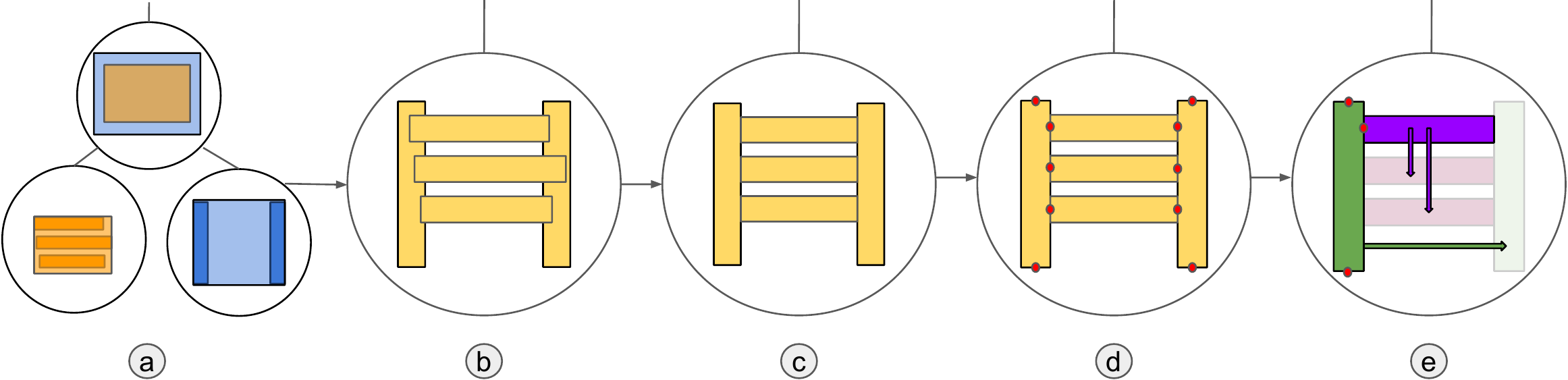}
        \caption{The steps of our program extraction pipeline. \emph{(a)} Fragment of an input hierarchical part graph showing chair back (parent node), chair back frame (blue child), and chair back surface (orange child). \emph{(b)} Locally flattening the hierarchy so that physically interacting leaf parts become siblings. \emph{(c)} Shortening leaf parts that intersect other leaf parts. \emph{(d)} Locating attachment points between parts. \emph{(e)} Forming leaf parts into symmetry groups.} 
\label{fig:progextract}
\end{figure*}

\subsubsection*{Semantics}

\dslname~has imperative semantics: every line of the program immediately takes effect and alters the state of the shape being constructed. Figure~\ref{fig:semantics} shows an example of a simple shape being imperatively constructed.
Declaring a cuboid instantiates a new piece of cuboid geometry with the requested dimensions, centered at the origin.
Invoking the \texttt{attach} command alters the cuboid, potentially translating, rotating, or resizing it in order to satisfy the attachment (see Appendix~\ref{sec:attach_semantics} for details). 
Higher-level macros expand into two or more \texttt{Cuboid} or \texttt{attach} lines, which are then immediately executed (see Appendix \ref{sec:macro_expansion} for details).

One distinct advantage of this imperative semantics, as opposed to an alternative formulation in which the program specifies constraints which are jointly optimized, is that the entire process of executing a program is end-to-end differentiable.
That is, it is possible to compute the gradient of the program's output geometry with respect to the continuous parameters in the text of the program (e.g., cuboid dimensions, attachment point locations).
We make use of this feature in results shown later in this paper.
\subsubsection*{Handling hierarchy}
Thus far, we have described a language that can generate flat assemblies of parts, but not hierarchical ones.
The extension to hierarchical shapes is straightforward: we represent hierarchical shapes by treating select non-leaf cuboids as the bounding box of another program (e.g., the contents of its ``BBlock'').
Figure~\ref{fig:dsl_examples} shows an example of a program in which cuboids expand into sub-programs. 

\section{Turning Shapes into Training Programs}
\label{sec:progextraction}

\dslname~allows us to write programs that generate new shapes.
However, we are interested in using the language to represent existing shapes in a dataset, so that we can learn to generate novel instances from the same underlying shape distribution. In this section, we describe how we accomplish this goal. Given an input shape, represented as a hierarchical part graph, the process divides into three steps: extracting program information, creating candidate programs, and checking program validity.  

\subsection{ Extracting Program Information}

To convert hierarchical part graphs into~\dslname~programs, we perform a series of data regularizations, record cuboid parameters, locate cuboid-to-cuboid attachments, and identify symmetry groups (Figure~\ref{fig:progextract}). We provide a high-level overview of the steps involved here, and a detailed description in Appendix \ref{sec:detailed_prog_extract}.

\subsubsection*{Regularization} Before we parse program attributes, we attempt to create more regularized part graphs through a series of data-cleaning steps. For instance, in the flattening phase, we restructure the part graph hierarchy so that leaf parts with spatial relationships are more often siblings. In the shortening phase, we decrease the dimensions of leaf cuboids that interpenetrate other leaf cuboids (to create more surface-to-surface part connections).

\subsubsection*{Cuboids} Ground truth cuboid dimensions are provided in the input part graphs. A cuboid is marked as aligned if its orientation matches its parent cuboid (with an allowable error of 5-degrees).

\subsubsection*{Attachment} To locate cuboid-to-cuboid attachments, we sample a uniform, dense point cloud on each cuboid in the scene. For each pair of cuboids, we compute the intersection of the point clouds. If the intersection set is non-zero, we record an attachment point within the volume formed by the intersection, with preference for locations on the centers of faces. For every cuboid, we then check if any of its parsed attachments could be represented as a squeeze relationship, and replace any that can.

\subsubsection*{Symmetry} To find symmetry groups, we identify collections of cuboids that share a reflectional or translational relationship about either the X, Y, or Z axis of their parent cuboid. For each collection, if all of the member cuboids have the same connectivity relationships, we form them into a symmetry group. Each symmetry group is represented by a transform applied to a single cuboid, and all other members are removed from the graph.

\subsection{Creating Candidate Programs}

Given the extracted program information, we know the content of the program, but not how the lines should be ordered. To make the task of learning a generative model of programs easier, we aim to extract only a single, ``canonical'' program for each shape. As the ordering of cuboid and symmetry lines doesn’t change the executed geometry, this consistency is enforced by ordering these lines according to the semantic label of each part involved in the line. Ties in this ordering between same part-type cuboids are broken by sorting on centroid position. 

Deciding on a single ordering of the attach and squeeze statements is more challenging.
Since~\dslname~has an imperative execution semantics, the order in which these commands are executed is significant: different orderings can potentially create different output geometries. To reduce the space of possible orderings, we only consider programs which follow a \emph{grounded} attachment order, which we define as follows:

\begin{itemize}
\denselist
    \item Initially, only the shape bounding box is grounded.
    \item The only valid attachments to perform are those which connect a cuboid to a grounded cuboid.
    \item After executing an attachment, the newly-attached cuboid becomes grounded.
\denselist
\end{itemize}

If there are multiple valid grounding orders, we first discard any orderings that produce worse geometric fits to the target shape. 
If ambiguities in the attachment ordering still remain, we break ties using (1) the semantic ordering of the cuboids involved in the attachment (2) preferring attachments from non-aligned to aligned cuboids and finally (3) preferring attachments from cuboid face-centers. 

\subsection{Validating Programs}
Once we extract a canonical ~\dslname~ program, we perform a series of checks to verify the results of our procedure. Programs must pass the following validation steps in order to be added to our training data:
\subsubsection*{Reconstruction} Executed programs should recreate the geometry of their respective ground truth part graph. To verify this, we sample point clouds from the surfaces of the ground truth shape and the geometry generated by executing the canonical program. These point clouds are compared using the F-score~\cite{TanksAndTemples} metric; a program is filtered out if it produces an F-score lower than 75.
\subsubsection*{Semantics} Programs must respect the semantics of ~\dslname. For instance, within each program, the connectivity graph of all parts should have only one component. Likewise, executed programs should not create geometry that extends beyond the bounding volumes they define.
\subsubsection*{Complexity} Programs that are overly complex (more than 12 leaf cuboid instantiations) are discarded.
Note that, when executed, programs can still produce more than 12 leaf cuboids through expanding symmetry macros.

\section{Learning to Generate Programs}
\label{sec:generation}

Given the programs extracted from our dataset, we now have the data we need to train a neural network to write novel hierarchical~\dslname~programs for us.
In this section, we describe the generative model architecture we use, our learning procedure, and how we sample new shapes from the learned model.

\subsection{Model Architecture}
\begin{figure*}[t!]
  \centering
  \includegraphics[width=\linewidth]{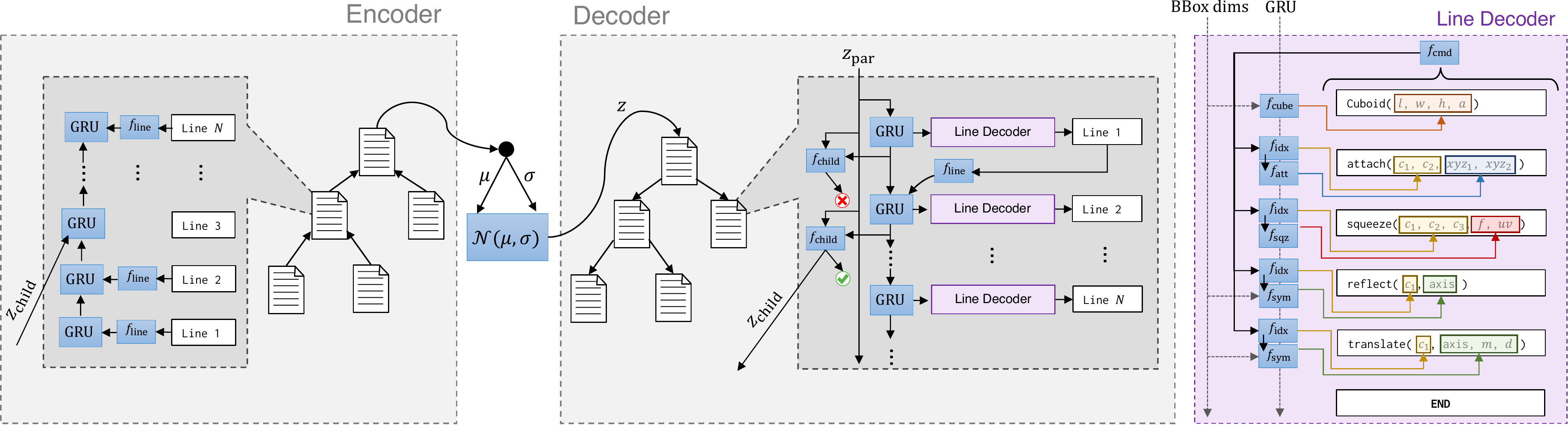}
\caption{
Architecture of our hierarchical sequence VAE for ~\dslname~ programs.
Given a ~\dslname~ program, the encoder ascends the hierarchy from the leaves to the root, encoding each sub-program into a latent $z$ vector. Given a latent code, the decoder recursively decodes a hierarchical ~\dslname~ program.
Within each hierarchy node, a recurrent neural network decodes each line of the program.
}
\label{fig:generative_model}
\end{figure*}

Figure~\ref{fig:generative_model} shows our generative model architecture. It is a hierarchical sequence VAE. The encoder branch embeds a hierarchical ~\dslname~ program into a latent space. The decoder branch converts a point in this latent space into a hierarchical~\dslname~program. The bottleneck of our network is parameterized by separate $\mu$ and $\sigma$ vectors in the standard variational autoencoder (VAE) setup. 

The dark grey callout in Figure~\ref{fig:generative_model} illustrates the operation of our decoder within a single node of the program hierarchy.
The decoder receives as input the latent code $z_\text{par}$ of its parent node (or the root latent code from the encoder, if it is the root node of the hierarchy).
This latent code is used to initialize the hidden state of a Gated Recurrent Unit (GRU), a recurrent language model which is responsible for constructing a representation of the program state. 
The output of the GRU cell is sent to the line decoder sub-routine, which predicts a line in the ~\dslname~ grammar, that is then passed as input back to the GRU cell at the next time step.

The purple callout in Figure~\ref{fig:generative_model} gives a detailed depiction of the line decoder sub-routine.
The line decoder receives the hidden state of the GRU cell, along with conditioning information about the size of the current bounding volume, and uses a collection of multilayer perceptrons (MLPs) to predict a 63-dimensional vector representing a single line in ~\dslname~. The sub-networks it uses are:

\begin{itemize}
\denselist
    \item \textbf{$f_\text{cmd}$: (7):} Predicts the type of command to execute. This is a one-hot vector whose seven entries correspond to \texttt{<start>} (the special program start token), \texttt{<stop>} (the special program stop token), \texttt{Cuboid}, \texttt{attach}, \texttt{squeeze}, \texttt{translate} and \texttt{reflect}.
    \item \textbf{$f_\text{cube}$: (4):} Predicts the length, width, height, and aligned flag for $\texttt{cuboid}$ lines, conditioned on the bounding volume dimensions.
    \item \textbf{$f_\text{idx}$: (11 $\times$ 3):} Predicts the indices of the cuboids involved in the line represented as 3 one-hot vectors, conditioned on the predicted command. We limit each node in the hierarchy to contain at most 10 children parts, so there are 11 choices (10 cuboids and the bounding volume). 
    \item \textbf{$f_\text{att}$: (3 $\times$ 2):} Predicts the $(x, y, z)$ coordinates involved in an $\texttt{attach}$ line, conditioned on the cuboids involved in the attach.
    \item \textbf{$f_\text{sqz}$: (8):} Predicts the the face involved in a $\texttt{squeeze}$ line as a one-hot vector in the first 6 indices. The last 2 indices predict the  $(u, v)$ coordinates. Both predictions are conditioned on the cuboids involved in the squeeze operation.
    \item \textbf{$f_\text{sym}$: (5):} Predicts the axis involved in a symmetry line as a one-hot vector in the first 3 indices. For $\texttt{translate}$ lines, the 4th index is the number of cuboids involved in the symmetry group, and the 5th index is the scale of the symmetry. All predictions are conditioned on the cuboid involved in the symmetry and the bounding volume dimensions.
\denselist
\end{itemize}

\subsubsection*{Hierarchical decoding}
To generate a hierarchical program, our decoder also includes a submodule $f_\text{child}$ which is executed after every predicted \texttt{Cuboid} command to determine whether that cuboid should be recursively expanded.
This is another MLP which takes as input both the current hidden state of the GRU as well as $z_\text{par}$, the overall latent code for this hierarchy node.
$f_\text{child}$ produces two outputs: a Boolean flag for whether the current cuboid should be expanded into a child program, and a new latent code $z_\text{child}$ which is used to initialize the decoder for this child program.

\subsection{Learning Procedure}

We implement our models in PyTorch\cite{paszke2017automatic}. All training is done with the Adam optimizer \cite{Kingma2014AdamAM}, with a learning rate of 0.0001 without batching. 

All multilayer perceptrons have 3 layers and use leaky ReLU~\cite{Maas2013RectifierNI} with $\alpha$ = 0.2. 

We train our model in a seq2seq fashion, where the ground truth input sequence is teacher forced to the model, and our model is tasked with predicting each subsequent line. During training, we use a program reconstruction loss that only considers entries of the predicted 63 dimensional vector that are relevant to the target line. For instance, when predicting a \texttt{Cuboid} line, no part of the reconstruction loss comes from the indices in the tensor associated with symmetry. The program reconstruction loss is comprised of a cross-entropy component for each one-hot prediction (with weight 1) and an l1 loss for each continuous component (with weight 50). Additionally we use a KL loss in the standard VAE setup with weight 0.1~\cite{kingma2014auto}.

\subsubsection*{Enforcing semantically-valid output}

As our model generates shape \emph{programs}, rather than raw shape geometry, we can use the semantics of the~\dslname~language to detect outputs that would be invalid, and prevent them from happening. For instance, attaches must be made in a grounded order. If a predicted \texttt{attach} line violates such a constraint, we use a backtracking procedure to find new `valid' parameter values whenever possible. During unconditional generation, if we cannot fix the line through backtracking, we reject the sample. During interpolation, if we cannot fix the line through backtracking we don't add the predicted line to the program.  Appendix~\ref{sec:validity_checks} describes the complete semantic validity procedure we enforce. We also note that this approach to forbidding the generation of invalid outputs is similar to that of the Grammar Variational Autoencoder~\cite{GrammarVAE}. However, that model only uses grammar \emph{syntax} to determine whether an output is valid, whereas as we use program \emph{semantics}.

%% file: 05-results.tex
\section{Results and Evaluation}
\label{sec:results}

\begin{figure*}[t!]
  \centering
  \setlength{\tabcolsep}{0.5pt}
  \small
    \begin{tabular}{c cc cc cc cc cc cc cc cc}
        \\
        \raisebox{2em}{\rotatebox{90}{Geom NN}} &
        \multicolumn{2}{c}{\includegraphics[trim={4cm 4cm 4cm 4cm},clip,width=.12\linewidth]{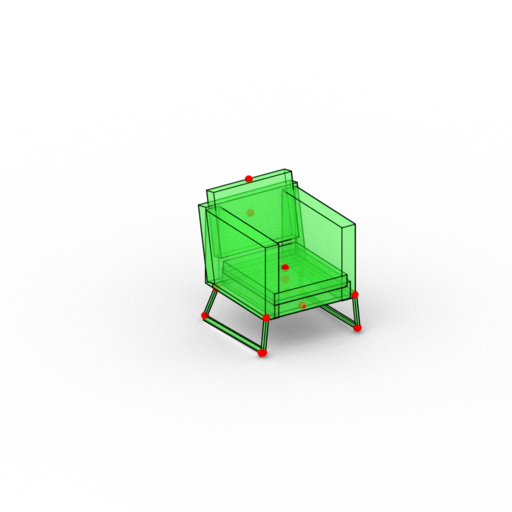}} &
        \multicolumn{2}{c}{\includegraphics[trim={4cm 4cm 4cm 4cm},clip,width=.12\linewidth]{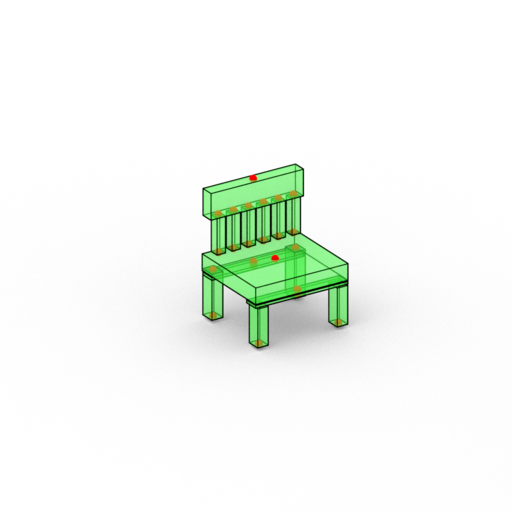}} &
        \multicolumn{2}{c}{\includegraphics[trim={4cm 4cm 4cm 4cm},clip,width=.12\linewidth]{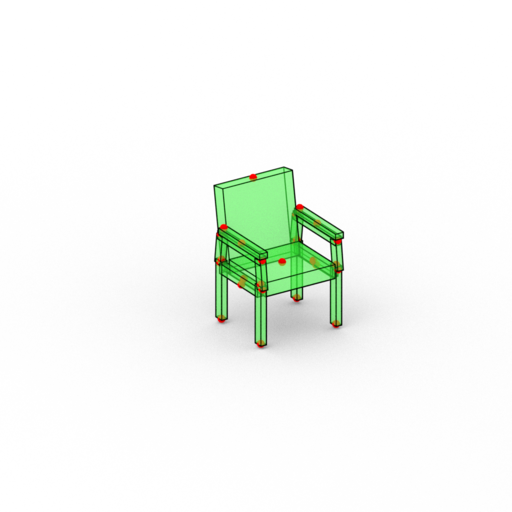}} &
        \multicolumn{2}{c}{\includegraphics[trim={4cm 4cm 4cm 4cm},clip,width=.12\linewidth]{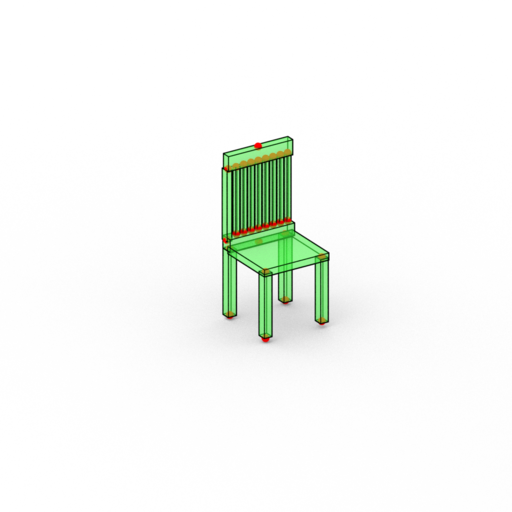}} &
        \multicolumn{2}{c}{\includegraphics[trim={4cm 4cm 4cm 4cm},clip,width=.12\linewidth]{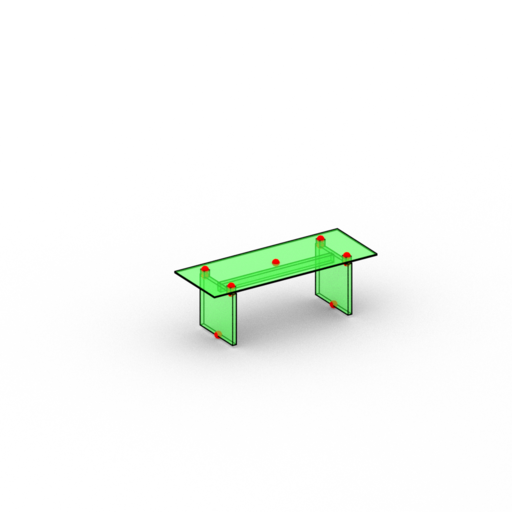}} &
        \multicolumn{2}{c}{\includegraphics[trim={4cm 4cm 4cm 4cm},clip,width=.12\linewidth]{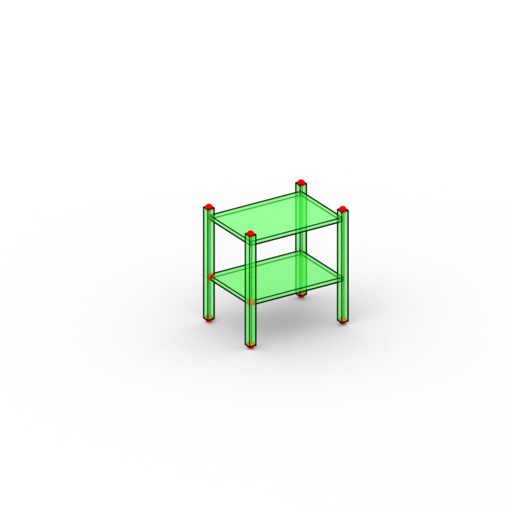}} &
        \multicolumn{2}{c}{\includegraphics[trim={4cm 4cm 4cm 4cm},clip,width=.12\linewidth]{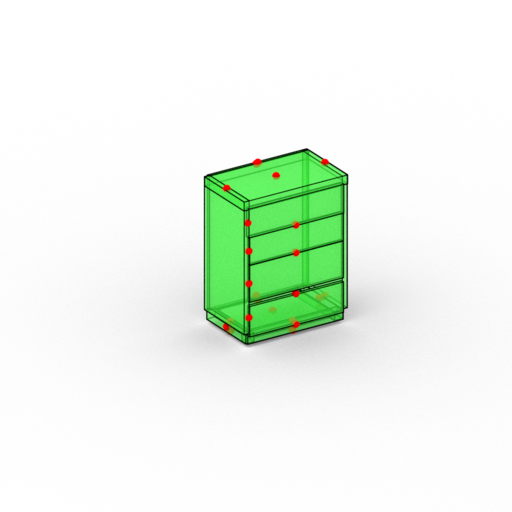}} &
        \multicolumn{2}{c}{\includegraphics[trim={4cm 4cm 4cm 4cm},clip,width=.12\linewidth]{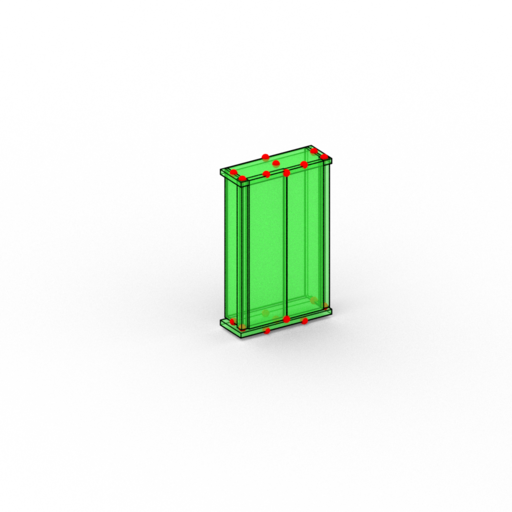}}
        \\
        \raisebox{2em}{\rotatebox{90}{Sample}} &
        \multicolumn{2}{c}{\includegraphics[trim={4cm 4cm 4cm 4cm},clip,width=.12\linewidth]{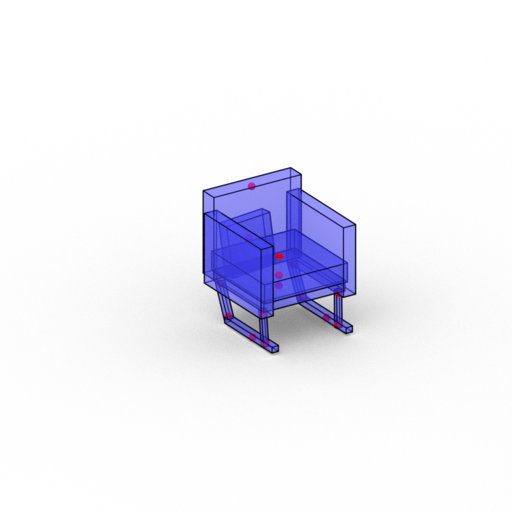}} &
        \multicolumn{2}{c}{\includegraphics[trim={4cm 4cm 4cm 4cm},clip,width=.12\linewidth]{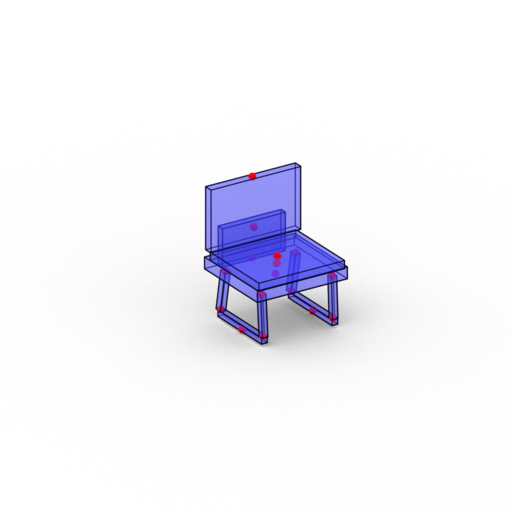}} &
        \multicolumn{2}{c}{\includegraphics[trim={4cm 4cm 4cm 4cm},clip,width=.12\linewidth]{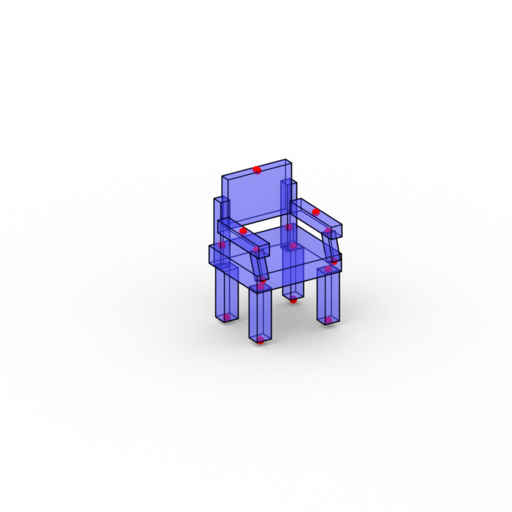}} &
        \multicolumn{2}{c}{\includegraphics[trim={4cm 4cm 4cm 4cm},clip,width=.12\linewidth]{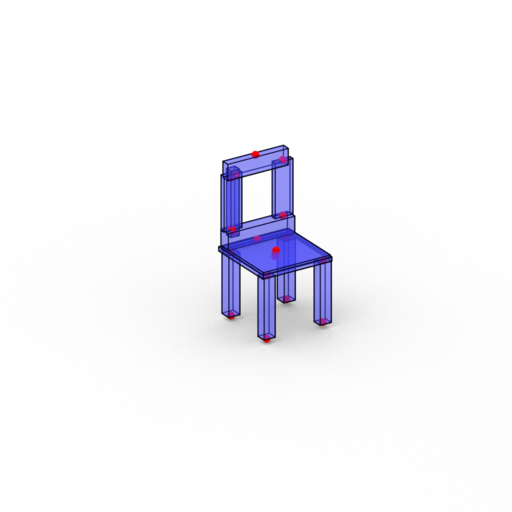}} &
        \multicolumn{2}{c}{\includegraphics[trim={4cm 4cm 4cm 4cm},clip,width=.12\linewidth]{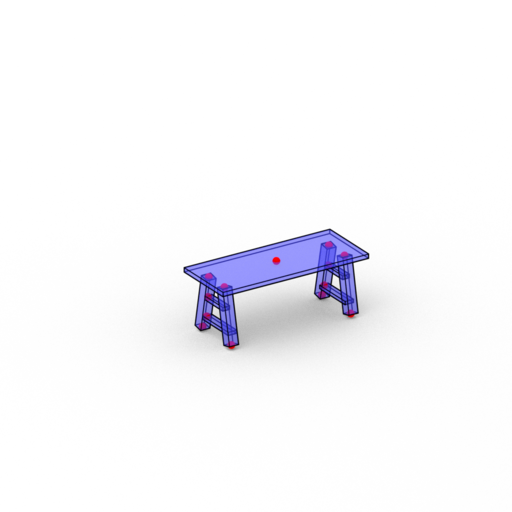}} &
        \multicolumn{2}{c}{\includegraphics[trim={4cm 4cm 4cm 4cm},clip,width=.12\linewidth]{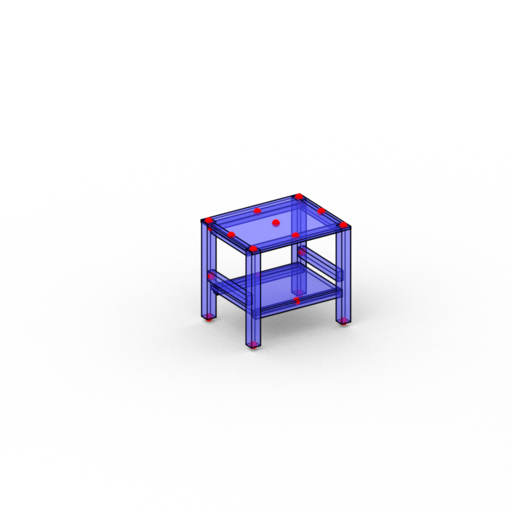}} &
        \multicolumn{2}{c}{\includegraphics[trim={4cm 4cm 4cm 4cm},clip,width=.12\linewidth]{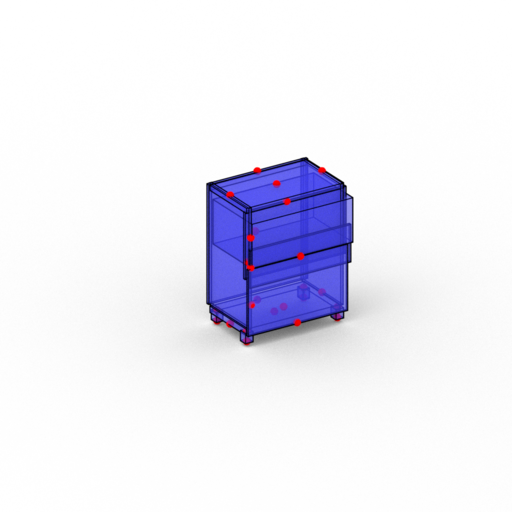}} &
        \multicolumn{2}{c}{\includegraphics[trim={4cm 4cm 4cm 4cm},clip,width=.12\linewidth]{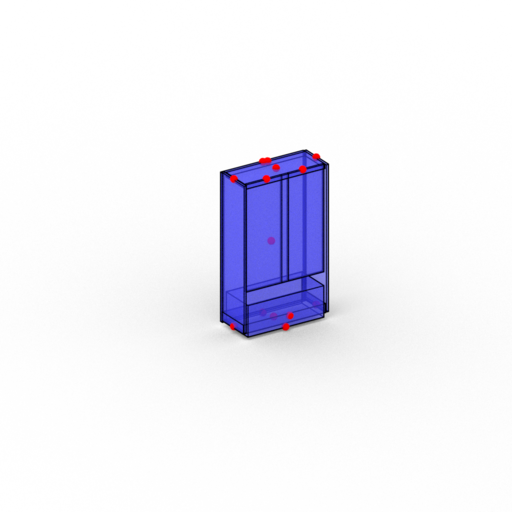}}
        \\
        \raisebox{2em}{\rotatebox{90}{Prog NN}} &
        \multicolumn{2}{c}{\includegraphics[trim={4cm 4cm 4cm 4cm},clip,width=.12\linewidth]{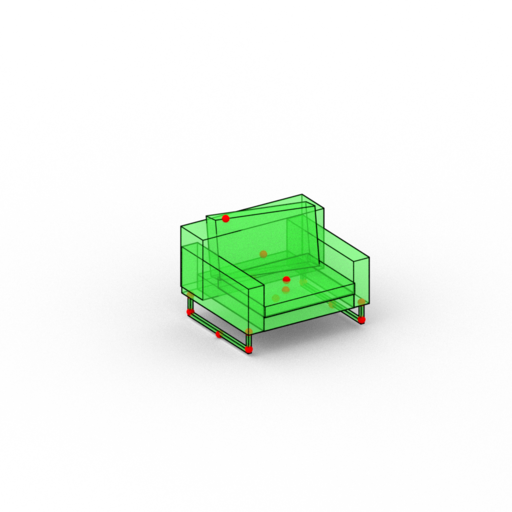}} &
        \multicolumn{2}{c}{\includegraphics[trim={4cm 4cm 4cm 4cm},clip,width=.12\linewidth]{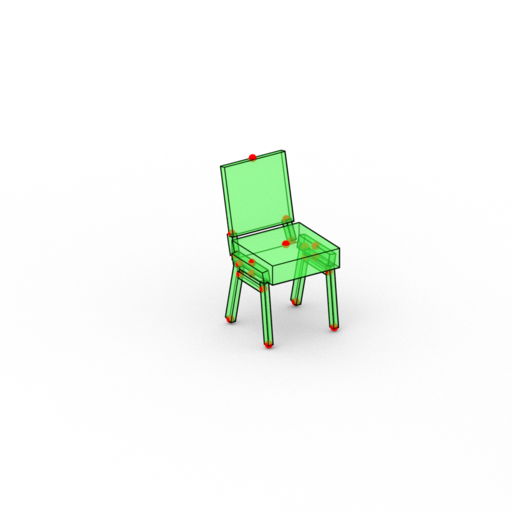}} &
        \multicolumn{2}{c}{\includegraphics[trim={4cm 4cm 4cm 4cm},clip,width=.12\linewidth]{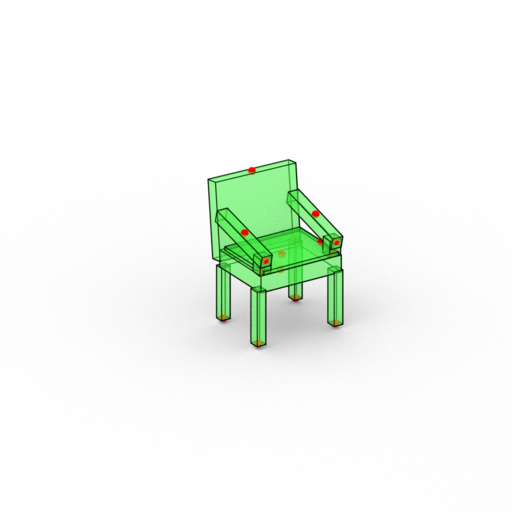}} &
        \multicolumn{2}{c}{\includegraphics[trim={4cm 4cm 4cm 4cm},clip,width=.12\linewidth]{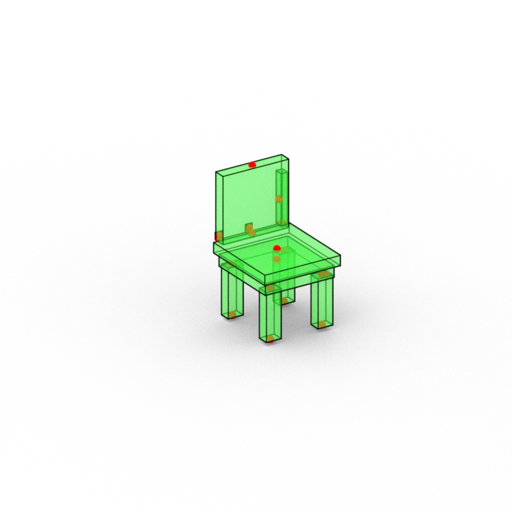}} &
        \multicolumn{2}{c}{\includegraphics[trim={4cm 4cm 4cm 4cm},clip,width=.12\linewidth]{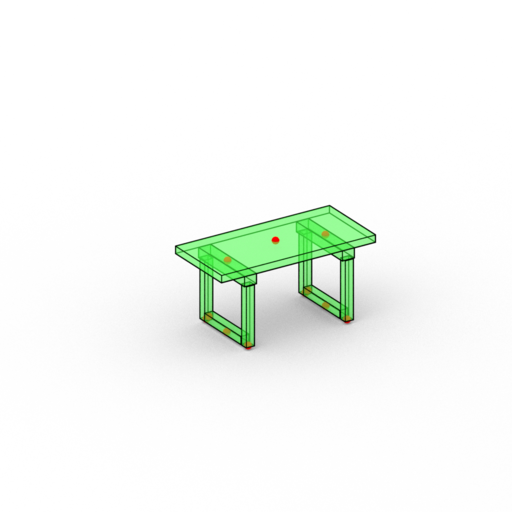}} &
        \multicolumn{2}{c}{\includegraphics[trim={4cm 4cm 4cm 4cm},clip,width=.12\linewidth]{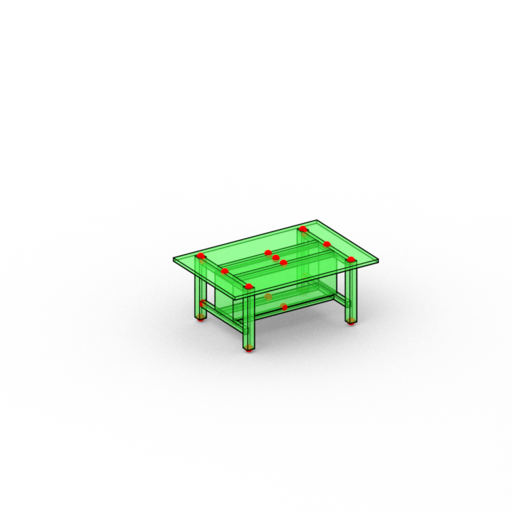}} &
        \multicolumn{2}{c}{\includegraphics[trim={4cm 4cm 4cm 4cm},clip,width=.12\linewidth]{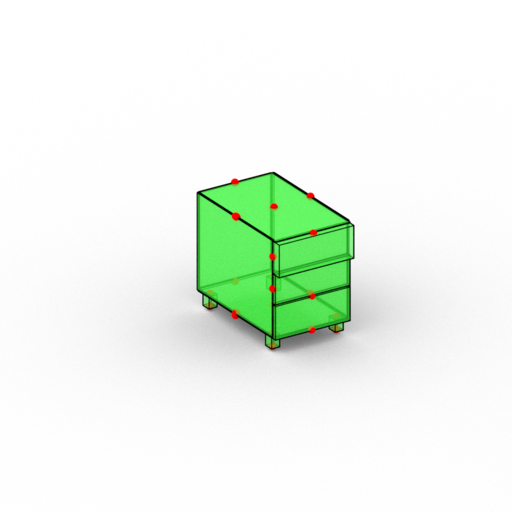}} &
        \multicolumn{2}{c}{\includegraphics[trim={4cm 4cm 4cm 4cm},clip,width=.12\linewidth]{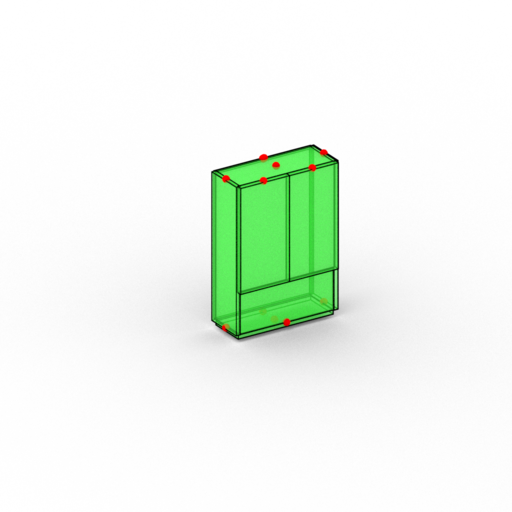}}
      
    \end{tabular}
\caption{ In the middle row, we show samples from our generative model of~\dslname~programs. In the top row, we show the nearest neighbor shape in the training set by Chamfer distance. In the bottom row, we show the nearest neighbor shape in the training set by program edit distance. Our method synthesizes interesting and high-quality structures that go beyond direct structural or geometric memorization. We quantitatively examine ~\dslname's generalization in Table~\ref{tab:variability}. Refer to the supplemental material for the corresponding program text. 
}
\label{fig:nn_qual}
\end{figure*}

In this section, we demonstrate our learned generative model's ability to synthesize high-quality hierarchical~\dslname~programs, and we compare it to alternative generative models of 3D shape structure. All of the experiments described were run on a GeForce RTX 2080 Ti GPU with an Intel i9-9900K CPU, and consumed 3GB of GPU memory.

We use objects from the PartNet dataset~\cite{PartNet} as our training data.
It contains 3D shapes in multiple categories, each with a hierarchical part segmentation and labeling.
For the experiments in this paper, we use the \emph{Chairs}, \emph{Tables}, and \emph{Storage} categories. After running the extraction procedure described in Section \ref{sec:progextraction}, we obtain 3835 \emph{Chair}, 6536  \emph{Table}, and 1551 \emph{Storage} ground truth programs.
\subsection{Novel Shape Synthesis}
In this section, we present both qualitative and quantitative evaluations of our method's ability to produce novel shape structures. 
Figure~\ref{fig:nn_qual} includes some unconditionally generated samples from our learned generative model for each of the three shape categories. 
Above each sample we show its nearest neighbor in the training data based on Chamfer distance. 
Additionally, below each sample we visualize its nearest neighbor in the training data based on program distance, the string edit distance of a tokenized version of our hierarchical programs. 
As shown, our method is able to generate complex and interesting structural variation without copying either the geometry or program structure of its training data. 

\begin{figure}[b!]
  \centering
  \setlength{\tabcolsep}{0.5pt}
  \small
    \begin{tabular}{c cc cc cc cc}
        \raisebox{2em}{\rotatebox{90}{Sample}} &

        \multicolumn{2}{c}{\includegraphics[trim={4cm 4cm 4cm 4cm},clip,width=.24\linewidth]{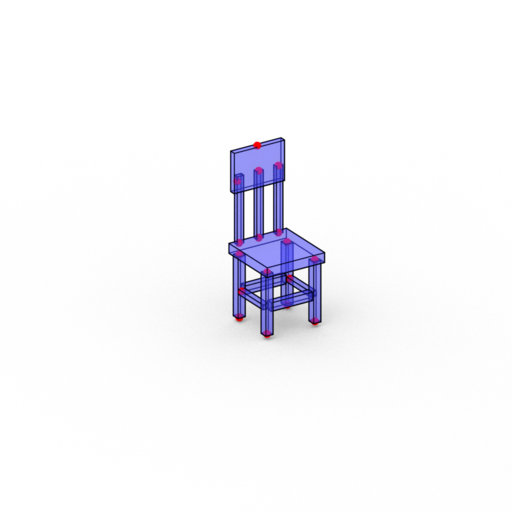}} &
        \multicolumn{2}{c}{\includegraphics[trim={4cm 4cm 4cm 4cm},clip,width=.24\linewidth]{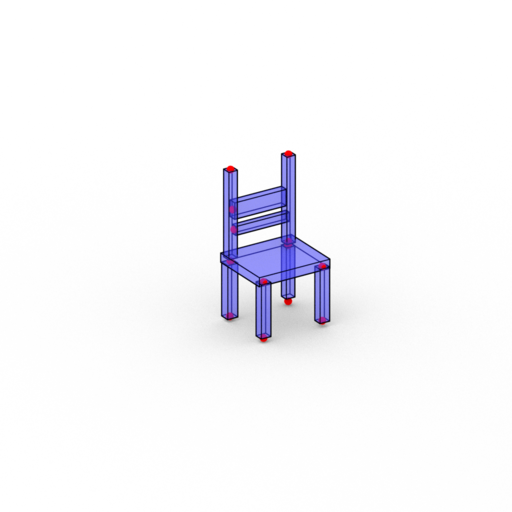}} &
        \multicolumn{2}{c}{\includegraphics[trim={4cm 4cm 4cm 4cm},clip,width=.24\linewidth]{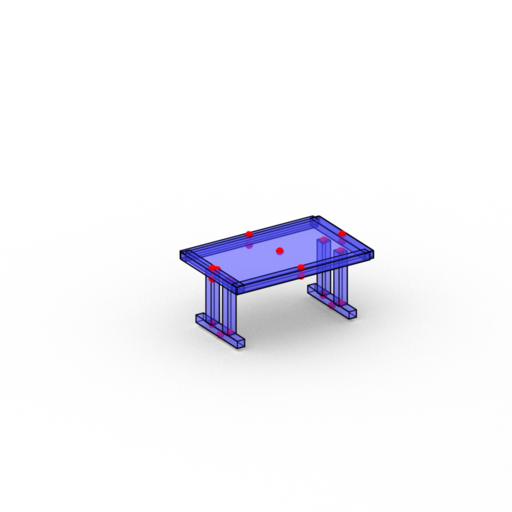}} &
        \multicolumn{2}{c}{\includegraphics[trim={4cm 4cm 4cm 4cm},clip,width=.24\linewidth]{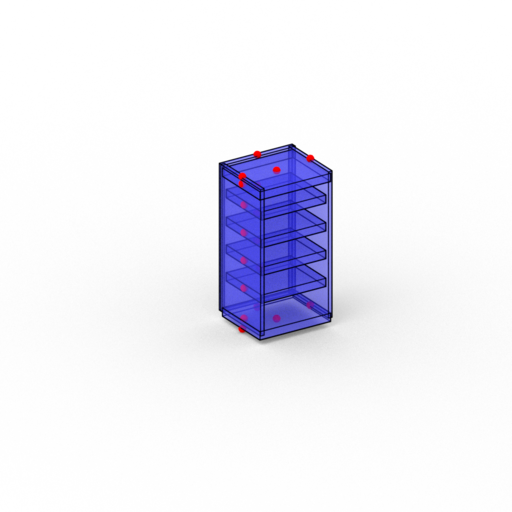}} 
        \\
        \raisebox{2em}{\rotatebox{90}{Variant}} &
        \multicolumn{2}{c}{\includegraphics[trim={4cm 4cm 4cm 4cm},clip,width=.24\linewidth]{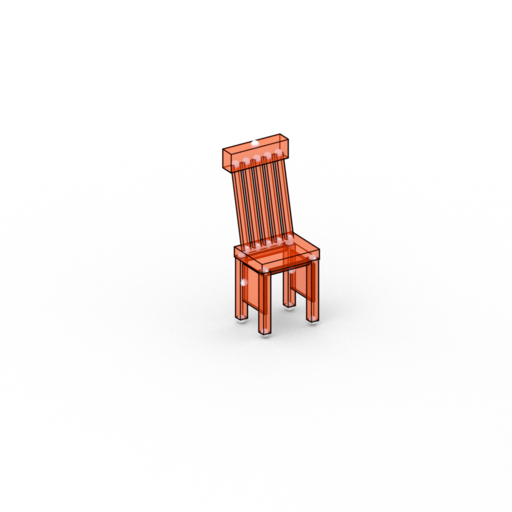}} &
        \multicolumn{2}{c}{\includegraphics[trim={4cm 4cm 4cm 4cm},clip,width=.24\linewidth]{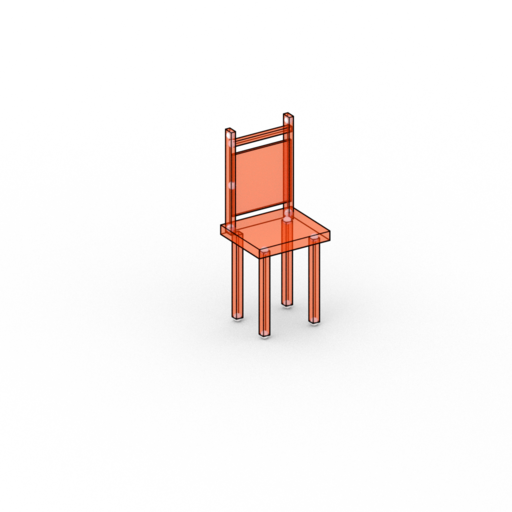}} &
        \multicolumn{2}{c}{\includegraphics[trim={4cm 4cm 4cm 4cm},clip,width=.24\linewidth]{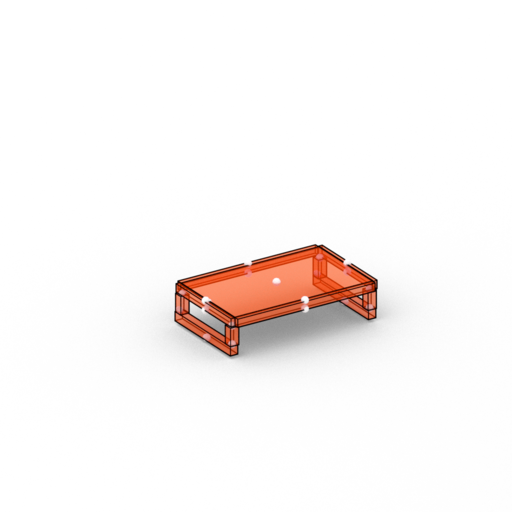}} &
        \multicolumn{2}{c}{\includegraphics[trim={4cm 4cm 4cm 4cm},clip,width=.24\linewidth]{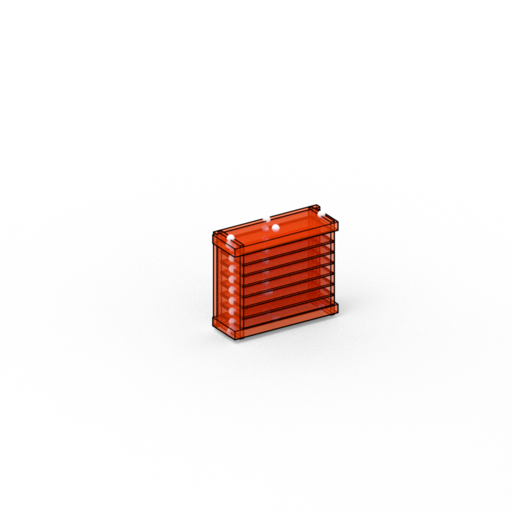}}
      
    \end{tabular}
\caption{Programs, by way of representational form, allow for easy semantic editing of generated output. Each column shows a sample from our model in the top row. In the bottom row we create a variant with the same structure, but different geometry, by editing only the continuous parameters of the program. Program text can be found in the supplemental material. 
} 
\label{fig:edit_qual}
\end{figure}

As our model directly generates programs, its outputs can be easily edited to produce variants. In Figure~\ref{fig:edit_qual} we demonstrate how by changing just the continuous parameters of programs generated by our model, we are able to create a wide variety of output geometry, all the while maintaining part-to-part attachment relationships.

We compare the generated results of our method against two baselines: 
\begin{itemize}
\denselist
\item \textbf{StructureNet} is a variational autoencoder that generates hierarchical part graphs with cuboids at each node ~\cite{StructureNet}.
\item \textbf{3D-PRNN} is a recurrent neural network that generates a sequence of cuboids ~\cite{zou20173d}. It enforces global bilateral symmetry by only generating cuboids with some part of their geometry on the negative side of the $x = 0$ plane, and then reflecting generated cuboids which fall entirely on that side of the plane.
\end{itemize}

We compare against the StructureNet models released by the authors. 
These were trained on the subset of PartNet that they were able to represent within the constraints of their problem formulation.
This is a heavily overlapping set, but not identical, with the shapes we were able to find valid ~\dslname~ programs for. 
In direct comparisons with StructureNet for reconstruction tasks, we only consider shapes that appear in the validation splits of both methods. 
We compare against a version of 3D-PRNN that was re-trained on the data we use for our generative model. 
Figure~\ref{fig:qualitative comparison} shows a qualitative comparison of unconditionally generated samples from each method. Our method is capable of generating diverse, structurally complex, 3D shape structures across multiple categories. 
Attachment as a primary operation provides a strong inductive bias for generating physically plausible shapes that maintain realistic part-to-part relationships. In contrast, both comparison methods that directly predict part placements in 3D space are prone to producing floating cuboids or jumbled collections of spatially colocated 
parts.

\subsubsection{Analysis of Shape Quality}

We also quantitatively compare the quality of the shape structures generated by different methods. 
Our desiderata for generated shape structures is that they should be physically plausible and come from the same distribution that the model was trained on. 
In order to asses the quality of generated output, we use the following metrics: 

\begin{itemize}

\denselist
    \item \textbf{Rootedness $\Uparrow$ (\% rooted):} The percentage of shapes for which a connected path exists between the ground and all leaf parts.
    \item \textbf{Stability $\Uparrow$ (\% stable):} The percentage of shapes which remain upright under gravity and small forces in a physical simulation. 
    \item \textbf{Realism $\Uparrow$ (\% fool):} The percentage of test set shapes classified as ``generated'' by a PointNet classifier trained to distinguish between generated shapes and shapes from the training dataset. 
    \item \textbf{Frechet Distance $\Downarrow$ (FD):} Measurement of distributional similarity between generated shapes and the training dataset using the feature space of a pre-trained PointNet model~\cite{FrechetInceptionDistance}
\denselist
\end{itemize}
Further details about these metrics are provided in Appendix \ref{sec:quant_metrics}.

We show results for these metrics on 1000 unconditional generated shapes in Table~\ref{tab:quality}. Our method largely outperforms 3D-PRNN and StructureNet across these metrics for three categories of shapes. While StructureNet achieves good rootedness scores, especially for the Storage category, our method performs better in the other three metrics along all categories. The samples from 3D-PRNN, achieve similar FD and \% fool scores with StructureNet, but perform markedly worse on the rootedness and stability metrics.

Additionally in this experiment we compare our model with a series of ablated versions:

\begin{itemize}
\denselist
    \item \textbf{Flat:} Training on programs with no hierarchies, only leaf parts.
    \item \textbf{No Order:} Training on programs without canonical ordering as described in Section~\ref{sec:progextraction}.
    \item \textbf{No Align:} Training on programs without an aligned flag for cuboids.
    \item \textbf{No Macros:} Training on programs without \texttt{squeeze}, \texttt{translate}, or \texttt{reflect} commands.
    \item \textbf{No Reject:} At generation time, discard unfixable, invalid program line predictions instead of rejecting the entire sample. 
\denselist
\end{itemize}

Training without hierarchy (Flat) slightly improves rootedness, but drastically lowers the quality of output as seen in the \% fool and FD columns. Training on programs without a canonical ordering (No Order) performs worse on every metric.
Removing the alignment flag (No Align) actually improves performance on the Chair category for \% rooted and \% fool, but drastically worsens the physical validity of generations for Tables and Storage, categories where parts are much more often aligned with their parent cuboid. 
Training without macros (No Macros) once again decreases the performance of all metrics, but not by a substantial margin.
Finally, we see that while the rejection sampling step does improve the quality of our generated samples, without it we still outperform 3D-PRNN and StructureNet by a wide margin.

\begin{table}[t!]
    \centering
    \small
    % \footnotesize
    \setlength{\tabcolsep}{2pt}
    %\vspace{-1em}
    \begin{tabular}{@{}llcccc@{}}
        \toprule
        \textbf{Category} & \textbf{Method} & \textbf{\% rooted}$\;\Uparrow$ & \textbf{\% stable}$\;\Uparrow$ & \textbf{\% fool}$\;\Uparrow$ & \textbf{FD}$\;\Downarrow$ \\
        \midrule
        \multirow{9}{*}{\emph{Chair}}
        & 3D-PRNN & 73.1 & 50.9 & 12.60 & 39.30 \\
        & StructureNet & 89.7 & 74.9 & 4.04 & 64.79 \\
        & Ours (Flat) & \textbf{95.0} & 60.0 & 11.58 & 77.45 \\
        & Ours (No Order) & 82.4 & 58.4 & 12.36 & 64.17 \\
        & Ours (No Align) & 94.6 & 84.6 & \textbf{28.68} & 29.32 \\
        & Ours (No Macros) & 92.0 & 77.9 & 19.56 & 36.78 \\
        & Ours (No Reject) &  92.9 & 79.7 & 23.36 & \textbf{20.63} \\
        & Ours & 94.5 & \textbf{84.7} & 25.06 & 22.34 \\
        & Ground Truth & 100 & 88.0 & --- & --- \\
        \midrule
        \multirow{9}{*}{\emph{Table}}
        & 3D-PRNN & 71.2 & 29.4 & 2.12 & 140.07 \\
        & StructureNet & 94.4 & 76.8 & 3.94 & 173.35 \\
        & Ours (Flat) & 87.0 & 66.0 & 29.84 & 148.63 \\
        & Ours (No Order) & 84.5 & 56.0 & 27.38 & 114.10 \\
        & Ours (No Align) & 92.2 & 61.5 & 23.64 & \textbf{46.64} \\
        & Ours (No Macros) &  95.9 & 85.0 & 33.16 & 53.21 \\
        & Ours (No Reject) & 94.1 & 76.4 & 29.20 & 52.78 \\
        & Ours & \textbf{96.2} & \textbf{85.9} & \textbf{33.21} & 49.07 \\
        & Ground Truth & 100 & 93.1 & --- & --- \\
        \midrule
        \multirow{9}{*}{\emph{Storage}}
        & 3D-PRNN & 44.8 & 20.8 & 4.62 & 94.08 \\
        & StructureNet & \textbf{96.2} & 75.0 & 5.04 & 92.85 \\
        & Ours (Flat) & 95.9 & 74.0 & 7.44 & 81.17 \\
        & Ours (No Order) & 87.9 & 63.4 & 8.70 & 107.42 \\
        & Ours (No Align) & 89.7 & 49.3 & 11.04 & \textbf{30.15} \\
        & Ours (No Macros) & 87.5 & 69.9 & 5.92 & 72.80 \\
        & Ours (No Reject) & 94.3 & 80.9 & 11.66 & 31.69 \\
        & Ours & 95.3 & \textbf{83.7} & \textbf{13.50} & 31.72 \\
        & Ground Truth & 100 & 87 & --- & --- \\
        \bottomrule
    \end{tabular}
    \caption{ Comparing the quality of generated samples. Our method outperforms other generative methods for 3D shape structure in terms of realism and physical validity. Through a series of ablation baselines, we validate various design decisions of our method.
    }
    \label{tab:quality}
\end{table}

\subsubsection{Analysis of Editability}

In this section, we quantitatively analyze our previous claim that directly predicting programs improves editability. We claim that a program is more editable if it is both compact and compromised of higher level functions. That is, a shorter program that uses higher-level constructs will be easier to understand and make changes to. 

As a strong baseline, we evaluate the editability of our programs against the generated outputs of 3D-PRNN and StructureNet. As 3D-PRNN and StructureNet do not directly produce ~\dslname~ programs, we use our extraction procedure described in Section \ref{sec:progextraction} in order to convert their generations into programs. As StructureNet predicts part graph hierarchies, the representational form our extraction procedure takes as input, we use our procedure without any of the data cleaning steps. As 3D-PRNN has no notion of hierarchy, we create single node part graphs out of their output samples, which are then run through our program extraction logic.

Table~\ref{tab:editability_compactness} shows results from an experiment where we compare the ~\dslname~ programs of each method's generations (directly predicted by our method, parsed programs from comparisons). The metrics we use are the number of lines in each program (as a coarse measure of compactness) and the percentage of lines which are macros (split by macro type). 

Compared with programs parsed from StructureNet, the programs generated by our model are much more compact and have higher rates of macro usage across all categories of shapes. While our method also has higher macro rate usage compared with 3D-PRNN, 3D-PRNN programs are more compact in the Chair and Table categories. Based on 3D-PRNN's poor performance within our shape quality experiments (Table~\ref{tab:quality}), and its significant deviation from the number of lines found in the ground truth programs (the cleanest set of ~\dslname~ programs we have access to), there is reason to believe that the compactness of its parsed programs more likely reflects shape simplicity rather than useful editability.

\begin{table}[t!]
    \centering
    \small
    % \footnotesize
    \setlength{\tabcolsep}{2pt}
    %\vspace{-1em}
    \begin{tabular}{@{}llccccc@{}}
        \toprule
        & & & \multicolumn{4}{c}{\rule[1.5pt]{2.5em}{0.5pt} Macros Per Line \rule[1.5pt]{2.5em}{0.5pt}}
        \\
        \textbf{Category} & \textbf{Method} & \textbf{Lines} $\Downarrow$ & \textbf{Refl} $\Uparrow$ & \textbf{Trans} $\Uparrow$ & \textbf{Squeeze} $\Uparrow$& \textbf{Total} $\Uparrow$
        \\
        \midrule
        \multirow{4}{*}{\emph{Chair}}
        & 3D-PRNN & \textbf{15.7} & \textbf{0.1100} & 0.0020 & 0.0240 & 0.1430 \\
        & StructureNet & 27.1 & 0.0600 & 0.0004 & 0.0700 & 0.1330 \\
        & Ours & 20.4 & 0.0880 & \textbf{0.0054} & \textbf{0.0920} & \textbf{0.1860} \\
        & Ground Truth & 24.4 & 0.0800 & 0.0090 & 0.1130 & 0.2070 \\
        \midrule
        \multirow{4}{*}{\emph{Table}}
        & 3D-PRNN &  \textbf{13.1} & \textbf{0.1300} & \textbf{0.0010} & 0.0680 & 0.1990 \\
        & StructureNet & 24.8 &  0.0270 & 0.0006 & 0.0620 & 0.0900 \\
        & Ours & 19.0 &  0.0990 & 0.0002 & \textbf{0.1440} & \textbf{0.2440} \\
        & Ground Truth & 20.0 & 0.0950 & 0.0050 & 0.1450 & 0.2460 \\
        \midrule
        \multirow{4}{*}{\emph{Storage}}
        & 3D-PRNN & 22.6 & 0.0170 & 0.0060 & 0.0530 & 0.0770 \\
        & StructureNet & 30.7 &  0.0390 & 0.0040 & 0.0770 & 0.1200 \\
        & Ours & \textbf{19.8} & \textbf{0.0820} & \textbf{0.0080} & \textbf{0.1440} & \textbf{0.2340} \\
        & Ground Truth & 24.7 & 0.0650 & 0.0147 & 0.1510 & 0.2320 \\
        \bottomrule
    \end{tabular}
    \small
    \caption{
     Markers of program editability for ~\dslname~ programs predicted by our generative model compared with  ~\dslname~ programs parsed from outputs of other generative methods. Training our model in the space of programs allows us to represent geometry more compactly. We find higher rates of macro functions per program line in our method's generations compared with extracting programs from other generative models' predictions. 
    }
    \label{tab:editability_compactness}
\end{table}

\begin{figure*}[t!]
    \centering
    \setlength{\tabcolsep}{1pt}
    \begin{tabular}{ccccc@{\hspace{2pt}}cccc@{\hspace{2pt}}ccc}
        % \multicolumn{11}{c}{\emph{Chair}}
        % \\
        & \multicolumn{3}{c}{3D-PRNN} && \multicolumn{3}{c}{Ours} && \multicolumn{3}{c}{StructureNet}
        \\
        \cmidrule{2-4} \cmidrule{6-8} \cmidrule{10-12}
        &       
        \includegraphics[trim={4.5cm 4.5cm 4.5cm 4.5cm},clip,width=.105\linewidth]{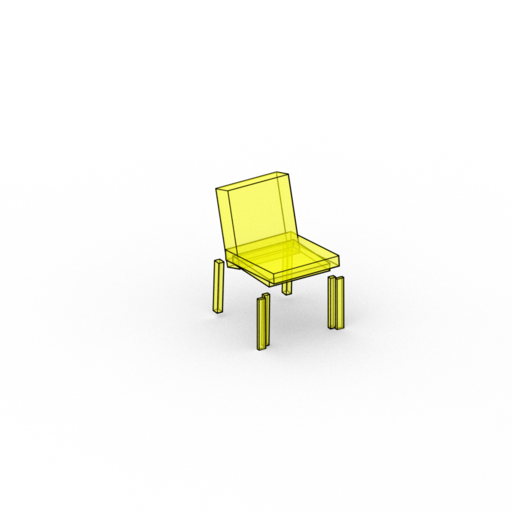} &
        \includegraphics[trim={4.5cm 4.5cm 4.5cm 4.5cm},clip,width=.105\linewidth]{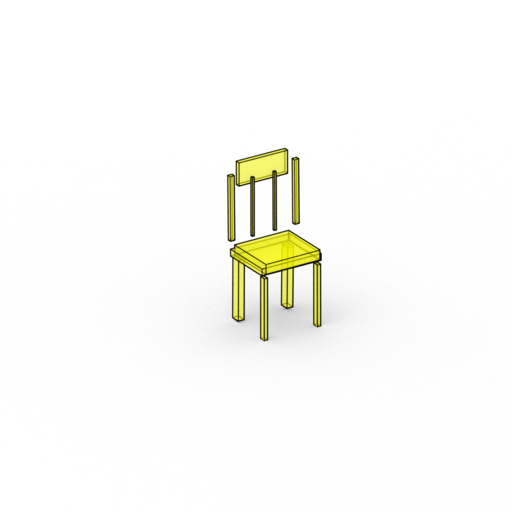} &
        \includegraphics[trim={4.5cm 4.5cm 4.5cm 4.5cm},clip,width=.105\linewidth]{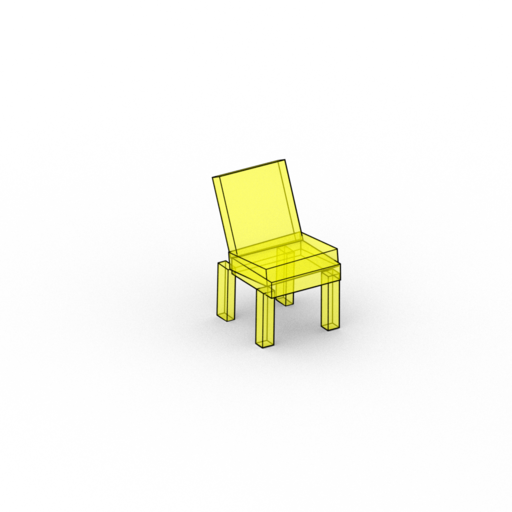} &&
        \includegraphics[trim={4.5cm 4.5cm 4.5cm 4.5cm},clip,width=.105\linewidth]{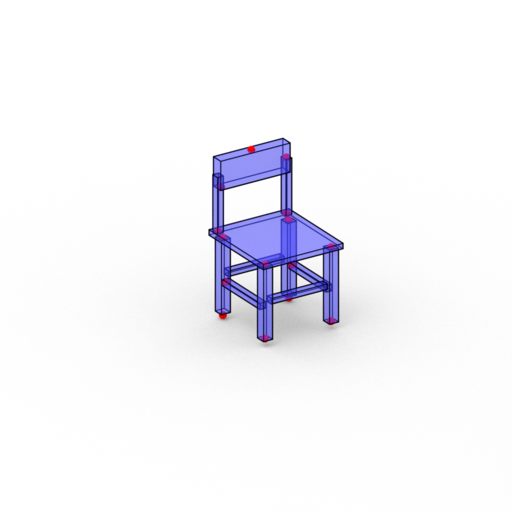} &
        \includegraphics[trim={4.5cm 4.5cm 4.5cm 4.5cm},clip,width=.105\linewidth]{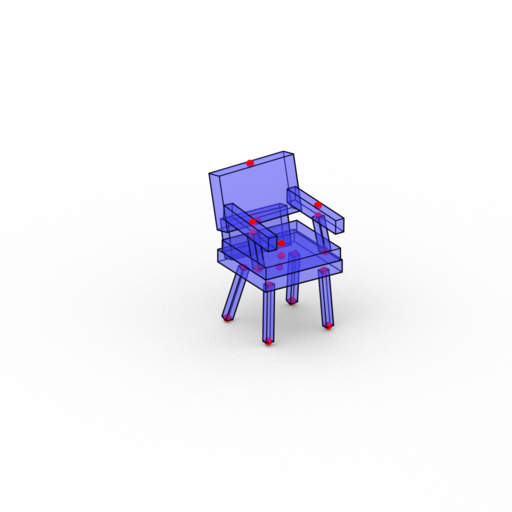} &
        \includegraphics[trim={4.5cm 4.5cm 4.5cm 4.5cm},clip,width=.105\linewidth]{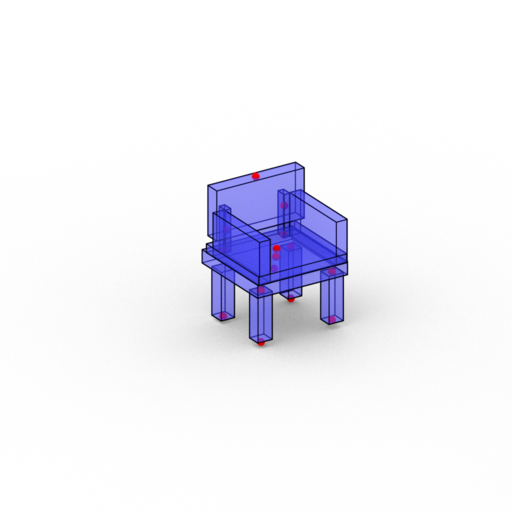} &&
        \includegraphics[trim={2.5cm 2.5cm 2.5cm 2.5cm},clip,width=.105\linewidth]{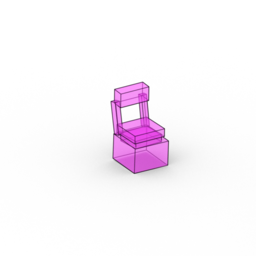} &
        \includegraphics[trim={2.5cm 2.5cm 2.5cm 2.5cm},clip,width=.105\linewidth]{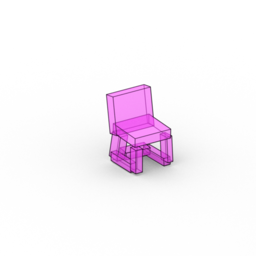} &
        \includegraphics[trim={2.5cm 2.5cm 2.5cm 2.5cm},clip,width=.105\linewidth]{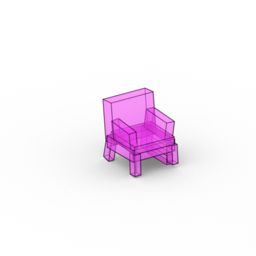}
        \\
        \raisebox{1.5em}{\rotatebox{90}{Chair}} &
        \includegraphics[trim={4.5cm 4.5cm 4.5cm 4.5cm},clip,width=.105\linewidth]{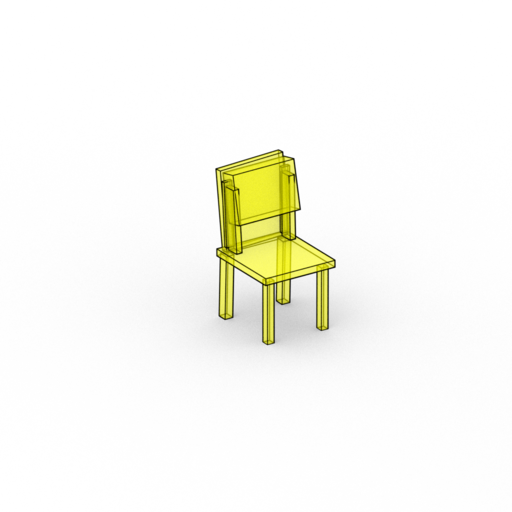} &
        \includegraphics[trim={4.5cm 4.5cm 4.5cm 4.5cm},clip,width=.105\linewidth]{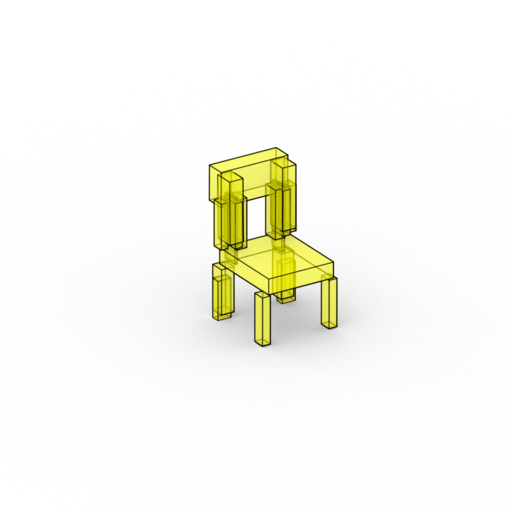} &
        \includegraphics[trim={4.5cm 4.5cm 4.5cm 4.5cm},clip,width=.105\linewidth]{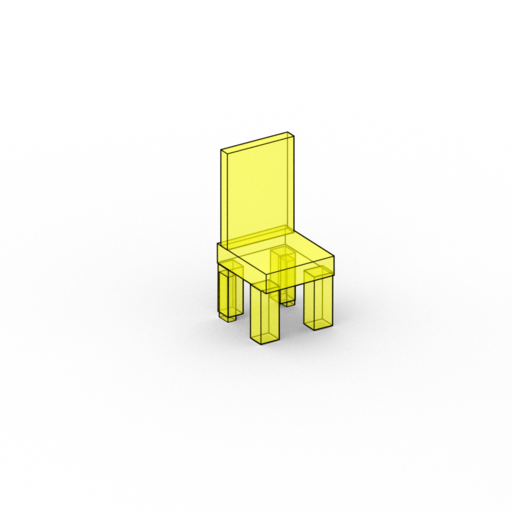} &&
        \includegraphics[trim={4.5cm 4.5cm 4.5cm 4.5cm},clip,width=.105\linewidth]{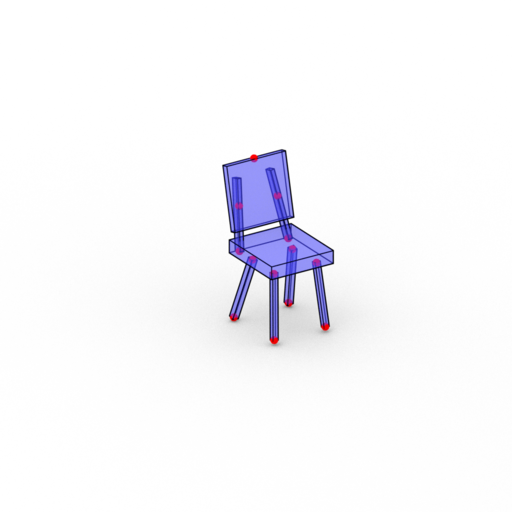} &
        \includegraphics[trim={4.5cm 4.5cm 4.5cm 4.5cm},clip,width=.105\linewidth]{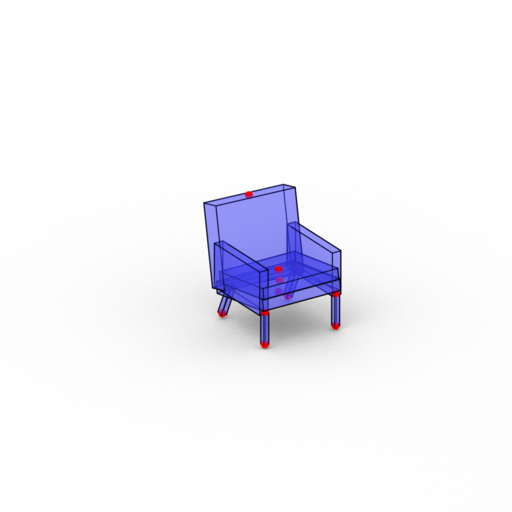} &
        \includegraphics[trim={4.5cm 4.5cm 4.5cm 4.5cm},clip,width=.105\linewidth]{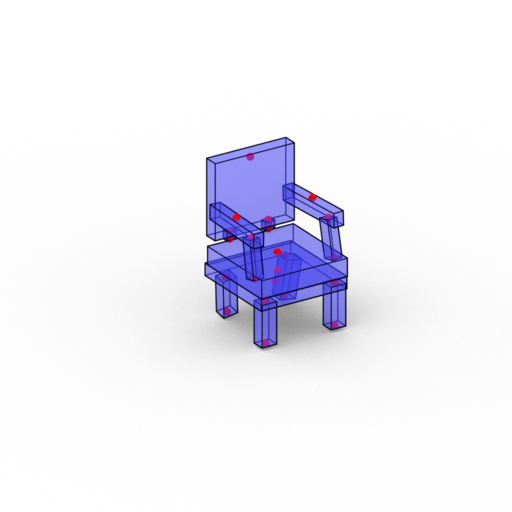} &&
        \includegraphics[trim={2.5cm 2.5cm 2.5cm 2.5cm},clip,width=.105\linewidth]{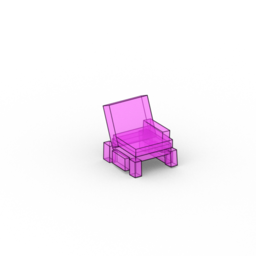} &
        \includegraphics[trim={4.5cm 4.5cm 4.5cm 4.5cm},clip,width=.105\linewidth]{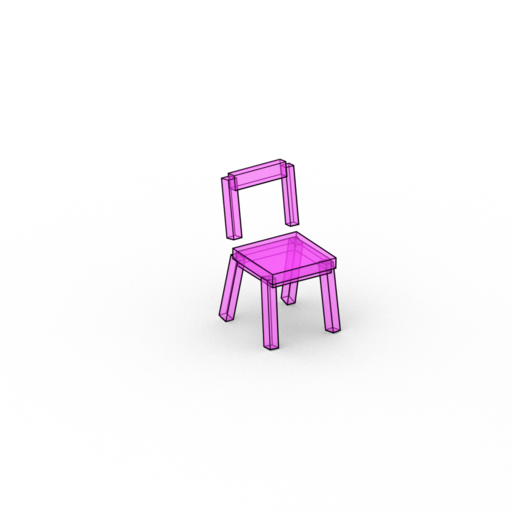} &
        \includegraphics[trim={2.5cm 2.5cm 2.5cm 2.5cm},clip,width=.105\linewidth]{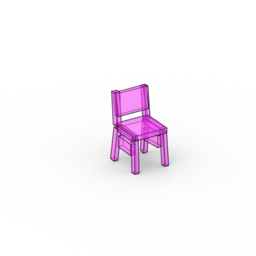}
        \\
        &
        \includegraphics[trim={4.5cm 4.5cm 4.5cm 4.5cm},clip,width=.105\linewidth]{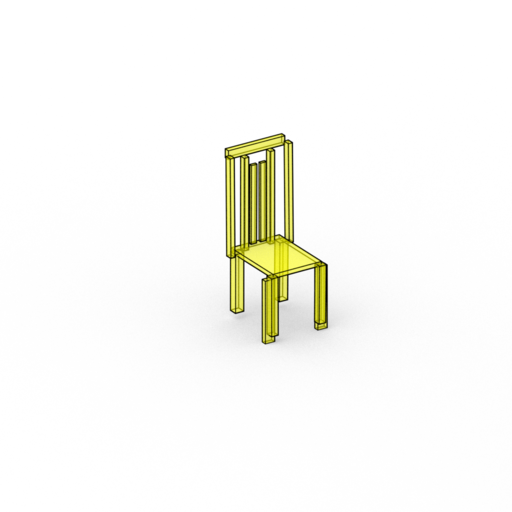} &
        \includegraphics[trim={4.5cm 4.5cm 4.5cm 4.5cm},clip,width=.105\linewidth]{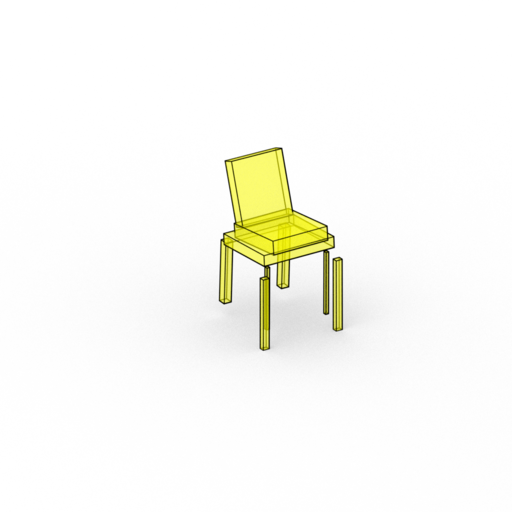} &
        \includegraphics[trim={4.5cm 4.5cm 4.5cm 4.5cm},clip,width=.105\linewidth]{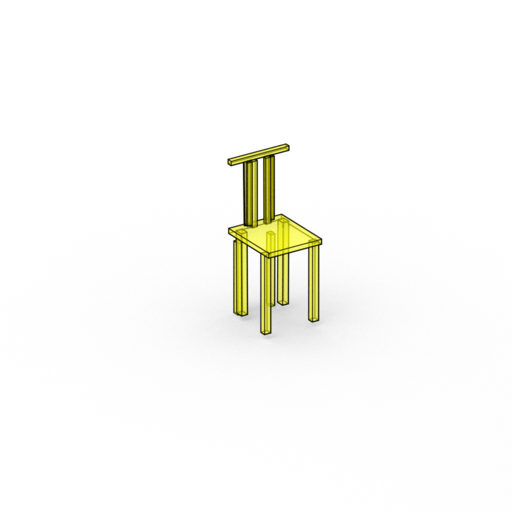}
        &&
        \includegraphics[trim={4.5cm 4.5cm 4.5cm 4.5cm},clip,width=.105\linewidth]{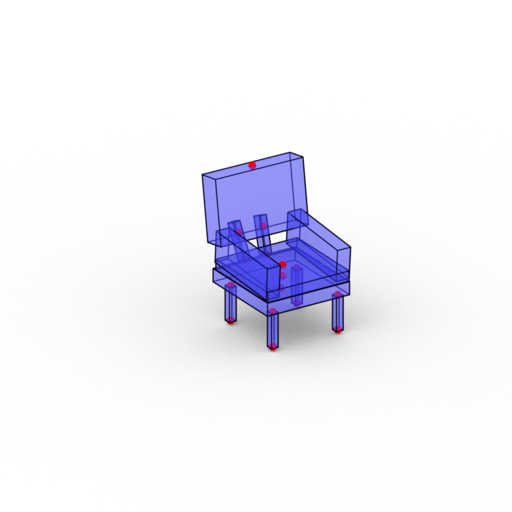} &
        \includegraphics[trim={4.5cm 4.5cm 4.5cm 4.5cm},clip,width=.105\linewidth]{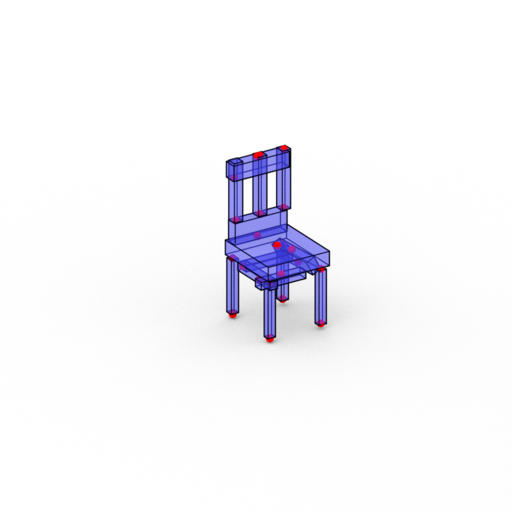} &
        \includegraphics[trim={4.5cm 4.5cm 4.5cm 4.5cm},clip,width=.105\linewidth]{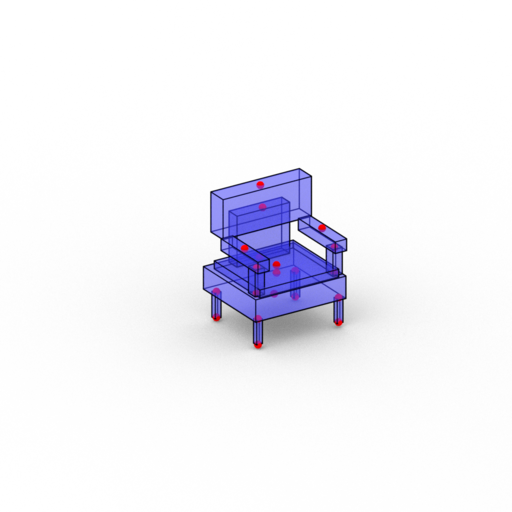} &&
        \includegraphics[trim={2.5cm 2.5cm 2.5cm 2.5cm},clip,width=.105\linewidth]{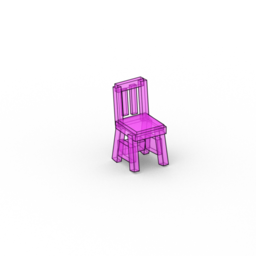} &
        \includegraphics[trim={2.5cm 2.5cm 2.5cm 2.5cm},clip,width=.105\linewidth]{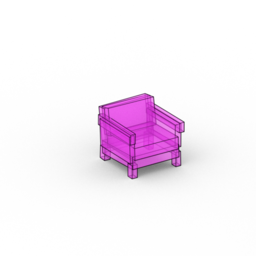} &
        \includegraphics[trim={2.5cm 2.5cm 2.5cm 2.5cm},clip,width=.105\linewidth]{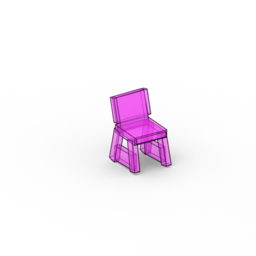}
        \\
        % \multicolumn{11}{c}{\emph{Table}}
        % \\
        & \multicolumn{3}{c}{3D-PRNN} && \multicolumn{3}{c}{Ours} && \multicolumn{3}{c}{StructureNet}
        \\
        \cmidrule{2-4} \cmidrule{6-8} \cmidrule{10-12}
        &
        \includegraphics[trim={4.5cm 4.5cm 4.5cm 4.5cm},clip,width=.105\linewidth]{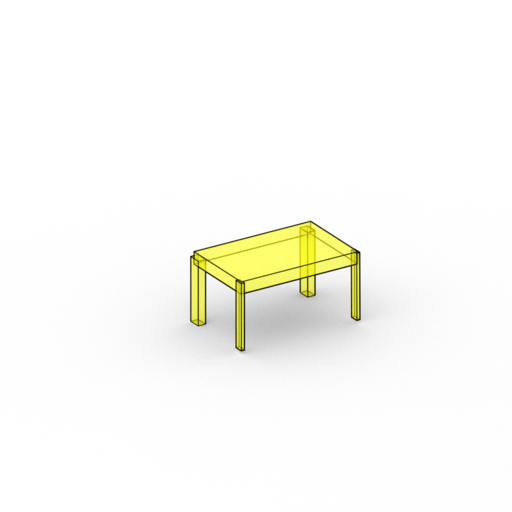} &
        \includegraphics[trim={4.5cm 4.5cm 4.5cm 4.5cm},clip,width=.105\linewidth]{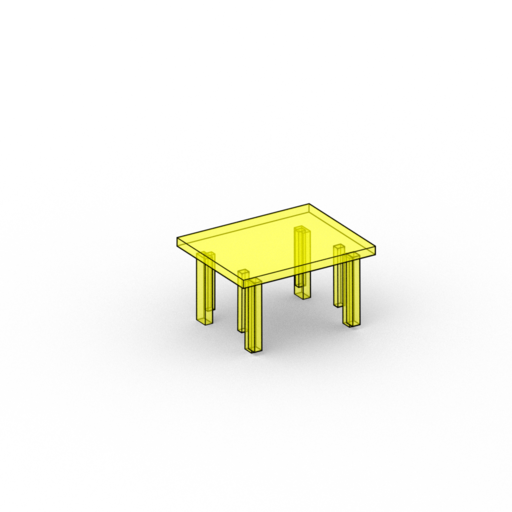} &
        \includegraphics[trim={4.5cm 4.5cm 4.5cm 4.5cm},clip,width=.105\linewidth]{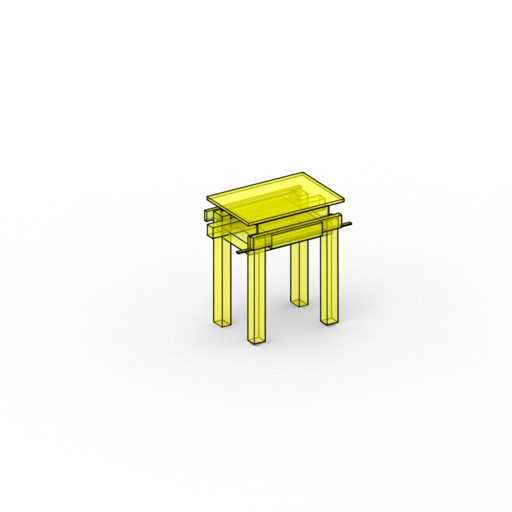} &&
        \includegraphics[trim={4.5cm 4.5cm 4.5cm 4.5cm},clip,width=.105\linewidth]{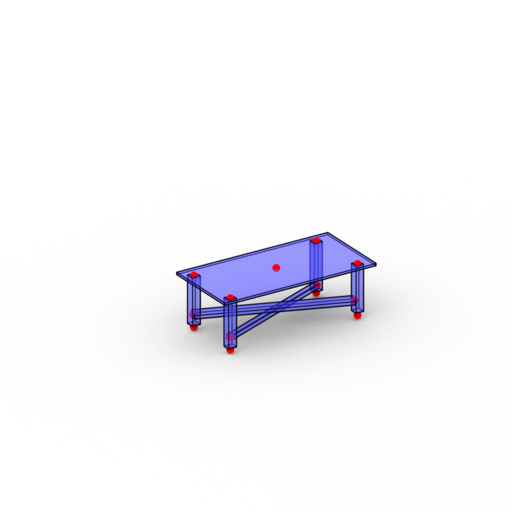} &
        \includegraphics[trim={4.5cm 4.5cm 4.5cm 4.5cm},clip,width=.105\linewidth]{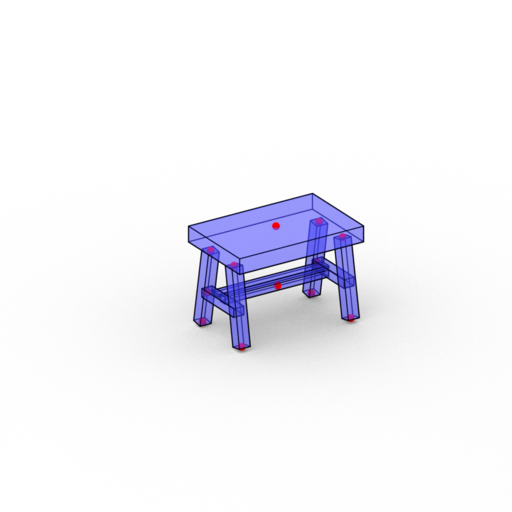} &
        \includegraphics[trim={4.5cm 4.5cm 4.5cm 4.5cm},clip,width=.105\linewidth]{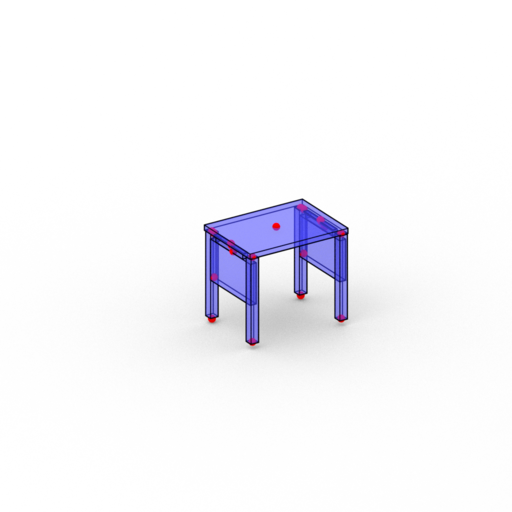} &&
        \includegraphics[trim={2.5cm 2.5cm 2.5cm 2.5cm},clip,width=.105\linewidth]{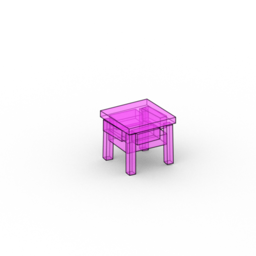} &
        \includegraphics[trim={2.5cm 2.5cm 2.5cm 2.5cm},clip,width=.105\linewidth]{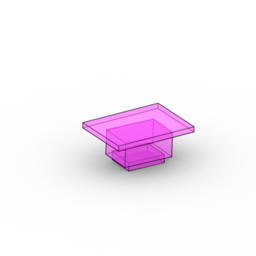} &
        \includegraphics[trim={2.5cm 2.5cm 2.5cm 2.5cm},clip,width=.105\linewidth]{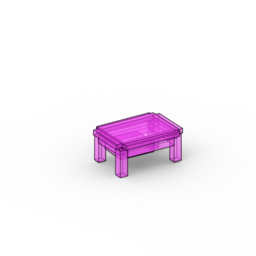}
        \\
        \raisebox{1.5em}{\rotatebox{90}{Table}} &
        \includegraphics[trim={4.5cm 4.5cm 4.5cm 4.5cm},clip,width=.105\linewidth]{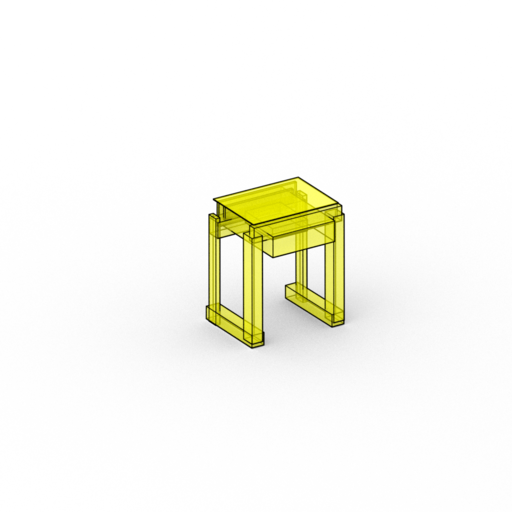} &
        \includegraphics[trim={4.5cm 4.5cm 4.5cm 4.5cm},clip,width=.105\linewidth]{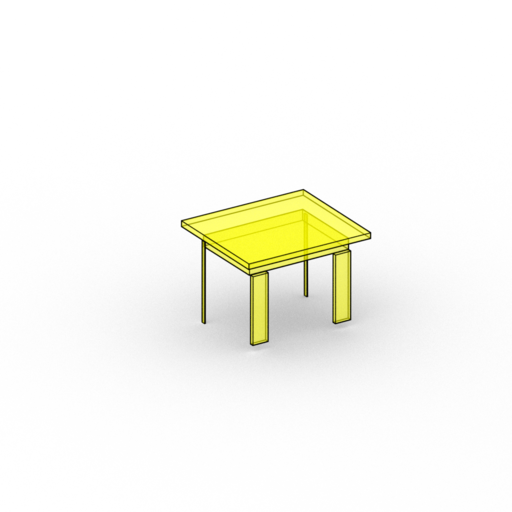} &
        \includegraphics[trim={4.5cm 4.5cm 4.5cm 4.5cm},clip,width=.105\linewidth]{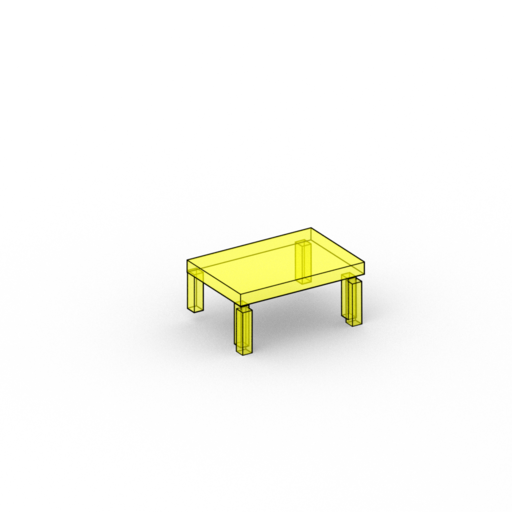} &&
        \includegraphics[trim={4.5cm 4.5cm 4.5cm 4.5cm},clip,width=.105\linewidth]{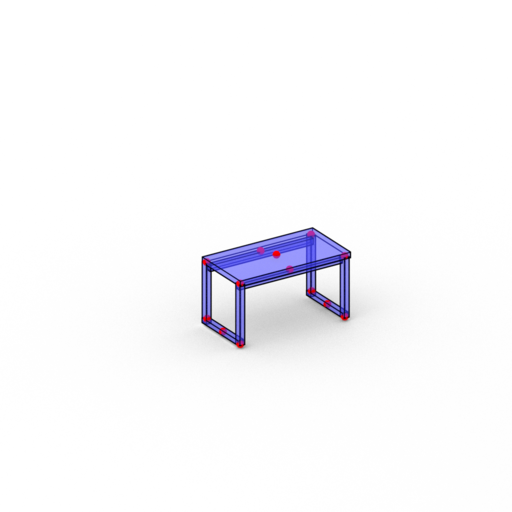} &
        \includegraphics[trim={4.5cm 4.5cm 4.5cm 4.5cm},clip,width=.105\linewidth]{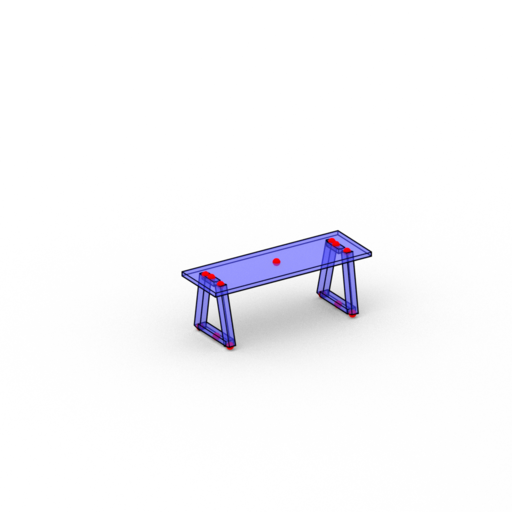} &
        \includegraphics[trim={4.5cm 4.5cm 4.5cm 4.5cm},clip,width=.105\linewidth]{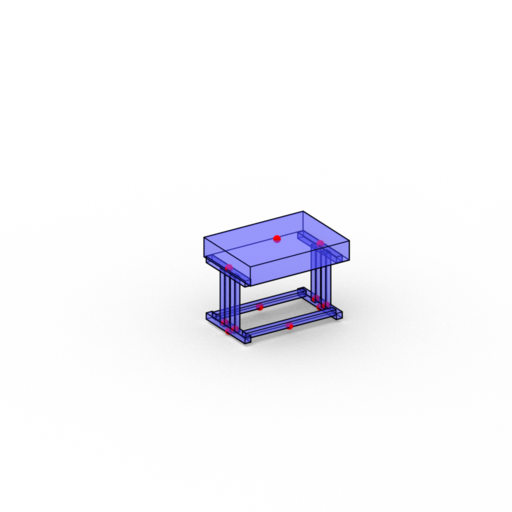} &&
        \includegraphics[trim={2.5cm 2.5cm 2.5cm 2.5cm},clip,width=.105\linewidth]{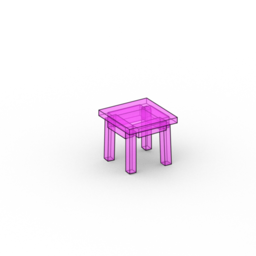} &
        \includegraphics[trim={2.5cm 2.5cm 2.5cm 2.5cm},clip,width=.105\linewidth]{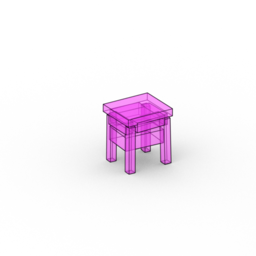} &
        \includegraphics[trim={2.5cm 2.5cm 2.5cm 2.5cm},clip,width=.105\linewidth]{figs/snet_samples/table/object-0005.png}
        \\
        &
        \includegraphics[trim={4.5cm 4.5cm 4.5cm 4.5cm},clip,width=.105\linewidth]{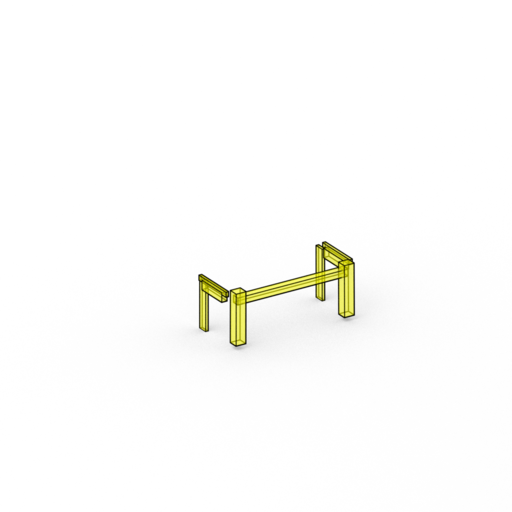} &
        \includegraphics[trim={4.5cm 4.5cm 4.5cm 4.5cm},clip,width=.105\linewidth]{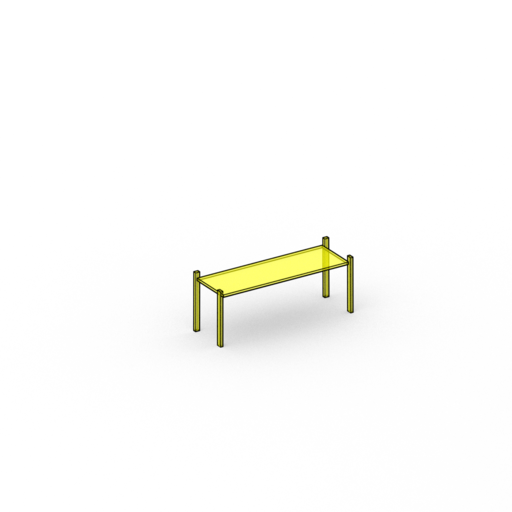} &
        \includegraphics[trim={4.5cm 4.5cm 4.5cm 4.5cm},clip,width=.105\linewidth]{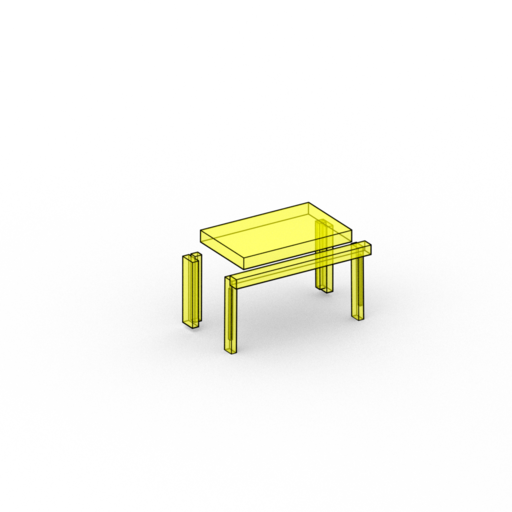}
        &&
        \includegraphics[trim={4.5cm 4.5cm 4.5cm 4.5cm},clip,width=.105\linewidth]{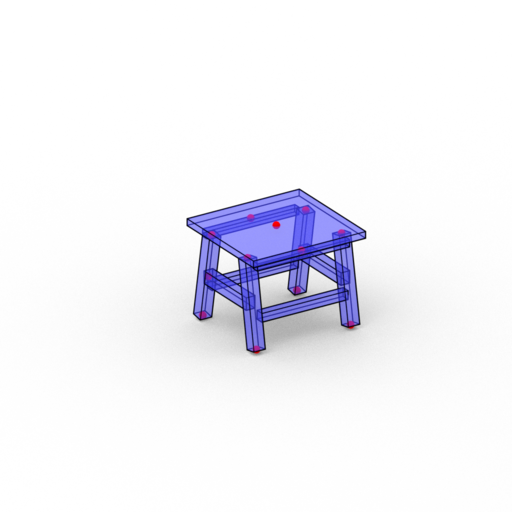} &
        \includegraphics[trim={4.5cm 4.5cm 4.5cm 4.5cm},clip,width=.105\linewidth]{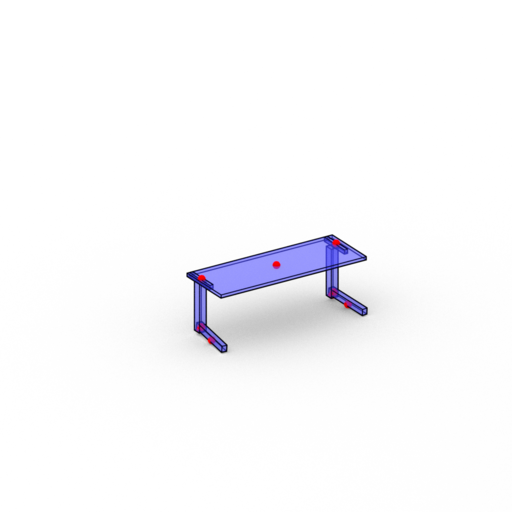} &
        \includegraphics[trim={4.5cm 4.5cm 4.5cm 4.5cm},clip,width=.105\linewidth]{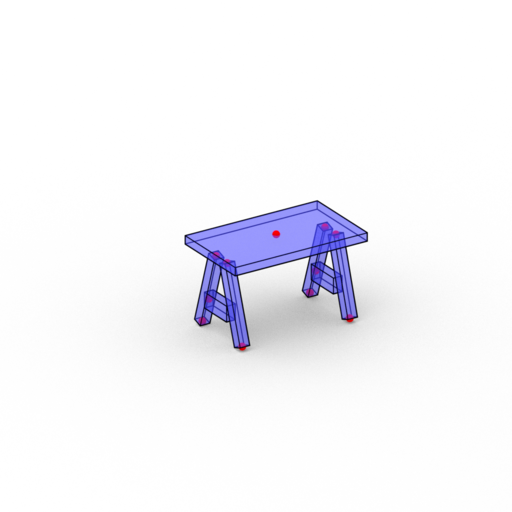} &&
        \includegraphics[trim={2.5cm 2.5cm 2.5cm 2.5cm},clip,width=.105\linewidth]{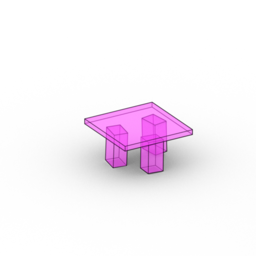} &
        \includegraphics[trim={2.5cm 2.5cm 2.5cm 2.5cm},clip,width=.105\linewidth]{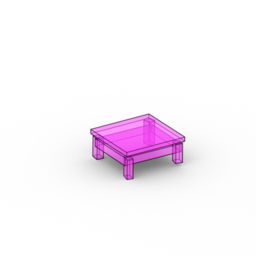} &
        \includegraphics[trim={4.5cm 4.5cm 4.5cm 4.5cm},clip,width=.105\linewidth]{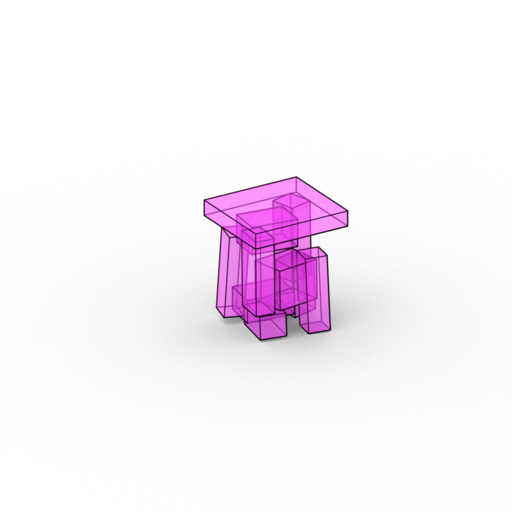}
        \\
        % \multicolumn{11}{c}{\emph{Storage}}
        % \\
        & \multicolumn{3}{c}{3D-PRNN} && \multicolumn{3}{c}{Ours} && \multicolumn{3}{c}{StructureNet}
        \\
        \cmidrule{2-4} \cmidrule{6-8} \cmidrule{10-12}
        &
        \includegraphics[trim={4.5cm 4.5cm 4.5cm 4.5cm},clip,width=.105\linewidth]{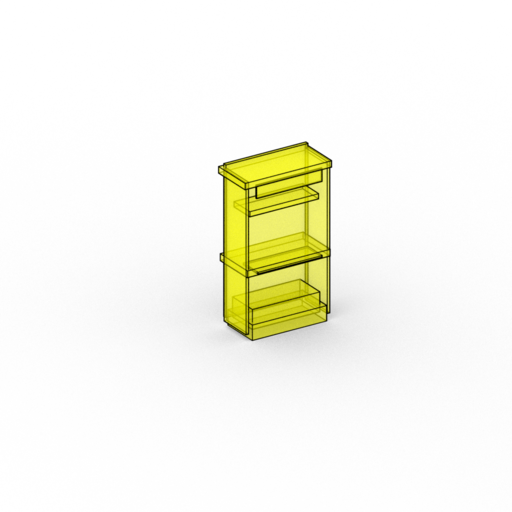} &
        \includegraphics[trim={4.5cm 4.5cm 4.5cm 4.5cm},clip,width=.105\linewidth]{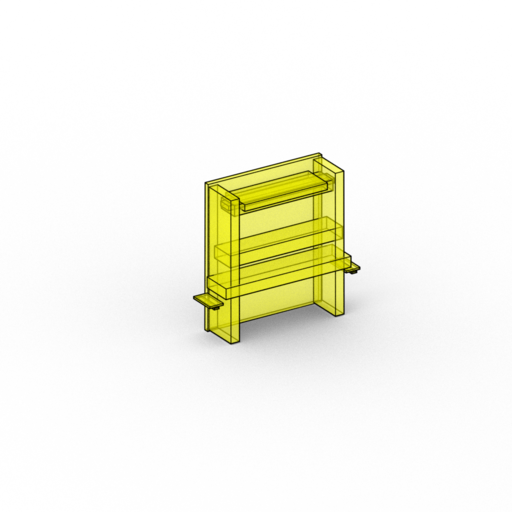} &
        \includegraphics[trim={4.5cm 4.5cm 4.5cm 4.5cm},clip,width=.105\linewidth]{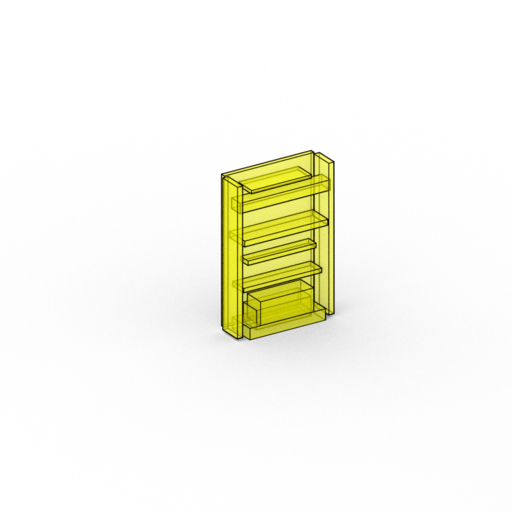} &&
        \includegraphics[trim={4.5cm 4.5cm 4.5cm 4.5cm},clip,width=.105\linewidth]{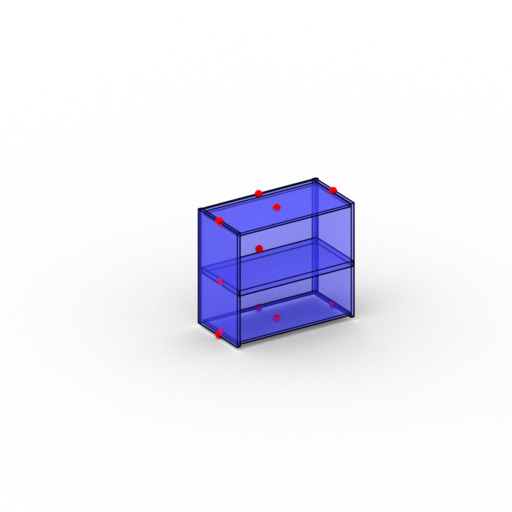} &
        \includegraphics[trim={4.5cm 4.5cm 4.5cm 4.5cm},clip,width=.105\linewidth]{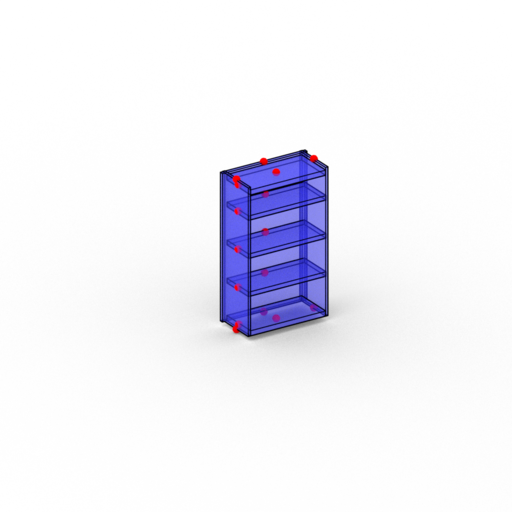} &
        \includegraphics[trim={4.5cm 4.5cm 4.5cm 4.5cm},clip,width=.105\linewidth]{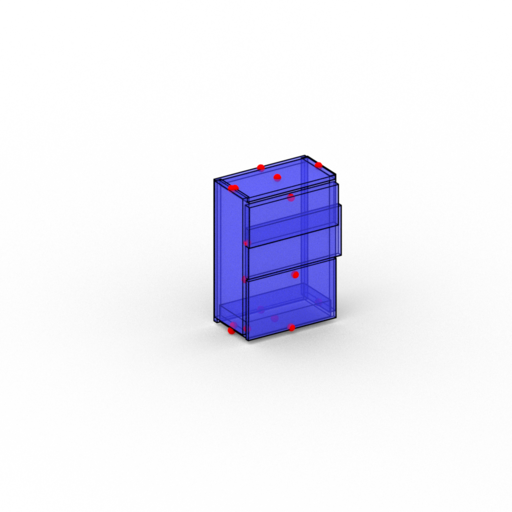} &&
        \includegraphics[trim={2.5cm 2.5cm 2.5cm 2.5cm},clip,width=.105\linewidth]{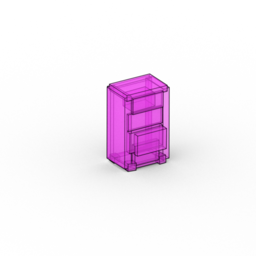} &
        \includegraphics[trim={2.5cm 2.5cm 2.5cm 2.5cm},clip,width=.105\linewidth]{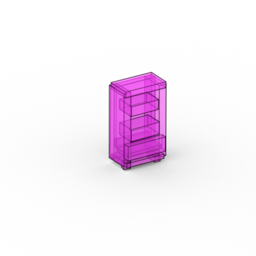} &
        \includegraphics[trim={2.5cm 2.5cm 2.5cm 2.5cm},clip,width=.105\linewidth]{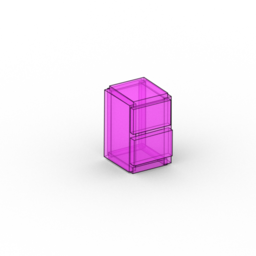}
        \\
        \raisebox{1.5em}{\rotatebox{90}{Storage}} &
        \includegraphics[trim={4.5cm 4.5cm 4.5cm 4.5cm},clip,width=.105\linewidth]{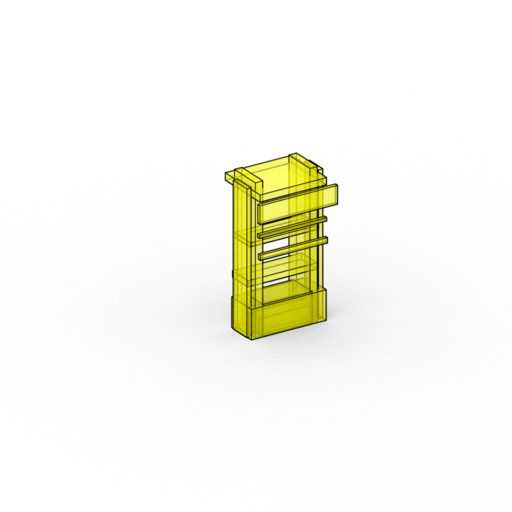} &
        \includegraphics[trim={4.5cm 4.5cm 4.5cm 4.5cm},clip,width=.105\linewidth]{figs/prnn_samples/storage/sample_19.png} &
        \includegraphics[trim={4.5cm 4.5cm 4.5cm 4.5cm},clip,width=.105\linewidth]{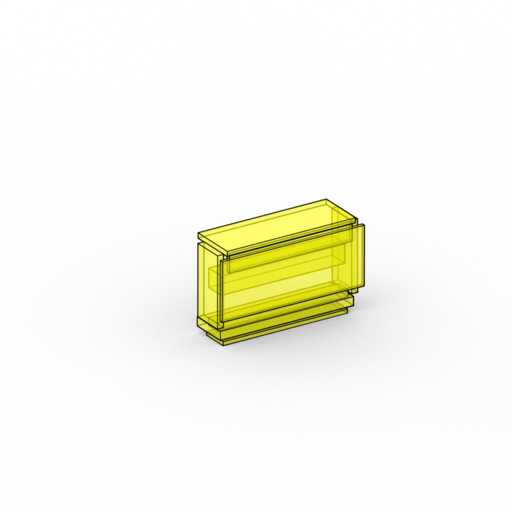} &&
        \includegraphics[trim={4.5cm 4.5cm 4.5cm 4.5cm},clip,width=.105\linewidth]{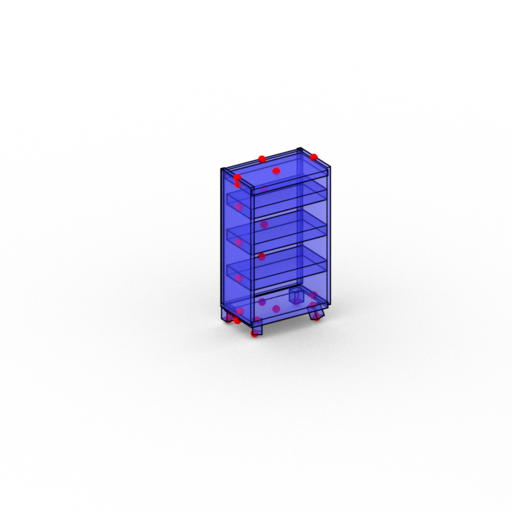} &
        \includegraphics[trim={4.5cm 4.5cm 4.5cm 4.5cm},clip,width=.105\linewidth]{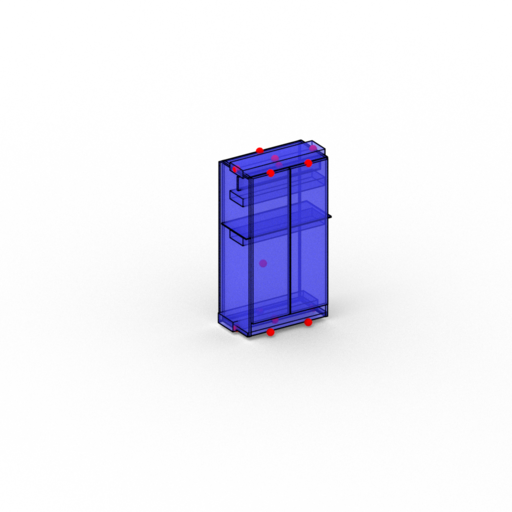} &
        \includegraphics[trim={4.5cm 4.5cm 4.5cm 4.5cm},clip,width=.105\linewidth]{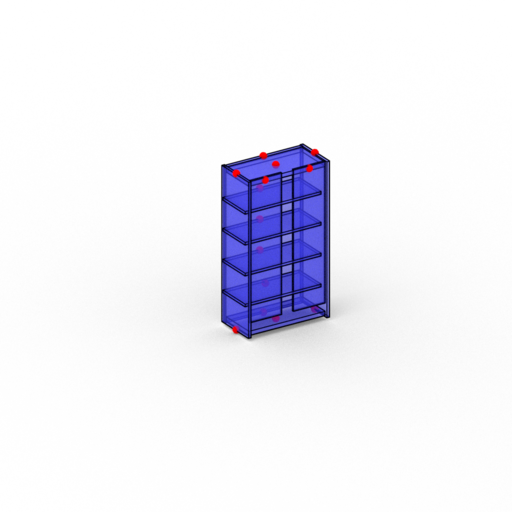} &&
        \includegraphics[trim={2.5cm 2.5cm 2.5cm 2.5cm},clip,width=.105\linewidth]{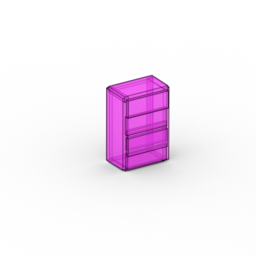} &
        \includegraphics[trim={2.5cm 2.5cm 2.5cm 2.5cm},clip,width=.105\linewidth]{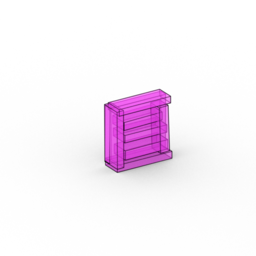} &
        \includegraphics[trim={2.5cm 2.5cm 2.5cm 2.5cm},clip,width=.105\linewidth]{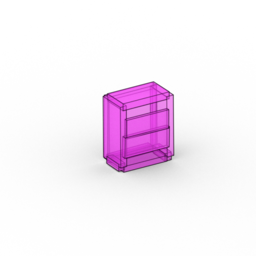}
        \\
        &
        \includegraphics[trim={4.5cm 4.5cm 4.5cm 4.5cm},clip,width=.105\linewidth]{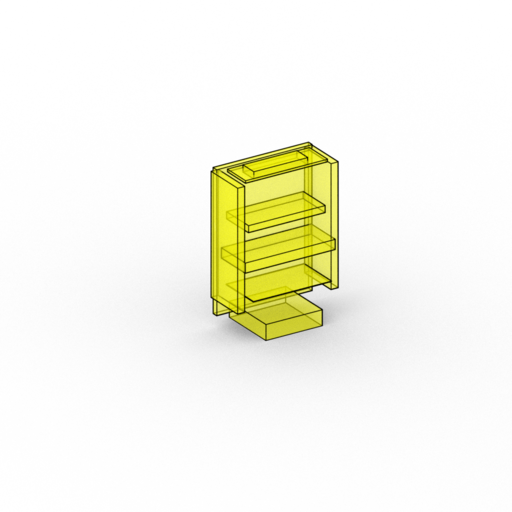} &
        \includegraphics[trim={4.5cm 4.5cm 4.5cm 4.5cm},clip,width=.105\linewidth]{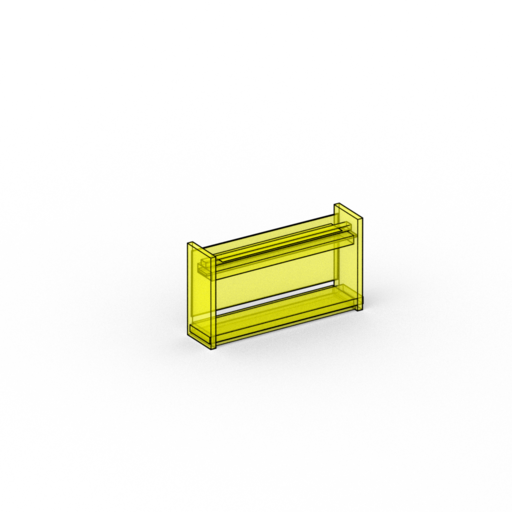} &
        \includegraphics[trim={4.5cm 4.5cm 4.5cm 4.5cm},clip,width=.105\linewidth]{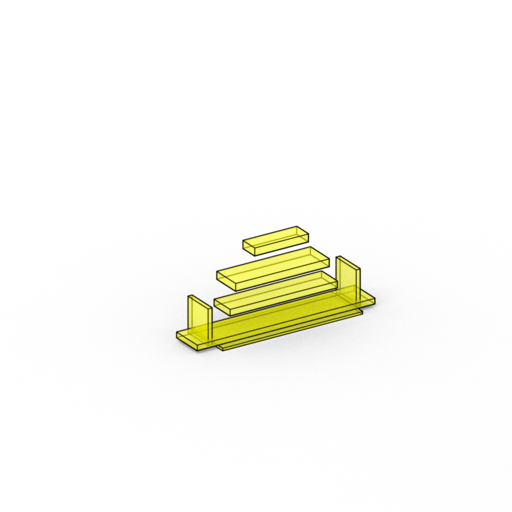}
        &&
        \includegraphics[trim={4.5cm 4.5cm 4.5cm 4.5cm},clip,width=.105\linewidth]{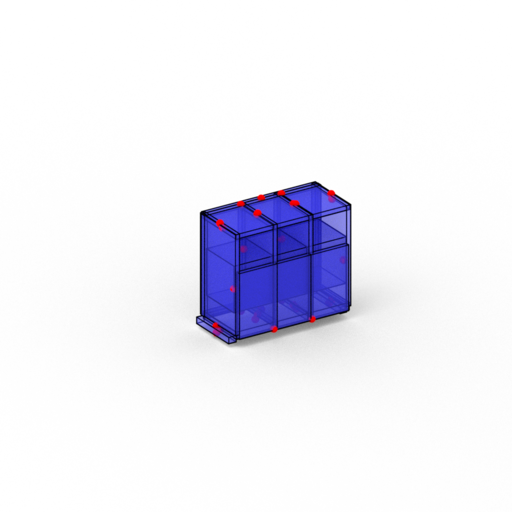} &
        \includegraphics[trim={4.5cm 4.5cm 4.5cm 4.5cm},clip,width=.105\linewidth]{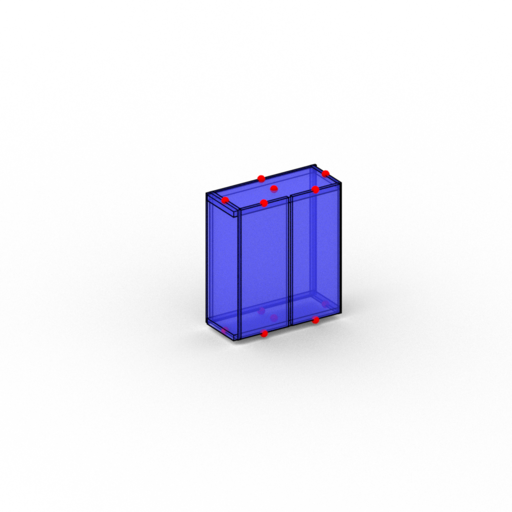} &
        \includegraphics[trim={4.5cm 4.5cm 4.5cm 4.5cm},clip,width=.105\linewidth]{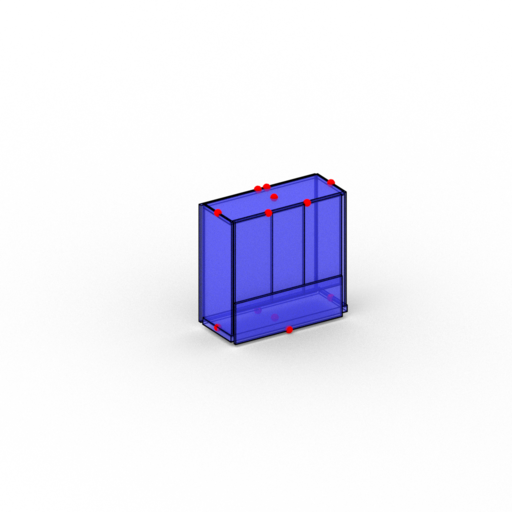} &&
        \includegraphics[trim={2.5cm 2.5cm 2.5cm 2.5cm},clip,width=.105\linewidth]{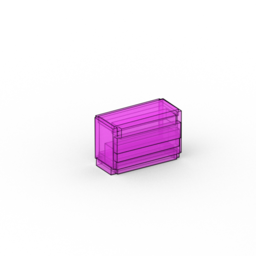} &
        \includegraphics[trim={2.5cm 2.5cm 2.5cm 2.5cm},clip,width=.105\linewidth]{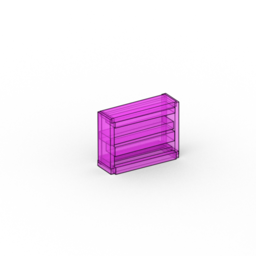} &
        \includegraphics[trim={4.5cm 4.5cm 4.5cm 4.5cm},clip,width=.105\linewidth]{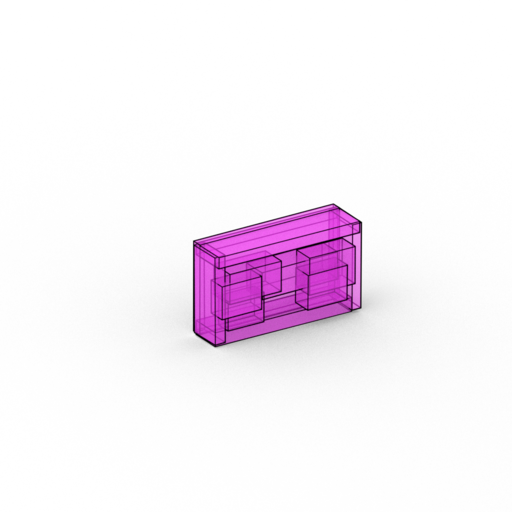}
    \end{tabular}
    \caption{Qualitative comparison between generated samples from our method, StructureNet, and 3D-PRNN.
    Across different categories, our method creates novel ~\dslname~ programs that, when executed, produce shape structures that maintain realistic and physically valid part-to-part relationships. Comparison methods that directly predict 3D shape geometry exhibit failure cases where parts become disconnected or intersect in an implausible manner. 
    }
    \label{fig:qualitative comparison}
\end{figure*}

\subsubsection{Analysis of Variability}

Beyond quality and editability, we also consider the variability of outputs of each method. Specifically, for generated shapes, we care about their novelty with respect to the training data, their complexity, and their variety. We present results of an experiment using Chamfer distance to quantify performance across these areas in Table \ref{tab:variability}. 

The \emph{Generalization} metric measures the average distance of each generated sample to its nearest neighbor in the training set. As all methods have higher generalization scores than the validation set, we can conclude that none of the methods appear to be overfitting. For our method specifically, this re-enforces the qualitative nearest neighbor results presented in Figure~\ref{fig:nn_qual}.

The \emph{Coverage} metric measures the average distance of each validation shape to its nearest neighbor in the set of generated shapes. Across all categories our method achieves the best results, and by a wide-margin for tables, which indicates that our generations have enough complexity to match the distribution of the validation shapes.

The \emph{Variety} metric measures the average distance of each generated shape to its nearest neighbor in the set of generated shapes besides itself. Once again, across all categories our method achieves top, or tied for top performance. 

Additionally, we look at average number of leaf parts as a coarse proxy for the complexity of a shape's structure, which is shown in Table~\ref{tab:complexity}. While our method has a similar number of leaf parts to the comparison methods for the Chair and Table categories, we do have fewer leaf parts on average for Storage. Qualitatively, these additional parts in the comparison methods often manifest as collections of spatially colocated cuboids, and not necessarily more complex shape structures.

In terms of the variability of programs generated by our method, we note that 65\% of Chair programs, 85\% of Table programs, and 53\% of Storage programs contained ~\dslname~ program structures not present in the training data. Thus our method not only exhibits novelty in the geometric domain, but also in the structural domain.

\subsubsection{Program Clustering}
Our approach is predicated on the assumption that a single program can represent a parametric family of multiple shapes, allowing for this shape space to be explored via manipulation of interpretable program parameters. To verify whether this is true, we cluster shapes that are represented by \emph{structurally-equivalent} programs (i.e. programs that are the same up to continuous parameter variations). Figure~\ref{fig:clustering_curves} shows program clustering results for the ground truth programs we parse from PartNet. These results demonstrate how the structure of a single ~\dslname~ program is able to represent related shapes through different parameterizations. The marked improvement in clustering when splitting by intermediate part programs compared with clustering on entire shape programs, provides additional support for our hierarchical approach; shape programs are more likely to share structure within a node of the hierarchy than they are to match entire hierarchies exactly.

\begin{table}[t!]
    \centering
    % \small
    \footnotesize
    \setlength{\tabcolsep}{1pt}
    %\vspace{-1em}
    \begin{tabular}{@{}llcccccc@{}}
        \toprule
        & & \multicolumn{1}{c}{\emph{Generalization}} & \multicolumn{1}{c}{\emph{Coverage}} & \multicolumn{1}{c}{\emph{Variety}}
        \\
        & & \multicolumn{1}{c}{NND to Train$\;\Uparrow$} & \multicolumn{1}{c}{NND from Val$\;\Downarrow$} & \multicolumn{1}{c}{NND to Self$\;\Uparrow$}
        \\
        \textbf{Category} & \textbf{Method} & 
        \textbf{CD} &
        \textbf{CD} & 
        \textbf{CD}
        \\
        \midrule
        \multirow{4}{*}{\emph{Chair}}

        & 3D-PRNN & \textbf{0.111} & 0.123 &  \textbf{0.104} \\
        & StructureNet & 0.104 & 0.119 & 0.087 \\
        & Ours & 0.108 &  \textbf{0.118} & \textbf{ 0.104} \\
        & Validation & 0.105 & --- & 0.114 \\
        \midrule
        \multirow{4}{*}{\emph{Table}}

        & 3D-PRNN & 0.095 & 0.130 & 0.086 \\
        & StructureNet & \textbf{0.129} & 0.141 & 0.0925 \\
        & Ours & 0.101 &  \textbf{0.108} &  \textbf{0.102} \\
        & Validation  & 0.09  & --- & 0.099 \\
        \midrule
        \multirow{4}{*}{\emph{Storage}}

        & 3D-PRNN & \textbf{0.134} &  0.132 & \textbf{0.119} \\
        & StructureNet & 0.129 & 0.135 & 0.107 \\
        & Ours  & 0.125  &  \textbf{0.129} &  \textbf{0.119} \\
        & Validation & 0.11 & --- & 0.125 \\
        \bottomrule
    \end{tabular}
    \caption{ We compare the geometric variability of generated shapes from different methods. In the first column, we measure generalization as the average nearest neighbor distance (NND) from generated samples to shapes in the training set. In the second column we measure coverage as the average NND from shapes in the validation set to generated samples. In the last column, we measure variety as the average NND from shapes in the generated samples to other generated shapes in the same set. Across three categories of shapes, our method performs the best on the coverage and variety metrics, while outperforming validation on generalization (demonstrating we are not overfitting).}
    \label{tab:variability}
\end{table}

\begin{table}[t!]
    \centering
    \small
    \setlength{\tabcolsep}{2pt}
    \begin{tabular}{@{}llcc@{}}
        \toprule
        \textbf{Category} & \textbf{Method} & \textbf{Avg \# Leaf Parts} \\
        \midrule
        \multirow{4}{*}{\emph{Chair}}
        & 3D-PRNN & 8.6 \\
        & StructureNet & 8.7   \\
        & Ours & 7.9  \\
        & Ground Truth & 9.7  \\
        \midrule
        \multirow{4}{*}{\emph{Table}}
        & 3D-PRNN & 7.07  \\
        & StructureNet & 8.16   \\
        & Ours & 7.84  \\
        & Ground Truth & 8.4   \\
        \midrule
        \multirow{4}{*}{\emph{Storage}}
        & 3D-PRNN & 10.6 \\
        & StructureNet & 12.3   \\
        & Ours & 8.4  \\
        & Ground Truth & 10.8 \\
        \bottomrule
    \end{tabular}
    \caption{We compare the average number of leaf parts in generated shapes, as a coarse proxy for complexity of shape structure. Our method generates similar numbers of leaf parts compared with other methods for Chairs and Tables, but fewer leaf parts for Storage. Qualitatively, the additional leaf parts measured in comparison methods often manifests as spurious overlapping cuboids, rather than more complex structural variety.}
    \label{tab:complexity}
\end{table}

\begin{table}[t!]
    \centering
    \small
    % \footnotesize
    \setlength{\tabcolsep}{2pt}
    %\vspace{-1em}
    \begin{tabular}{@{}llcc@{}}
        \toprule
        & & \multicolumn{2}{c}{Avg. Step Size$\;\Downarrow$}\\
        \textbf{Category} & \textbf{Method} & \textbf{Geo} & \textbf{Prog} \\
        \midrule
        \multirow{2}{*}{\emph{Chair}}
        & StructureNet &  \textbf{0.0384} & 3.90 \\
        & Ours & \textbf{0.0384} & \textbf{1.33} \\
        \midrule
        \multirow{2}{*}{\emph{Table}}
        & StructureNet & 0.0474 & 4.75 \\
        & Ours & \textbf{0.0389} & \textbf{2.48} \\
        \midrule    
        \multirow{2}{*}{\emph{Storage}}
        & StructureNet & 0.0512 & 4.29 \\
        & Ours & \textbf{0.0482} & \textbf{2.6} \\
        \bottomrule
    \end{tabular}
    \caption{
    We measure smoothness along random high-frequency interpolation sequences in each method's latent space. The Geo column measures smoothness with Chamfer distance, while the Prog column measures smoothness with program edit distance. Note that 3D-PRNN is missing because it is not a latent variable model and thus does not support interpolation.}
    \label{tab:interpolation}
\end{table}

\subsubsection{Synthesizing Surface Geometry}

\begin{figure}[t!]
    \centering
    \setlength{\tabcolsep}{1pt}
    \begin{tabular}{ccccc}
        \raisebox{.5em}{\rotatebox{90}{Part Cuboids}} &
        \includegraphics[trim={4.5cm 4.5cm 4.5cm 4.5cm},clip,width=.24\linewidth]{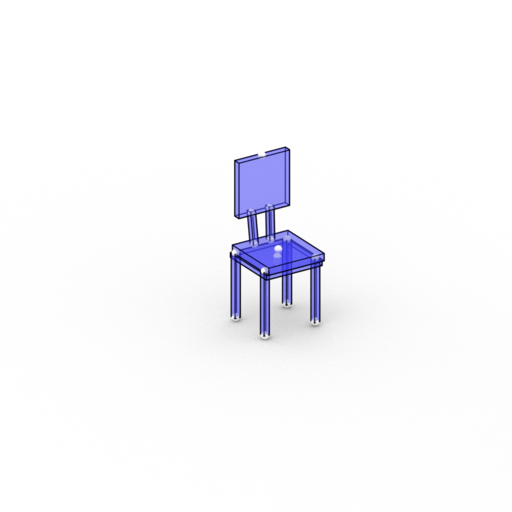} &
        \includegraphics[trim={4.5cm 4.5cm 4.5cm 4.5cm},clip,width=.24\linewidth]{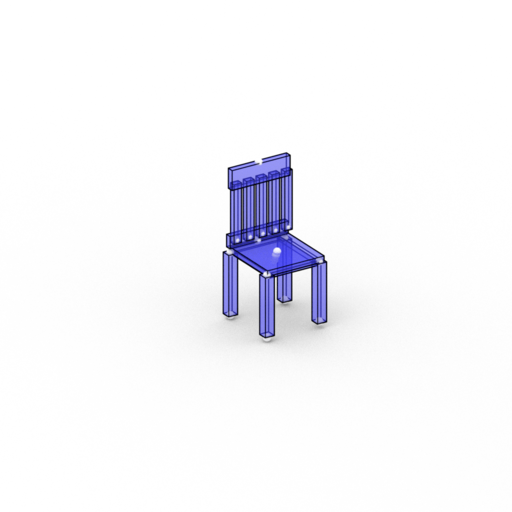} &
        \includegraphics[trim={4.5cm 4.5cm 4.5cm 4.5cm},clip,width=.24\linewidth]{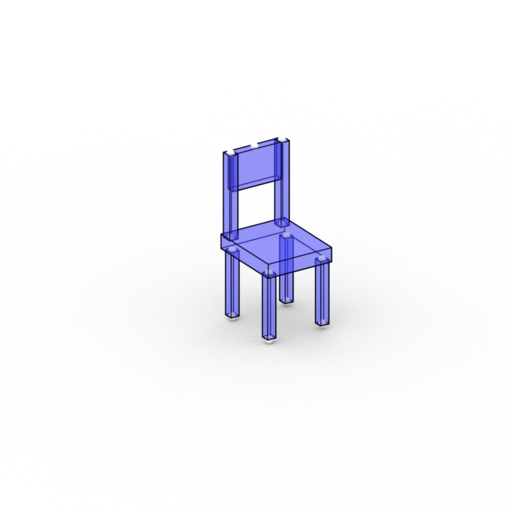} &
        \includegraphics[trim={4.5cm 4.5cm 4.5cm 4.5cm},clip,width=.24\linewidth]{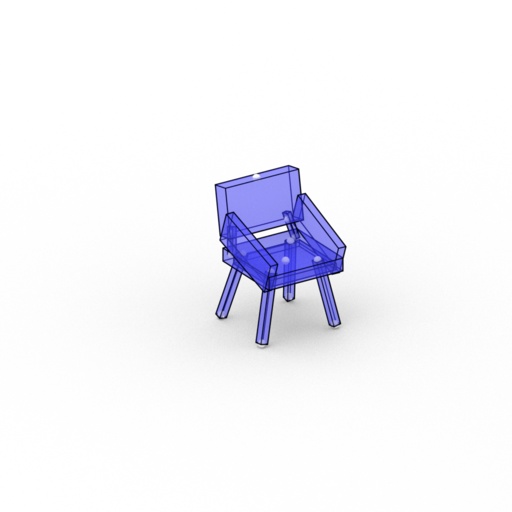} 
        \\
        \raisebox{.25em}{\rotatebox{90}{Surface Points}} &
        \includegraphics[trim={4.5cm 4.5cm 4.5cm 4.5cm},clip,width=.24\linewidth]{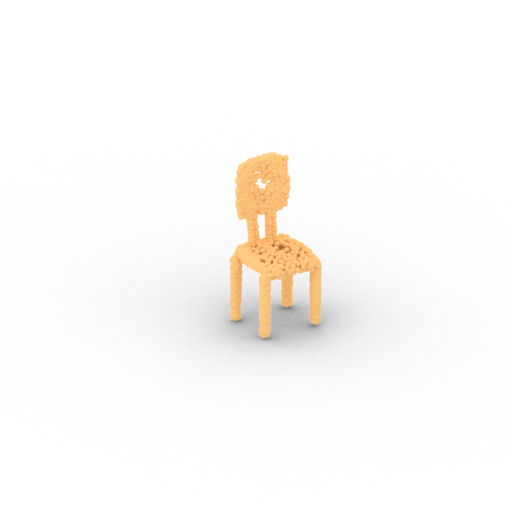} &
        \includegraphics[trim={4.5cm 4.5cm 4.5cm 4.5cm},clip,width=.24\linewidth]{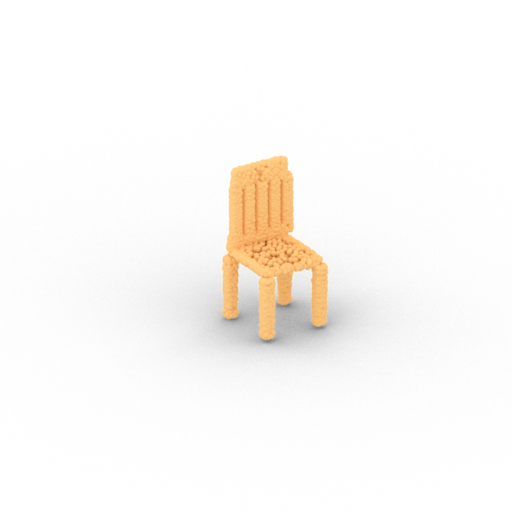} &
        \includegraphics[trim={4.5cm 4.5cm 4.5cm 4.5cm},clip,width=.24\linewidth]{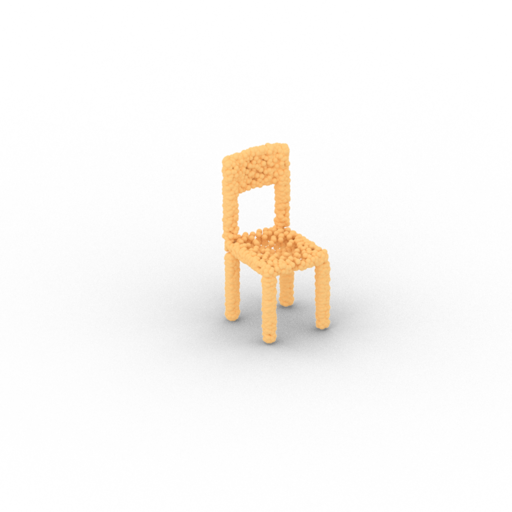} &
        \includegraphics[trim={4.5cm 4.5cm 4.5cm 4.5cm},clip,width=.24\linewidth]{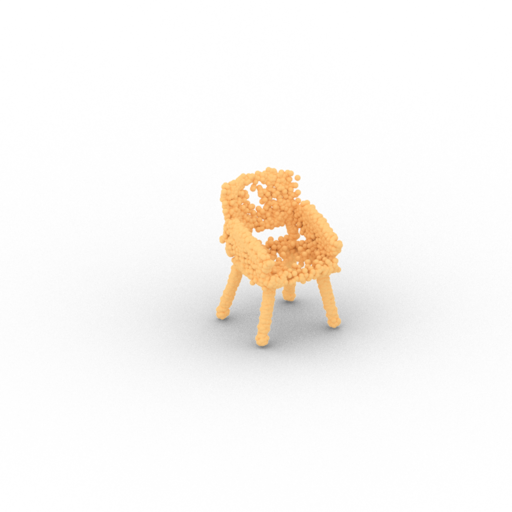}
    \end{tabular}
    \caption{Converting generated ~\dslname programs into dense point clouds. We use a point cloud decoder to predict the surface geometry of each leaf part proxy in our 3D shape structure. In this process, geometric details begin to take form, at the cost of some artifacts. We discuss a method for improving this procedure in section \ref{sec:conclusion}.}
    \label{fig:point_cloud_surface_geo}
\end{figure}

While collections of part proxies are a useful modeling representation for 3D shape structures, they do not directly attempt to capture the wide range of intra-part variability present in man-made objects. 
We demonstrate how ~\dslname~ programs can additionally be used to model parts at finer levels of detail by turning ~\dslname~ programs into dense point clouds. As a proof of concept, we augment our generative model with a point cloud encoder that consumes dense point cloud samples of ground truth leaf parts, and a point cloud decoder that generates dense point clouds for every leaf part within its predicted bounding volume. Figure~\ref{fig:point_cloud_surface_geo} shows some qualitative results of our method, trained on point clouds sampled from the dense geometry of Chairs found in PartNet. These generated surfaces provide additional detail over the geometry specified by their cuboid part proxies, as evidenced by both the rounding in the legs and back slats, and also in the curvature of the chair back surfaces.

\begin{figure}[t!]
  \centering
  \setlength{\tabcolsep}{0.5pt}
  \small
    \begin{tabular}{cc cc}
        \includegraphics[width=.5\linewidth]{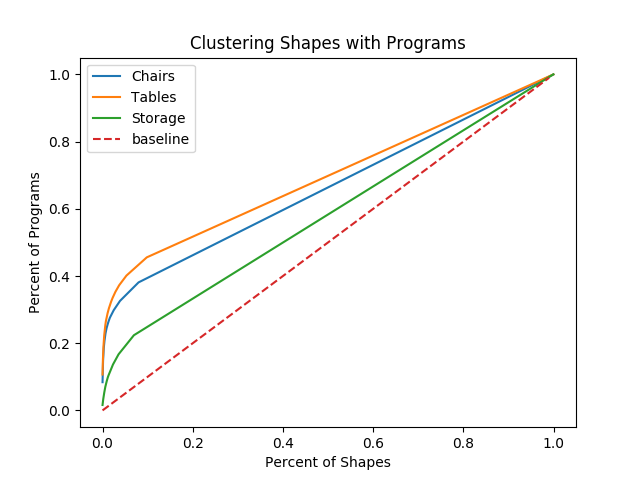} &
        \includegraphics[width=.5\linewidth]{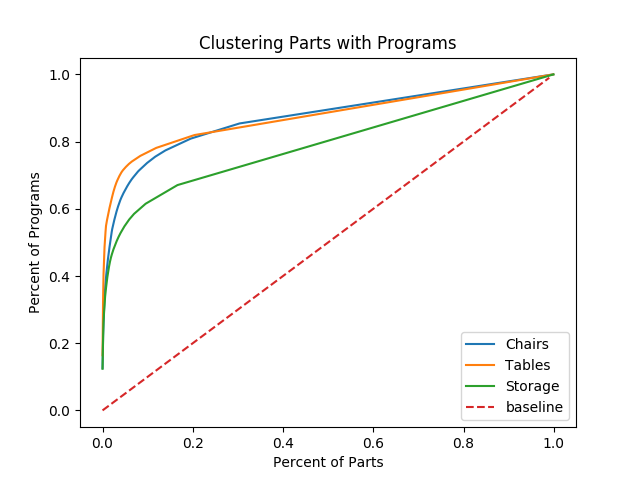} 
    \end{tabular}
\caption{ Clustering results that demonstrate how the structure of a single ~\dslname~ program is capable of capturing a family of related shapes. Using ground truth programs found with our program extraction procedure, in the left graph we plot the percentage of shapes captured as we consider more program structures extracted from the data. In the right graph we show the same plot but with parts (nodes) instead of shapes (full hierarchy). }
\label{fig:clustering_curves}
\end{figure}

\begin{figure*}[t!]
    \centering
    \setlength{\tabcolsep}{1pt}
    \begin{tabular}{ccccccccc}
        & Source Shape &
        \multicolumn{5}{c}{\rule[1.5pt]{15em}{0.7pt} Interpolations \rule[1.5pt]{15em}{0.7pt}} &
        Target Shape
        \\
        \raisebox{1em}{\rotatebox{90}{StructureNet}} &
        \raisebox{3em}{\multirow{2}{*}{\includegraphics[trim={4.5cm 4.5cm 4.5cm 4.5cm},clip,width=.13\linewidth]{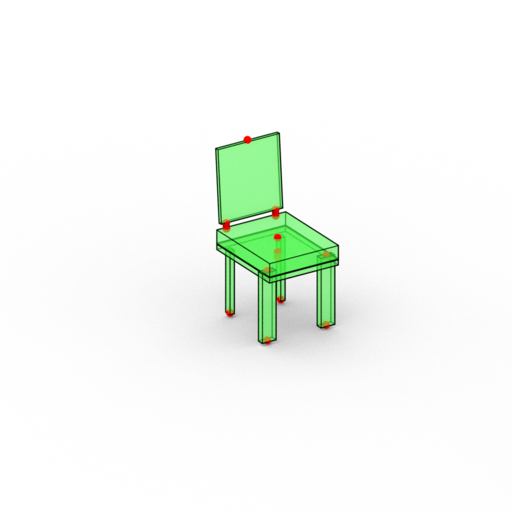}}} &
        \includegraphics[trim={4.5cm 4.5cm 4.5cm 4.5cm},clip,width=.13\linewidth]{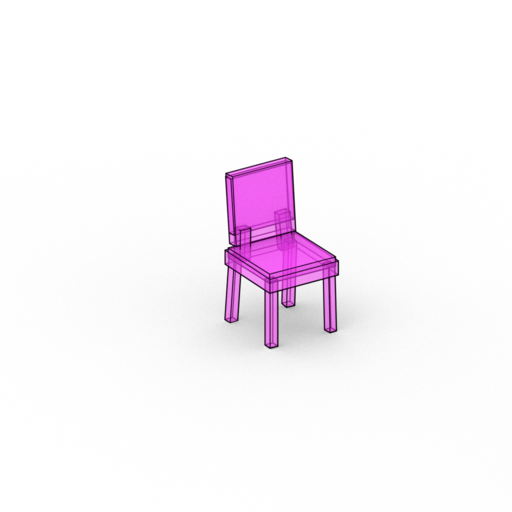} &
        \includegraphics[trim={4.5cm 4.5cm 4.5cm 4.5cm},clip,width=.13\linewidth]{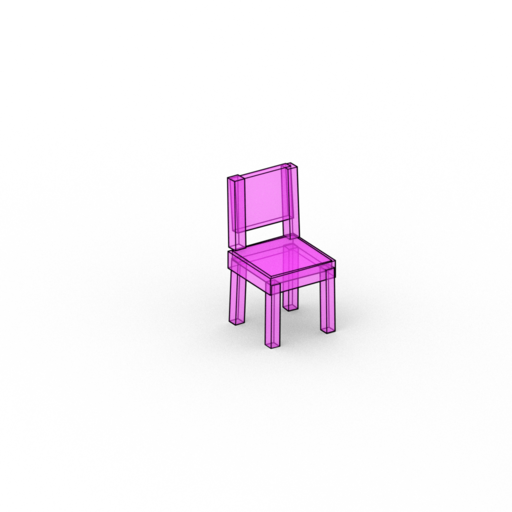} &
        \includegraphics[trim={4.5cm 4.5cm 4.5cm 4.5cm},clip,width=.13\linewidth]{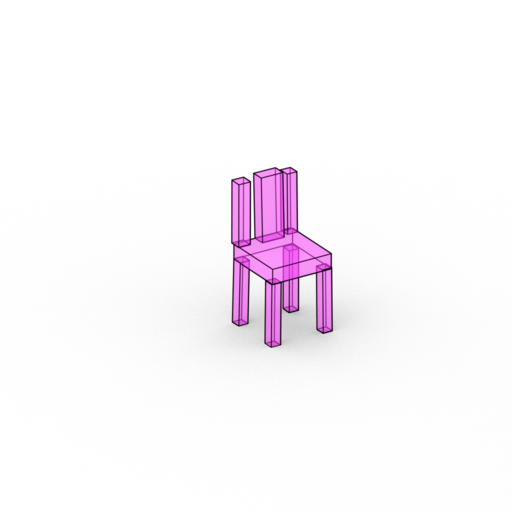} &
        \includegraphics[trim={4.5cm 4.5cm 4.5cm 4.5cm},clip,width=.13\linewidth]{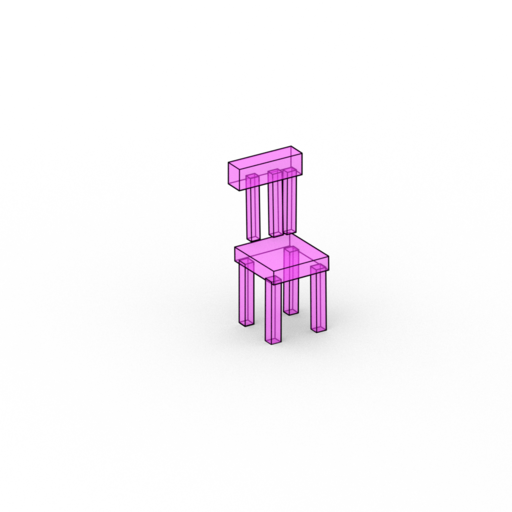} &
        \includegraphics[trim={4.5cm 4.5cm 4.5cm 4.5cm},clip,width=.13\linewidth]{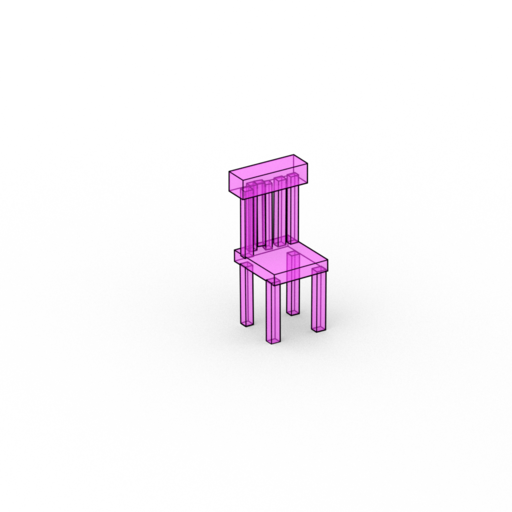} &
        \raisebox{3em}{\multirow{2}{*}{\includegraphics[trim={4.5cm 4.5cm 4.5cm 4.5cm},clip,width=.13\linewidth]{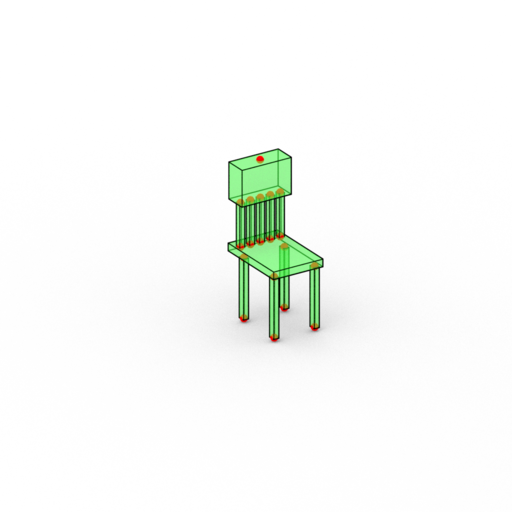}}} 
        \\
        \raisebox{3em}{\rotatebox{90}{Ours}} &
        &
        \includegraphics[trim={4.5cm 4.5cm 4.5cm 4.5cm},clip,width=.13\linewidth]{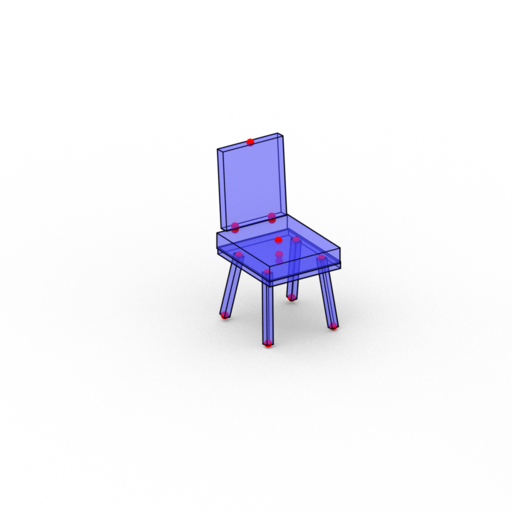} &
        \includegraphics[trim={4.5cm 4.5cm 4.5cm 4.5cm},clip,width=.13\linewidth]{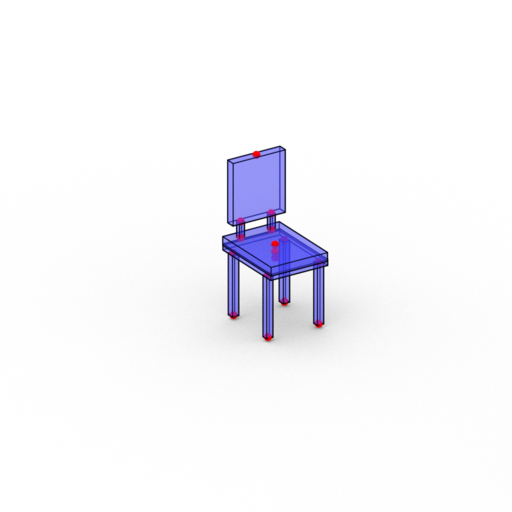} &
        \includegraphics[trim={4.5cm 4.5cm 4.5cm 4.5cm},clip,width=.13\linewidth]{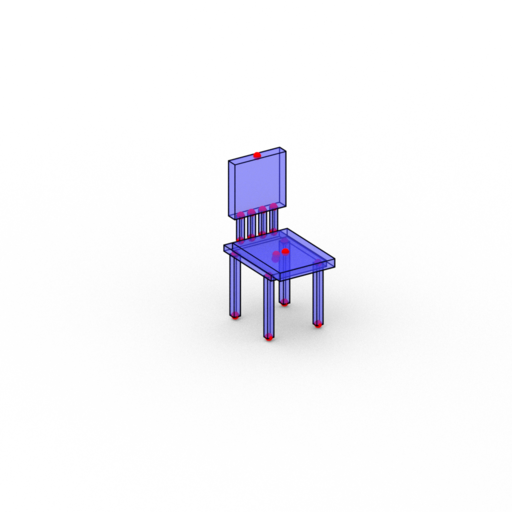} &
        \includegraphics[trim={4.5cm 4.5cm 4.5cm 4.5cm},clip,width=.13\linewidth]{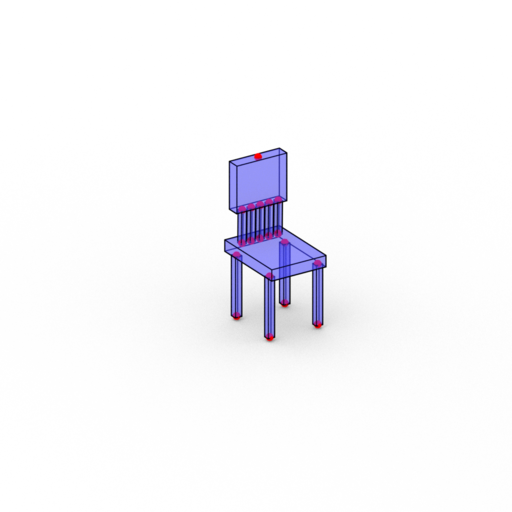} &
        \includegraphics[trim={4.5cm 4.5cm 4.5cm 4.5cm},clip,width=.13\linewidth]{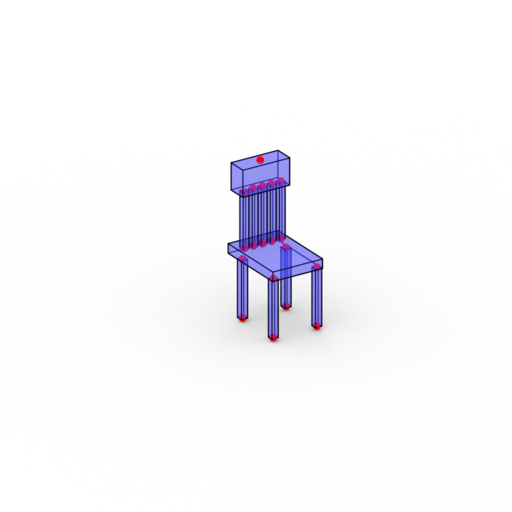} 
        \\
        \raisebox{1em}{\rotatebox{90}{StructureNet}} &
        \raisebox{3em}{\multirow{2}{*}{\includegraphics[trim={4.5cm 4.5cm 4.5cm 4.5cm},clip,width=.13\linewidth]{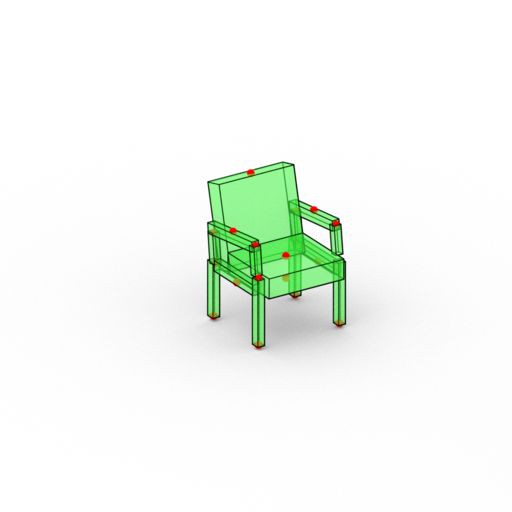}}} &
        \includegraphics[trim={4.5cm 4.5cm 4.5cm 4.5cm},clip,width=.13\linewidth]{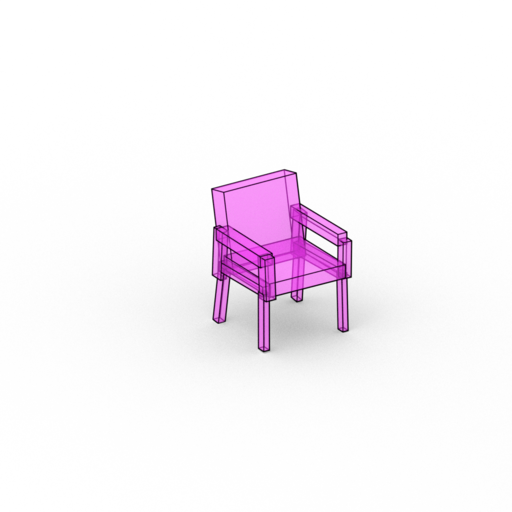} &
        \includegraphics[trim={4.5cm 4.5cm 4.5cm 4.5cm},clip,width=.13\linewidth]{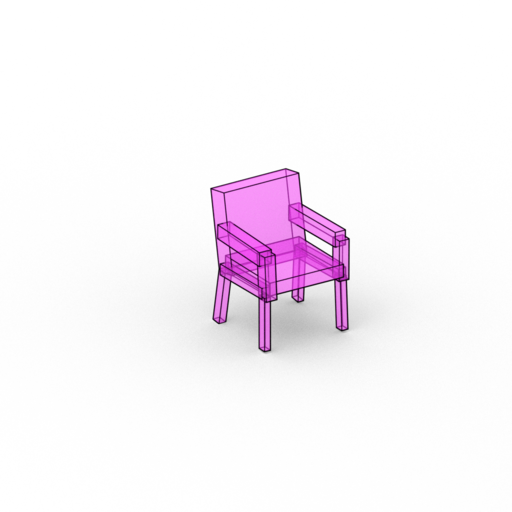} &
        \includegraphics[trim={4.5cm 4.5cm 4.5cm 4.5cm},clip,width=.13\linewidth]{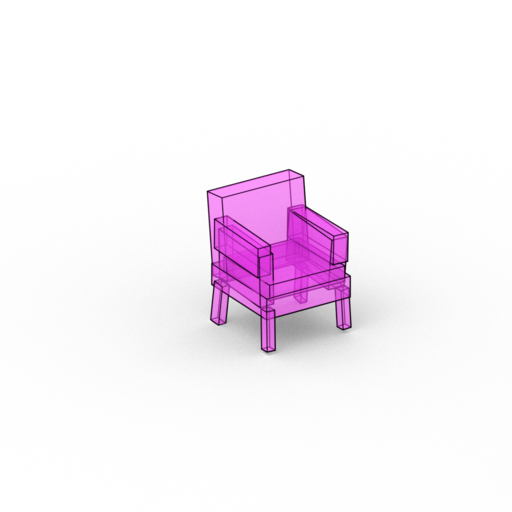} &
        \includegraphics[trim={4.5cm 4.5cm 4.5cm 4.5cm},clip,width=.13\linewidth]{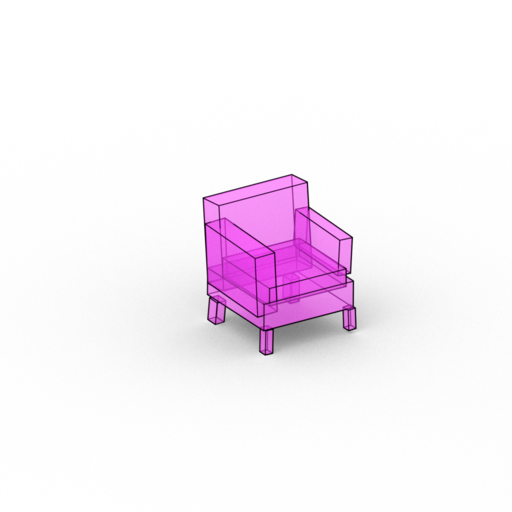} &
        \includegraphics[trim={4.5cm 4.5cm 4.5cm 4.5cm},clip,width=.13\linewidth]{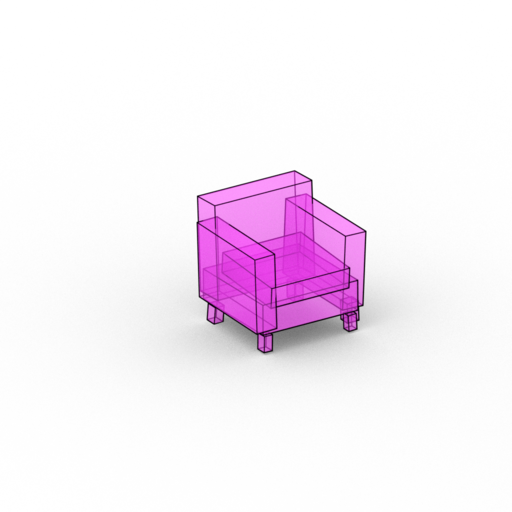} &
        \raisebox{3em}{\multirow{2}{*}{\includegraphics[trim={4.5cm 4.5cm 4.5cm 4.5cm},clip,width=.13\linewidth]{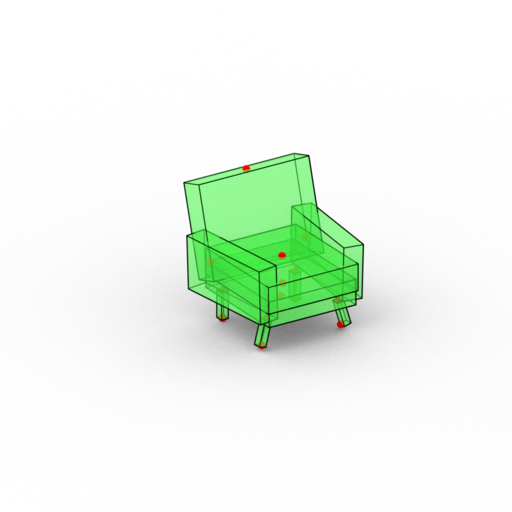}}} 
        \\
        \raisebox{3em}{\rotatebox{90}{Ours}} &
        &
        \includegraphics[trim={4.5cm 4.5cm 4.5cm 4.5cm},clip,width=.13\linewidth]{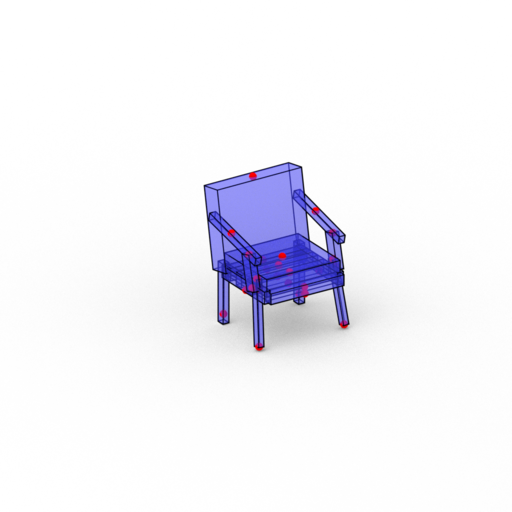} &
        \includegraphics[trim={4.5cm 4.5cm 4.5cm 4.5cm},clip,width=.13\linewidth]{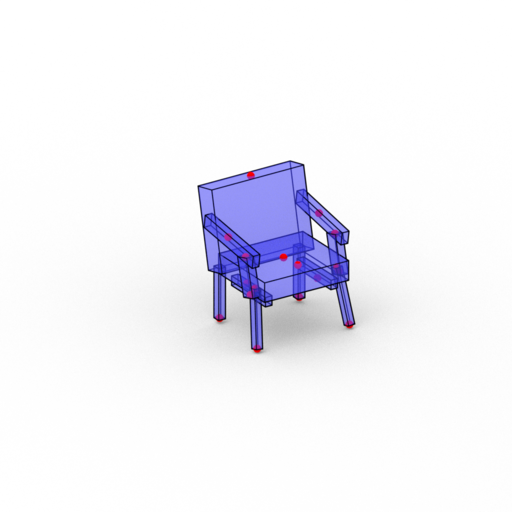} &
        \includegraphics[trim={4.5cm 4.5cm 4.5cm 4.5cm},clip,width=.13\linewidth]{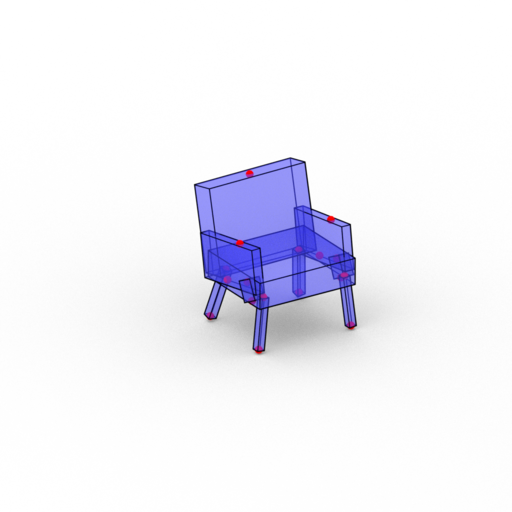} &
        \includegraphics[trim={4.5cm 4.5cm 4.5cm 4.5cm},clip,width=.13\linewidth]{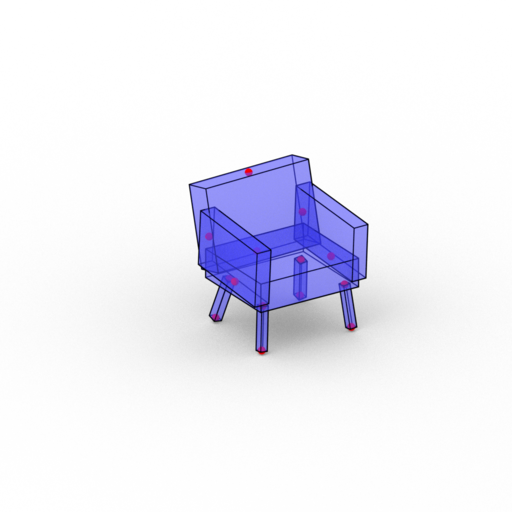} &
        \includegraphics[trim={4.5cm 4.5cm 4.5cm 4.5cm},clip,width=.13\linewidth]{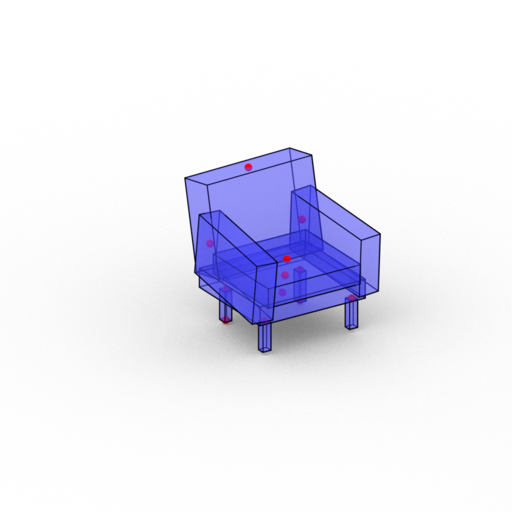} 
        \\
        \raisebox{1em}{\rotatebox{90}{StructureNet}} &
        \raisebox{3em}{\multirow{2}{*}{\includegraphics[trim={4.5cm 4.5cm 4.5cm 4.5cm},clip,width=.13\linewidth]{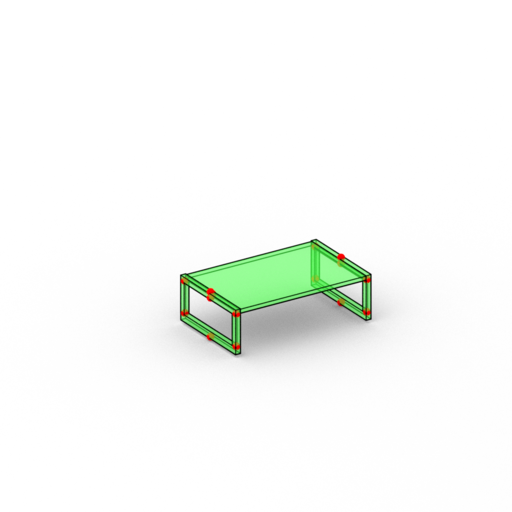}}} &
        \includegraphics[trim={4.5cm 4.5cm 4.5cm 4.5cm},clip,width=.13\linewidth]{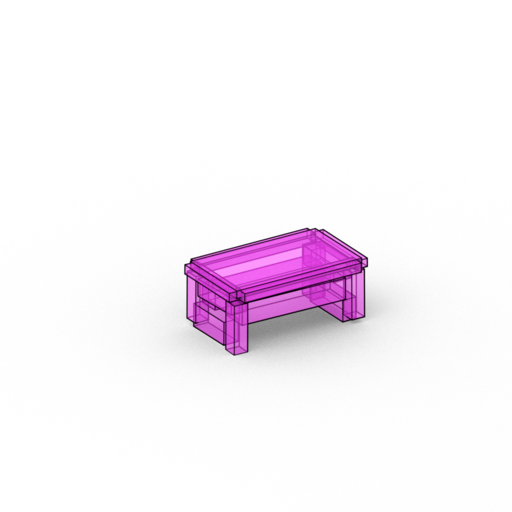} &
        \includegraphics[trim={4.5cm 4.5cm 4.5cm 4.5cm},clip,width=.13\linewidth]{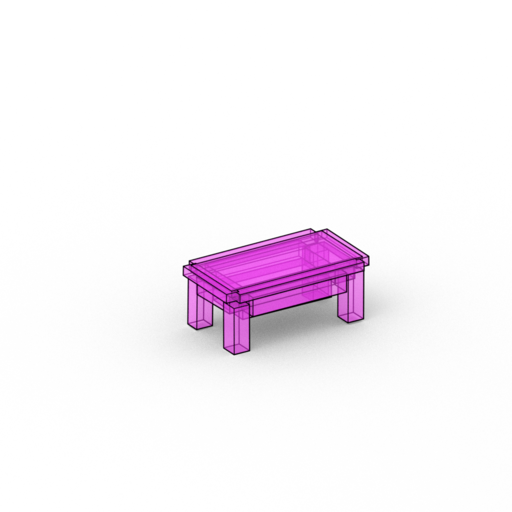} &
        \includegraphics[trim={4.5cm 4.5cm 4.5cm 4.5cm},clip,width=.13\linewidth]{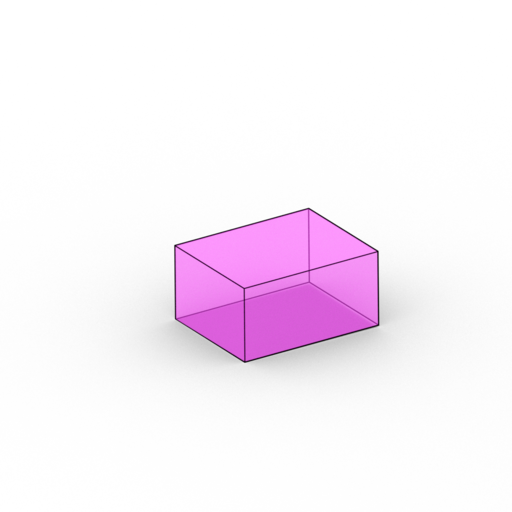} &
        \includegraphics[trim={4.5cm 4.5cm 4.5cm 4.5cm},clip,width=.13\linewidth]{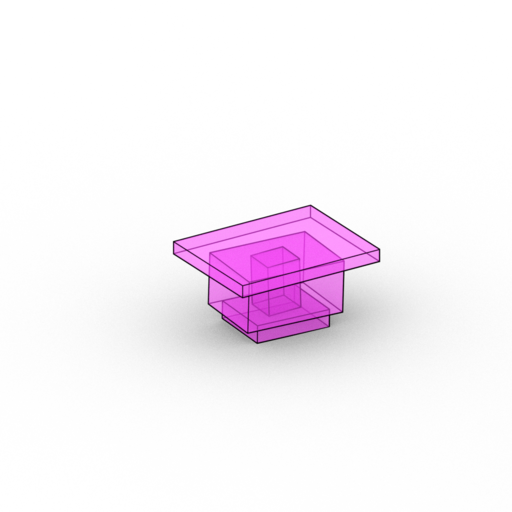} &
        \includegraphics[trim={4.5cm 4.5cm 4.5cm 4.5cm},clip,width=.13\linewidth]{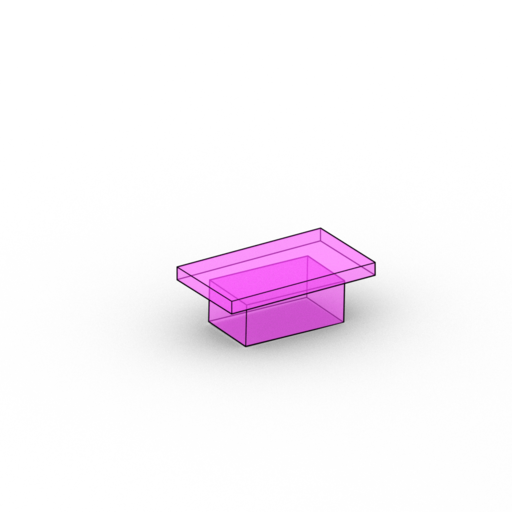} &
        \raisebox{3em}{\multirow{2}{*}{\includegraphics[trim={4.5cm 4.5cm 4.5cm 4.5cm},clip,width=.13\linewidth]{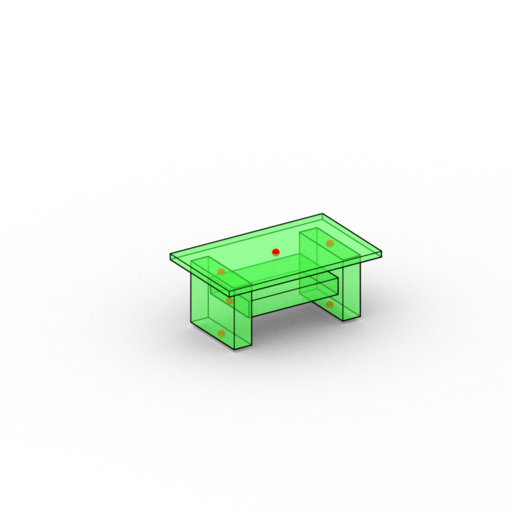}}} 
        \\
        \raisebox{3em}{\rotatebox{90}{Ours}} &
        &
        \includegraphics[trim={4.5cm 4.5cm 4.5cm 4.5cm},clip,width=.13\linewidth]{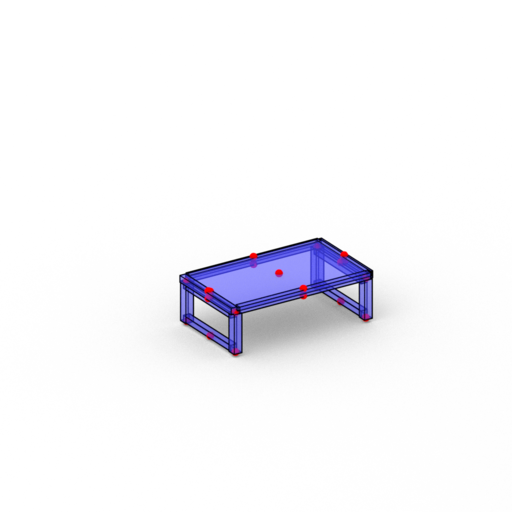} &
        \includegraphics[trim={4.5cm 4.5cm 4.5cm 4.5cm},clip,width=.13\linewidth]{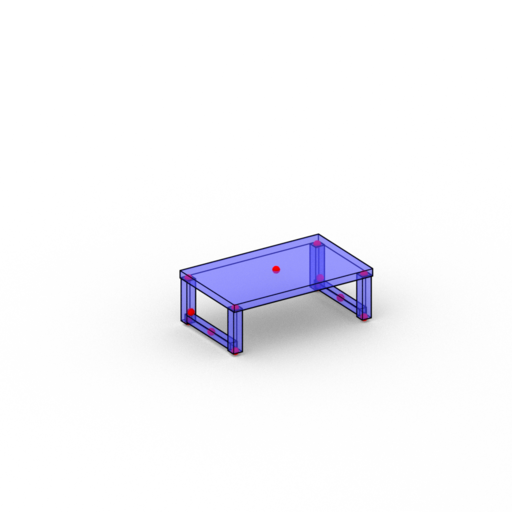} &
        \includegraphics[trim={4.5cm 4.5cm 4.5cm 4.5cm},clip,width=.13\linewidth]{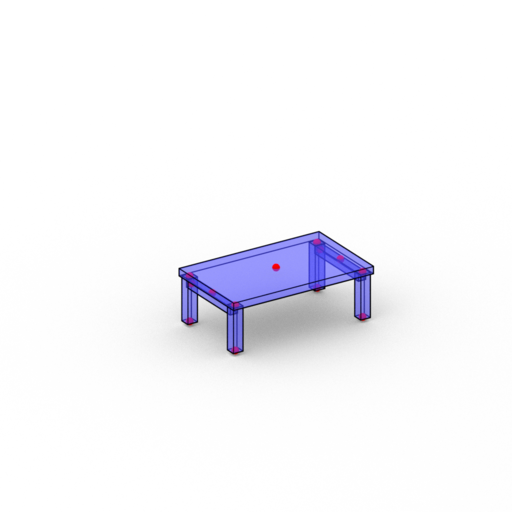} &
        \includegraphics[trim={4.5cm 4.5cm 4.5cm 4.5cm},clip,width=.13\linewidth]{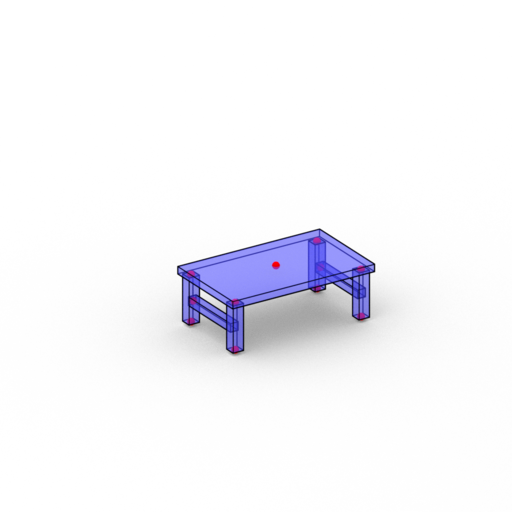} &
        \includegraphics[trim={4.5cm 4.5cm 4.5cm 4.5cm},clip,width=.13\linewidth]{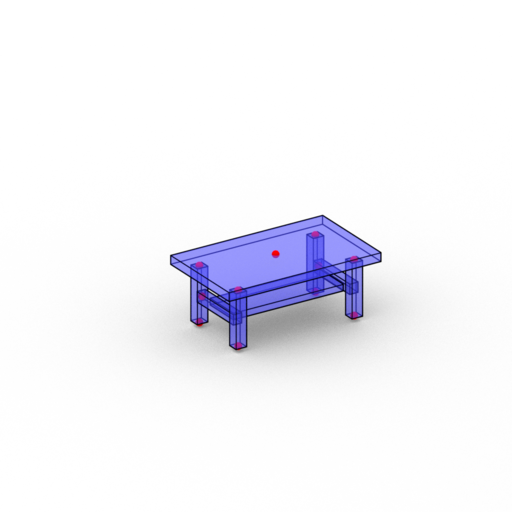} 
        \\
        \raisebox{1em}{\rotatebox{90}{StructureNet}} &
        \raisebox{3em}{\multirow{2}{*}{\includegraphics[trim={4.5cm 4.5cm 4.5cm 4.5cm},clip,width=.13\linewidth]{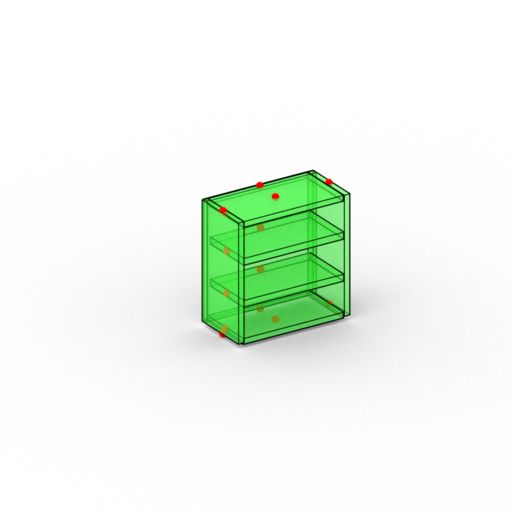}}} &
        \includegraphics[trim={4.5cm 4.5cm 4.5cm 4.5cm},clip,width=.13\linewidth]{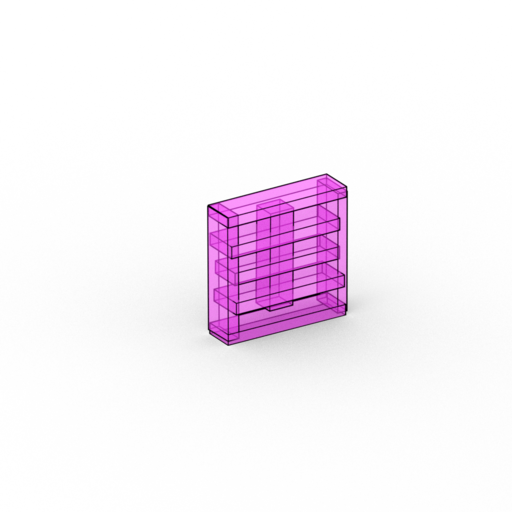} &
        \includegraphics[trim={4.5cm 4.5cm 4.5cm 4.5cm},clip,width=.13\linewidth]{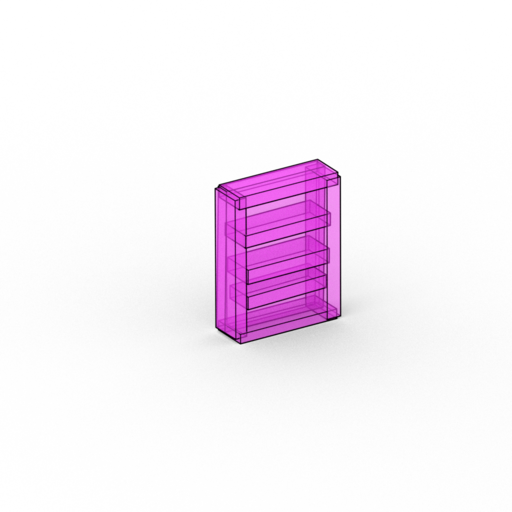} &
        \includegraphics[trim={4.5cm 4.5cm 4.5cm 4.5cm},clip,width=.13\linewidth]{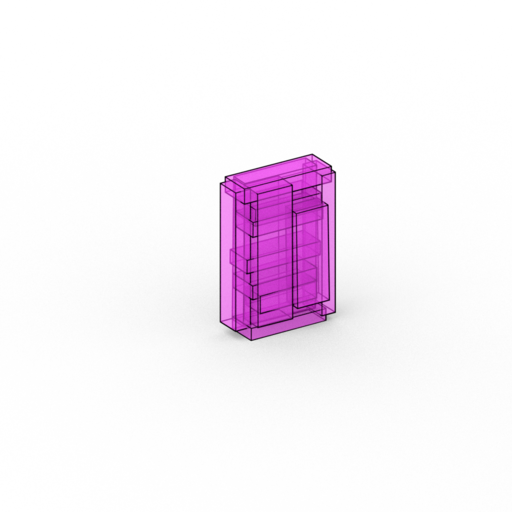} &
        \includegraphics[trim={4.5cm 4.5cm 4.5cm 4.5cm},clip,width=.13\linewidth]{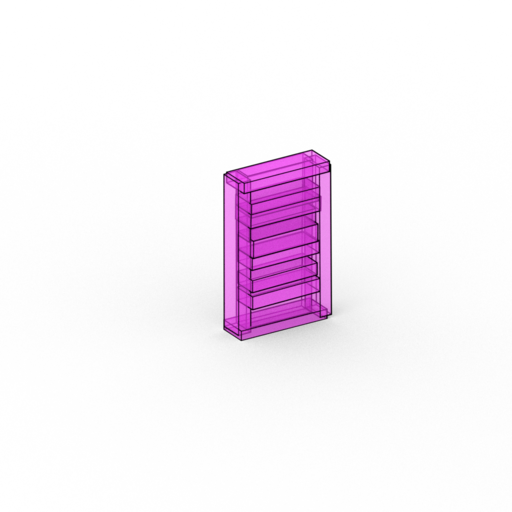} &
        \includegraphics[trim={4.5cm 4.5cm 4.5cm 4.5cm},clip,width=.13\linewidth]{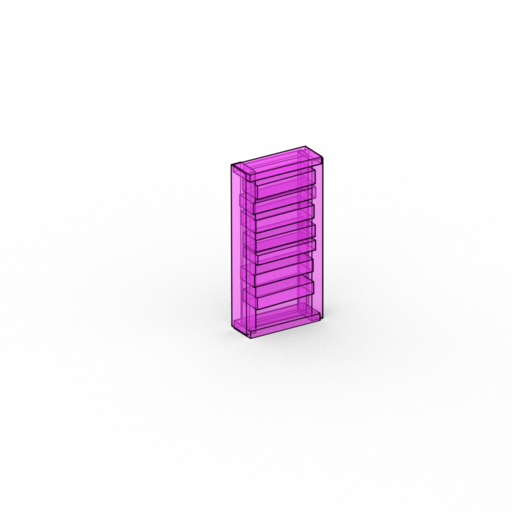} &
        \raisebox{3em}{\multirow{2}{*}{\includegraphics[trim={4.5cm 4.5cm 4.5cm 4.5cm},clip,width=.13\linewidth]{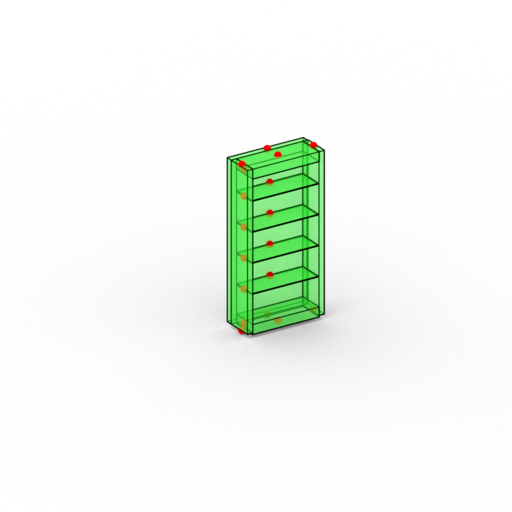}}} 
        \\
        \raisebox{3em}{\rotatebox{90}{Ours}} &
        &
        \includegraphics[trim={4.5cm 4.5cm 4.5cm 4.5cm},clip,width=.13\linewidth]{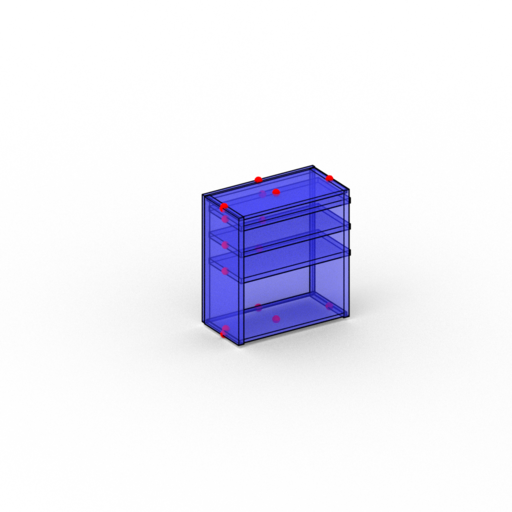} &
        \includegraphics[trim={4.5cm 4.5cm 4.5cm 4.5cm},clip,width=.13\linewidth]{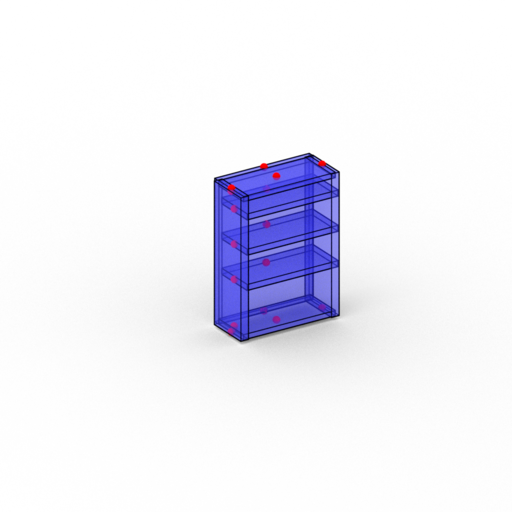} &
        \includegraphics[trim={4.5cm 4.5cm 4.5cm 4.5cm},clip,width=.13\linewidth]{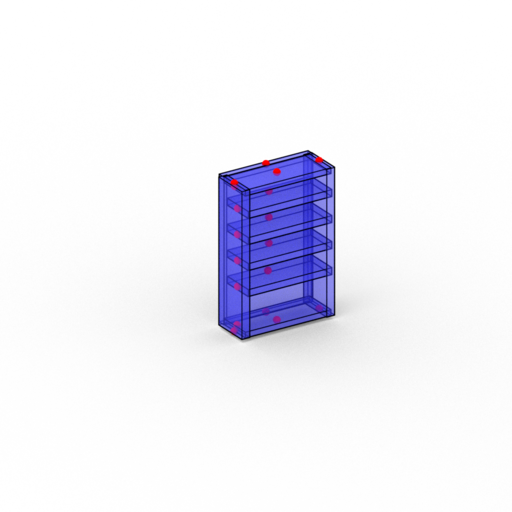} &
        \includegraphics[trim={4.5cm 4.5cm 4.5cm 4.5cm},clip,width=.13\linewidth]{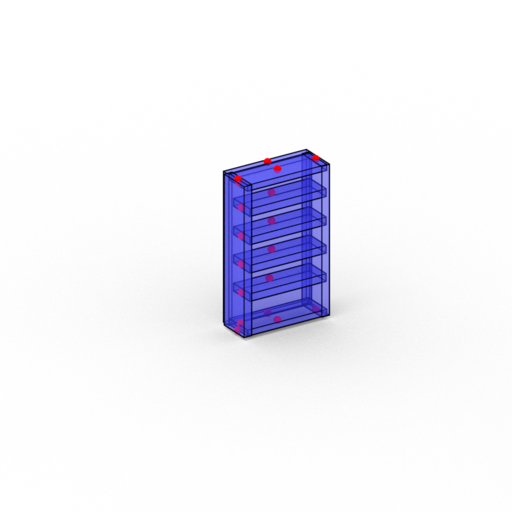} &
        \includegraphics[trim={4.5cm 4.5cm 4.5cm 4.5cm},clip,width=.13\linewidth]{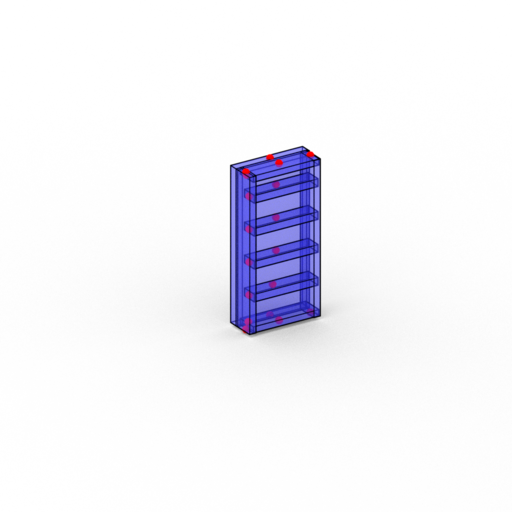} 
    \end{tabular}
    \caption{A qualitative comparison of latent space interpolation between our method and StructureNet on shapes from the validation set. Our method's interpolations within program space produce sequences that combine smooth continuous variation with discrete structural transitions. 
    }
    \label{fig:interpolation_examples}
\end{figure*}

\subsection{Latent Space Interpolation}

Beyond novel shape generation, we evaluate the ability of our method to interpolate between two points in our latent space. 
The presence of smooth, semantic transitions between end-points indicates a well-formed latent space. 
In Figure~\ref{fig:interpolation_examples} we qualitatively compare our method with StructureNet on the task of interpolating between shapes in the validation sets of both models.
Our interpolations demonstrate both geometrically smooth and semantically consistent transitions. 
For instance, in the top interpolation sequence, the surface of the chair back in the source shape gradually shrinks vertically until in the target shape it is just a horizontal bar. 
At the same time, the number of vertical slats in the chair back gradually increases from 2, to 4, to 5.

In Table~\ref{tab:interpolation}, we attempt to quantify the smoothness along random interpolation sequences within the latent space of each generative model. 
In this experiment, 100 interpolation sequences were computed from sources to targets that were randomly sampled in each model's latent space, with 100 interpolation steps per sequence. 
Each method's geometric smoothness is computed by taking the average Chamfer distance (normalized by shape scale) between each interpolation step. 
The lower geometric smoothness of our method, compared to StructureNet in the Table and Storage categories, demonstrates the quality of the latent space learned by our method. 
Moreover, using our procedure to turn StructureNet outputs into~\dslname~programs, we can measure the program smoothness along these interpolation paths. 
Each method's program smoothness is computed by taking the average tokenized program edit distance between each interpolation step. 
As a measure for structural change throughout the transitions of an interpolation sequence, our lower program smoothness metric again shows how our method benefits by operating within the space of 3D shape programs.

\subsection{Synthesis from Unstructured Geometry}

Another way to inspect the structure of a generative model's latent space is through performing ``synthesis from X",
by projecting X into the latent space of the generative model. 
As an application for 3D reconstruction, we are able perform such a projection with point clouds, demonstrating how our generative model's latent space can synthesize~\dslname~programs from unstructured geometry. 

Specifically, we train a PointNet$++$ encoder~\cite{qi2017pointnet++} to map point clouds sampled on dense mesh geometry to the latent space learned by our generative model. These latent codes are then converted into programs by our trained decoder. 

In Table~\ref{tab:pc_recon}, we show an experiment comparing our method against StructureNet on the task of reconstructing point cloud samplings of dense geometry on the intersection of each method's validation set for Chairs in Partnet (463 shapes total). We evaluate reconstruction accuracy with F-score~\cite{TanksAndTemples}, and the physical validity of reconstructions with the rootedness and stability metrics. When projecting point clouds into the latent space of each method (top two rows), our method outperforms StructureNet on both reconstruction accuracy and maintaining physical validity. This demonstrates, once again, the well-structured nature of our method's latent space.

Moreover, as the~\dslname~ interpreter is differentiable, we can further refine the continuous parameters of a program by minimizing the Chamfer distance between executed geometry and a target point cloud with a gradient-based optimizer.
We compare this procedure \textbf{(Ours + Opt Program)} against the following conditions:
\begin{itemize}
\denselist
    \item \textbf{SN + Opt Cuboids}: Starting with StructureNet's reconstruction, then directly optimizing predicted cuboids to minimize Chamfer distance to the target point cloud.
    \item \textbf{SN + Opt Program}: Parsing StructureNet's reconstruction into a~\dslname~program, then optimizing the program to minimize Chamfer distance to the target point cloud.
    \item \textbf{Ours + Opt Cuboids}: Starting with our reconstruction, directly optimizing predicted cuboids to minimize Chamfer distance to the target point cloud.
\denselist
\end{itemize}

We show results for this experiment in the last four rows of Table~\ref{tab:pc_recon}. All of the optimization procedures improve reconstruction accuracy at the cost of physical validity. However, Ours + Opt Program is the only condition that achieves a desirable trade-off in this exchange, gaining much more reconstruction accuracy improvement than it loses in physical validity.

We show some qualitative results of this experiment in Figure~\ref{fig:point_cloud_reconstruct}. Through latent space projection, our model is able to output the rough 3D shape structure (column 1) of an input unstructured point cloud (column 0). Through our differentiable interpreter, we are able to find continuous parameters for the predicted program structure that ultimately lead to better reconstruction fits (column 3). Shape programs place a strong structural regularization prior over unstructured 3D data, and thus our presented method is less prone to ``losing'' semantic parts, such as small legs, in comparison to the other conditions.

\begin{table}[t!]
    \centering
    \small
    % \footnotesize
    \setlength{\tabcolsep}{2pt}
    %\vspace{-1em}
    \begin{tabular}{@{}lccc@{}}
        \toprule
        \textbf{Method} & \textbf{F1}$\;\Uparrow$ & \textbf{\% rooted}$\;\Uparrow$ & \textbf{\% stable}$\;\Uparrow$  \\
        \midrule
        StructureNet & 24.3 & 95.1 & 78.4 \\
        Ours & \textbf{31.1} & \textbf{95.5} & \textbf{84.4} \\
        \midrule
        \midrule
        SN + Opt Cuboids & \textbf{80.0} & 92.9 & 72.7 \\
        SN + Opt Program & 77.4 & 90.0 & 71.9 \\
        Ours + Opt Cuboids &  77.6 & 93.1 & 72.9 \\
        Ours + Opt Program & 75.8 & \textbf{95.3} & \textbf{80.2} \\
        \bottomrule
    \end{tabular}
    \caption{
    Results from our point cloud reconstruction experiment. Our model's well-formed latent space allows for more accurate and physically valid reconstructions without further optimization. With additional optimization, using the reconstructed program from our method and our differentiable interpreter finds the best trade-off between reconstruction accuracy and maintaining physical validity. }
    \label{tab:pc_recon}
\end{table}

\begin{figure*}[t!]
    \centering
    \setlength{\tabcolsep}{1pt}
    \small
    \begin{tabular}{ccccccc}
        Input Points & Ours & Ours + Opt Cuboids & Ours + Opt Program & SN & SN + Opt Cuboids & SN + Opt Program
        \\
        \includegraphics[trim={4.5cm 4.5cm 4.5cm 4.5cm},clip,width=.14\linewidth]{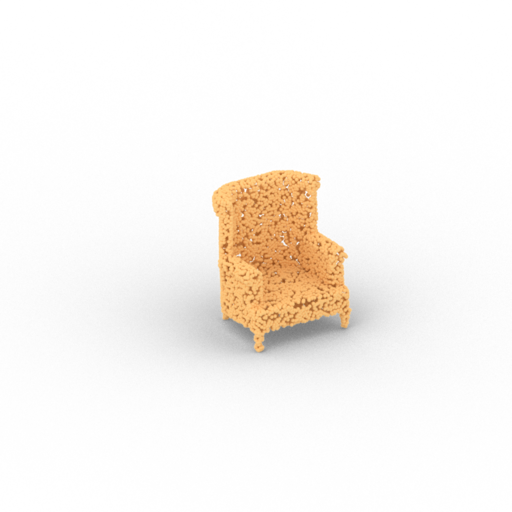} &
        \includegraphics[trim={4.5cm 4.5cm 4.5cm 4.5cm},clip,width=.14\linewidth]{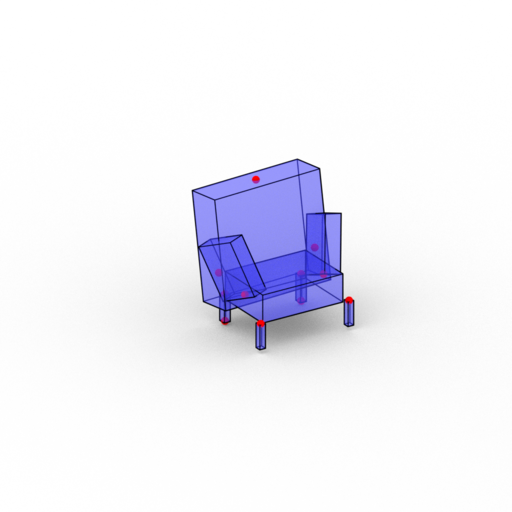} &
        \includegraphics[trim={4.5cm 4.5cm 4.5cm 4.5cm},clip,width=.14\linewidth]{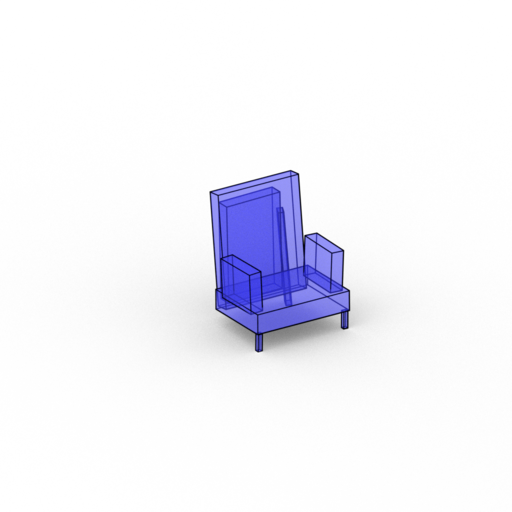} &
        \includegraphics[trim={4.5cm 4.5cm 4.5cm 4.5cm},clip,width=.14\linewidth]{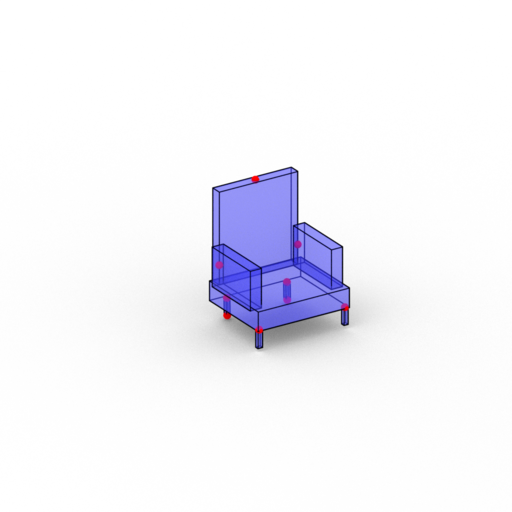} &
        \includegraphics[trim={4.5cm 4.5cm 4.5cm 4.5cm},clip,width=.14\linewidth]{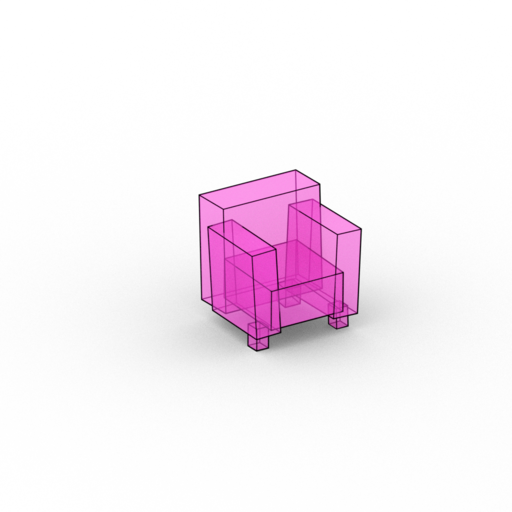} &
        \includegraphics[trim={4.5cm 4.5cm 4.5cm 4.5cm},clip,width=.14\linewidth]{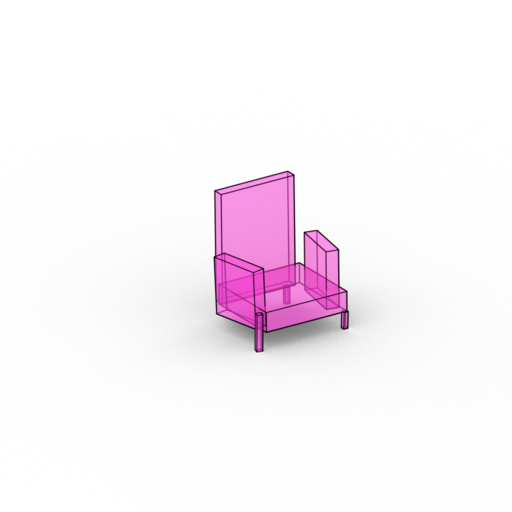} &
        \includegraphics[trim={4.5cm 4.5cm 4.5cm 4.5cm},clip,width=.14\linewidth]{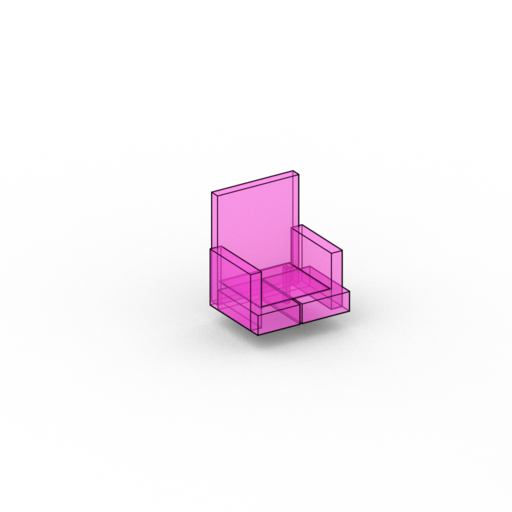}

        \\
        \includegraphics[trim={4.5cm 4.5cm 4.5cm 4.5cm},clip,width=.14\linewidth]{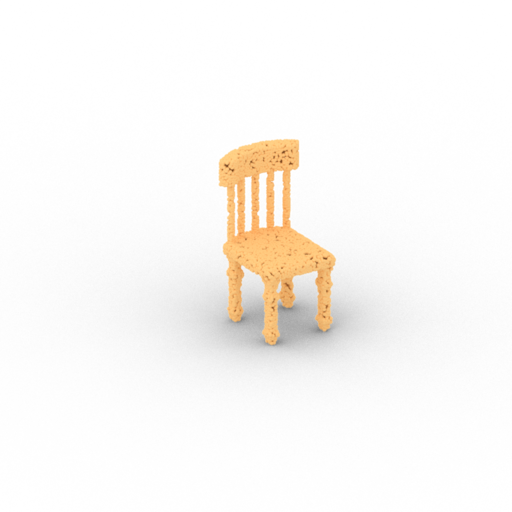} &
        \includegraphics[trim={4.5cm 4.5cm 4.5cm 4.5cm},clip,width=.14\linewidth]{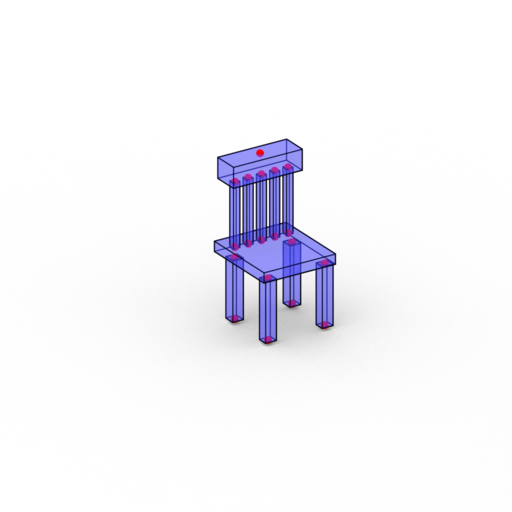} &
        \includegraphics[trim={4.5cm 4.5cm 4.5cm 4.5cm},clip,width=.14\linewidth]{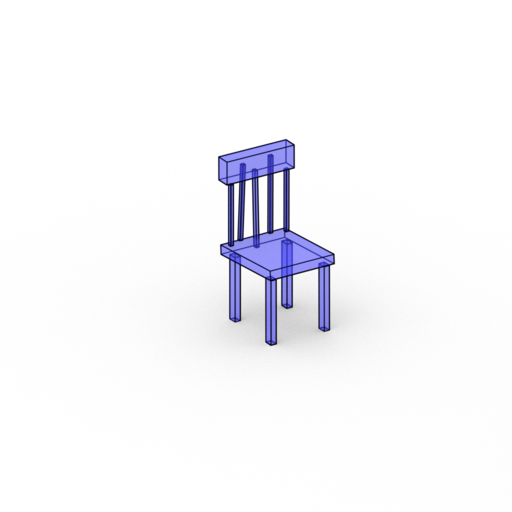} &
        \includegraphics[trim={4.5cm 4.5cm 4.5cm 4.5cm},clip,width=.14\linewidth]{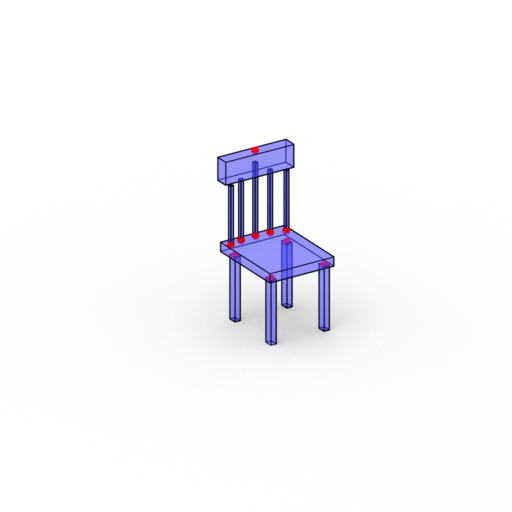} &
        \includegraphics[trim={4.5cm 4.5cm 4.5cm 4.5cm},clip,width=.14\linewidth]{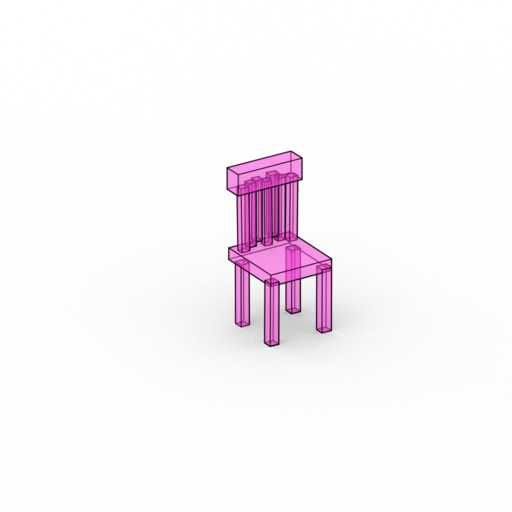} &
        \includegraphics[trim={4.5cm 4.5cm 4.5cm 4.5cm},clip,width=.14\linewidth]{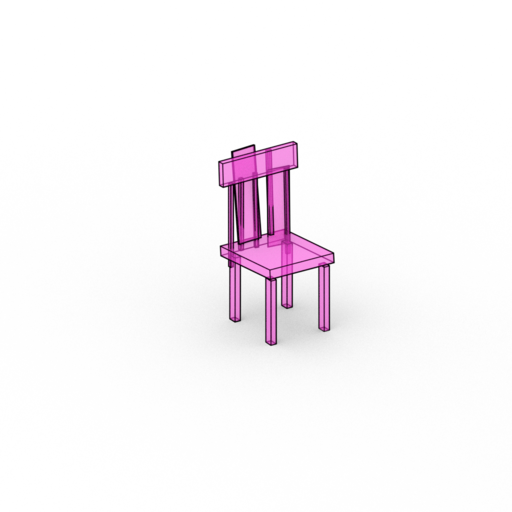} &
        \includegraphics[trim={4.5cm 4.5cm 4.5cm 4.5cm},clip,width=.14\linewidth]{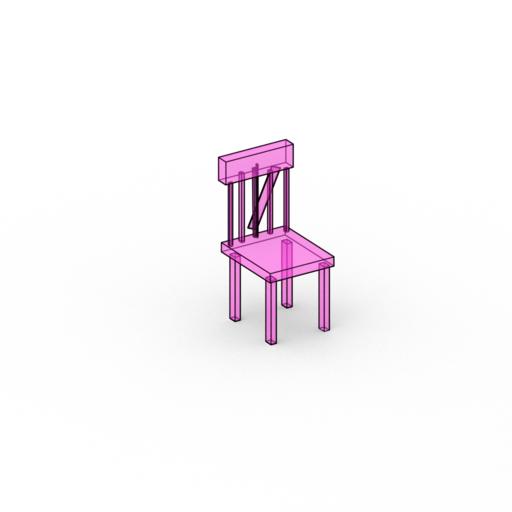}
        \\
        \includegraphics[trim={4.5cm 4.5cm 4.5cm 4.5cm},clip,width=.14\linewidth]{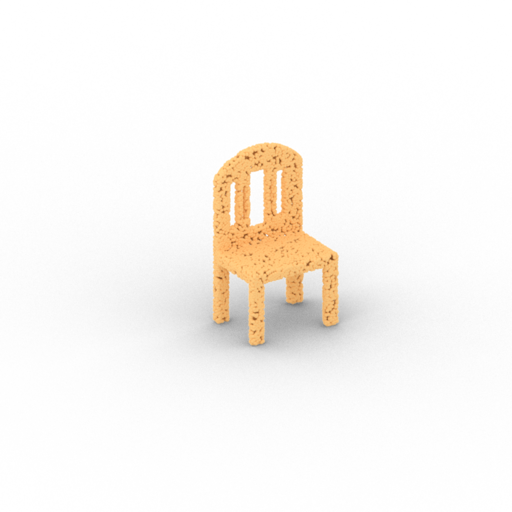} &
        \includegraphics[trim={4.5cm 4.5cm 4.5cm 4.5cm},clip,width=.14\linewidth]{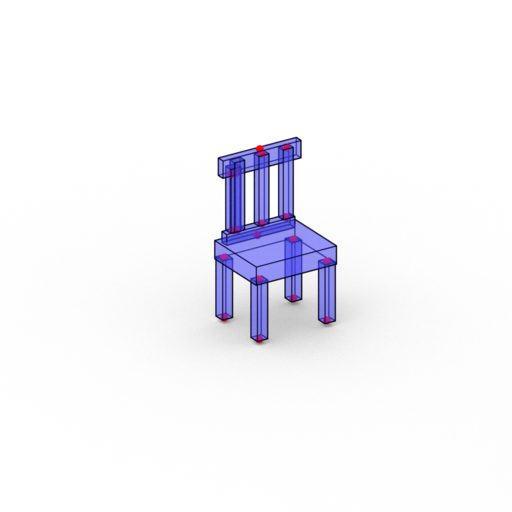} &
        \includegraphics[trim={4.5cm 4.5cm 4.5cm 4.5cm},clip,width=.14\linewidth]{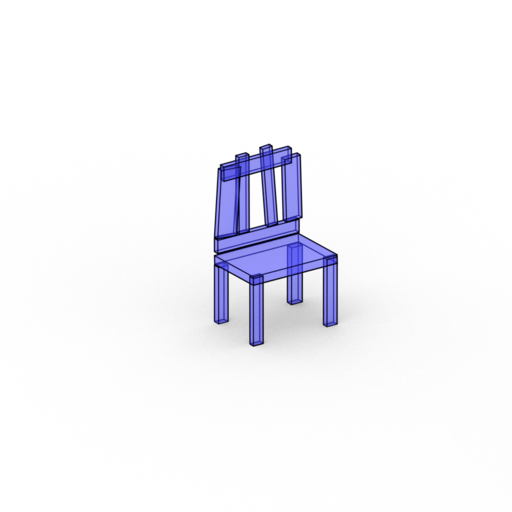} &
        \includegraphics[trim={4.5cm 4.5cm 4.5cm 4.5cm},clip,width=.14\linewidth]{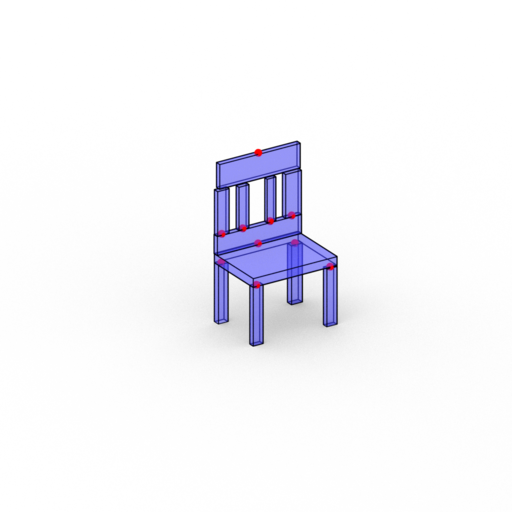} &
        \includegraphics[trim={4.5cm 4.5cm 4.5cm 4.5cm},clip,width=.14\linewidth]{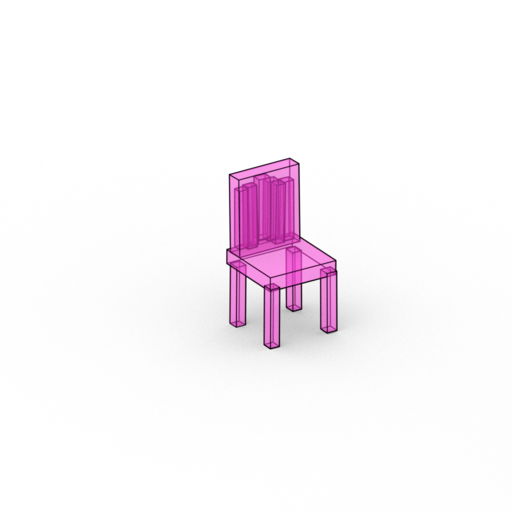} &
        \includegraphics[trim={4.5cm 4.5cm 4.5cm 4.5cm},clip,width=.14\linewidth]{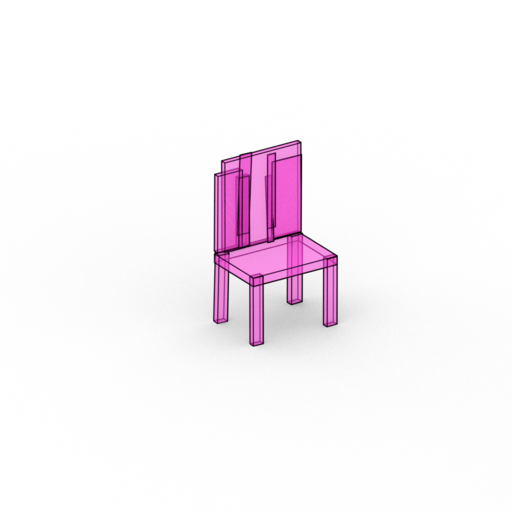} &
        \includegraphics[trim={4.5cm 4.5cm 4.5cm 4.5cm},clip,width=.14\linewidth]{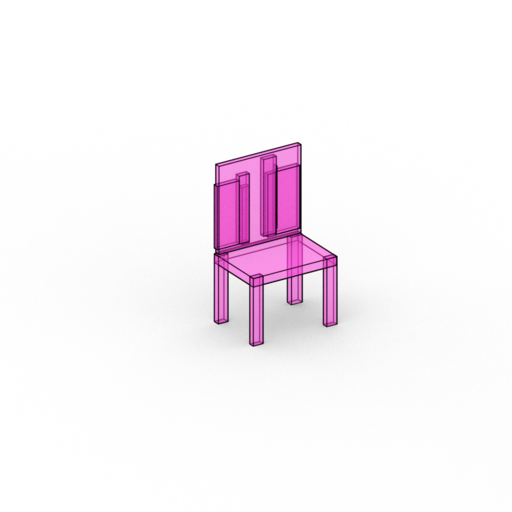} 
    \end{tabular}
    \caption{Qualitative comparison of synthesis from point clouds of our method against StructureNet (SN). Our method is able to infer good program structures that match well with the unstructured geometry. The continuous parameters of this program structure can be further refined through an optimization procedure in order to better fit the target point cloud without creating artifacts.
    }
    \label{fig:point_cloud_reconstruct}
\end{figure*}

%% file: 06-conclusion.tex
\section{Conclusion and Future Work}
\label{sec:conclusion}

In this paper, we took a first step toward marrying the complementary strengths of neural and procedural 3D shape generative models by introducing a hybrid neural-procedural approach for synthesizing novel 3D shape structures.
We introduced~\dslname, a low-level ``assembly language'' for shape structures, in which shapes are constructed by declaring cuboidal parts and attaching them to one another.
We also introduced a differentiable interpreter for~\dslname, allowing the optimization of program parameters to produce desired output geometry.
After describing how to extract consistent programs from existing shape structures in the PartNet dataset, we then defined a deep generative model for~\dslname~programs, effectively training a neural network to write novel shape programs for us.
We evaluated the quality of the generative model along several axes, showing that it produces more plausible and physically-valid shapes, and that its latent space is better-structured than that of other generative models of shape structure.
We also found that directly generating shape programs leads to more compact, editable programs than extracting programs from shapes generated by methods that directly output 3D geometry.

\begin{figure}[t!]
   \centering
   \setlength{\tabcolsep}{1pt}
     \begin{tabular}{ccc}
         \includegraphics[width=.5\linewidth]{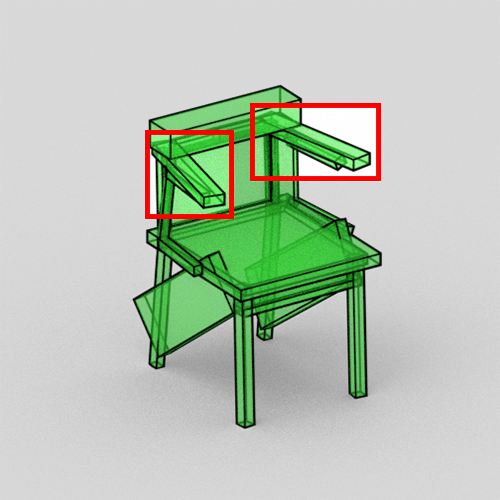} &
      \includegraphics[width=.5\linewidth]{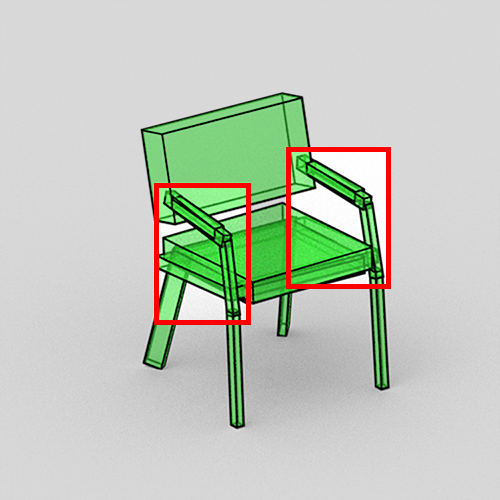}
     \end{tabular}
 \caption{Examples of PartNet shapes that contain parts whose orientations cannot be inferred from part-to-part attachments alone. While these shapes can be represented with ~\dslname~ programs that attach parts to ``floating'' points within the bounding volume, such programs are not added to our training data during our program extraction phase. As a result, our generative model never learns to produce shapes that require this type of attachment pattern. }
 \label{fig:global_attach}
 \end{figure}

\subsubsection*{Limitations}
As mention in Section~\ref{sec:progextraction}, we do not successfully extract training programs from every shape in our dataset.

For instance, our program extraction procedure assumes that the orientation of all parts can be specified through solely part-to-part attachments, yet as demonstrated in Figure ~\ref{fig:global_attach}, this does not hold for all shapes.
While it is possible to reconstruct these shapes with~\dslname~programs (through attaching parts to ``floating'' points in space via the bounding volume) such programs will never be added to our training data, and thus our generative model won't learn to produce such constructs. 
Our design decision to discard training programs with more than 12 total \texttt{Cuboid} declarations has a similar effect: it limits our generative model from synthesizing the most complex of shape structures that exist in our dataset. 
We impose such strict criteria in order to make our training programs exhibit more regularity, simplifying the learning task for our neural network at the expense of its potential expressivity.

This highlights a central tradeoff: higher variability in the training programs may result in lower quality shapes synthesized by a generative model.
This phenomenon is not unique to our setting: it is well-known that e.g., image generative models perform better on very-regularly-structured domains, such as human faces.
The question, looking forward, is how to capture more data variability while keeping a high-degree of regularity in the input data representation?
We believe that using programs as a data representation is the best avenue of attack, here.
As we have shown in our work, a single program can capture a wide range of parametrically related shapes.
One program, many shapes; strong regularity, but also high variability.
We are excited to investigate extensions of~\dslname, as well as other shape-generating languages, which can capture even more shape variability with highly-regular structures.

While~\dslname~has a strong inductive bias for generating physically-connected shapes, it is not guaranteed to do so.
Hierarchical part structures which are locally connected everywhere may occasionally still exhibit disconnected leaf cuboids.
This is more likely to happen with very non-axis-aligned structures that result in loose bounding cuboids at the intermediate levels of the hierarchy.
It is worth investigating mechanisms to guarantee that~\dslname~programs maintain leaf-to-leaf connectivity.

\subsubsection*{Future work} In Section~\ref{sec:results}, we showed an example of refining our generated hierarchical cuboid structures with point cloud surface geometry.
This is not a new idea; other recent related work on part-based shape generation takes a similar approach to refining high-level part structures~\cite{GRASS,StructureNet,gao2019sdmnet}.
However, there is more work to be done at the intersection of structure generation and surface generation.
These two paradigms could be much more closely married than they have been thus far, as existing part-surface generation has been explored largely in an independent, part-by-part fashion.
What would it look like to swap out the ``surface style'' code for a shape while retaining its ``structure'' code?
Our procedural representation may confer distinct advantages here, as the attachments explicitly specify where and how part geometries must connect.

It would also be interesting to move beyond cuboids as the proxy geometry used for atomic parts, as not all atomic parts are well-approximated by rectilinear geometry.
In some cases, spherical, cylindrical, or more general curvilinear geometry would be a better choice.
Pursuing this direction would help push more shape variability into the procedural representation, so that we do not lean so heavily on the neural network to capture it.

Another way to push knowledge from the learned latent space into the programs would be to make the programs include constraints on their parameters: either independent bounds, or correlations between parameters.
For instance, it is non-semantic to make a chair leg too thin, or to make a chair back much narrower than the seat to which it is attached.
It should be possible to mine shape datasets for this information, and to include it in the data used to train the generative model.

There are also more opportunities to apply generative models of shape programs to ``synthesis from X'' applications.
While we showed translation from point clouds to shape programs, there are many more exciting possibilities in terms of linking 3D geometry, 2D images, and shape programs, and seamlessly using the three modalities to author different forms of shape manipulations. 

Finally, if we aim for our generated shapes to be useful in embodied AI applications, they should also be equipped with information about kinematics and/or dynamics.
For example, a program which specifies a cabinet could also specify the type of hinge with which the door attaches to the body, and how far that hinge opens.
Ultimately, we believe that shape programs, and generative models which produce them, are the right fundamental representation for both human creative tasks and AI analysis tasks involving part-based 3D shapes.

%% file: 07-acknowledgments.tex
\begin{acks}
We would like to thank the anonymous reviewers for their helpful suggestions.
Renderings of part cuboids and point clouds were produced using the Blender Cycles renderer.
This research was supported by the National Science Foundation (\#1753684, \#1941808), a Brown University Presidential Fellowship, gifts from the University of College London AI Center and Adobe Research, and by GPU donations from NVIDIA.
Daniel Ritchie is an advisor to Geopipe, Inc. and owns equity in the company.
Geopipe is a start-up that is developing 3D technology to build immersive virtual copies of the real world with applications in various fields, including games and architecture.
\end{acks}

%% file: appendix.tex
\appendix

\section{Semantics of the \texttt{attach} Command}
\label{sec:attach_semantics}

\begin{figure}
    \centering
    \includegraphics[width=\linewidth]{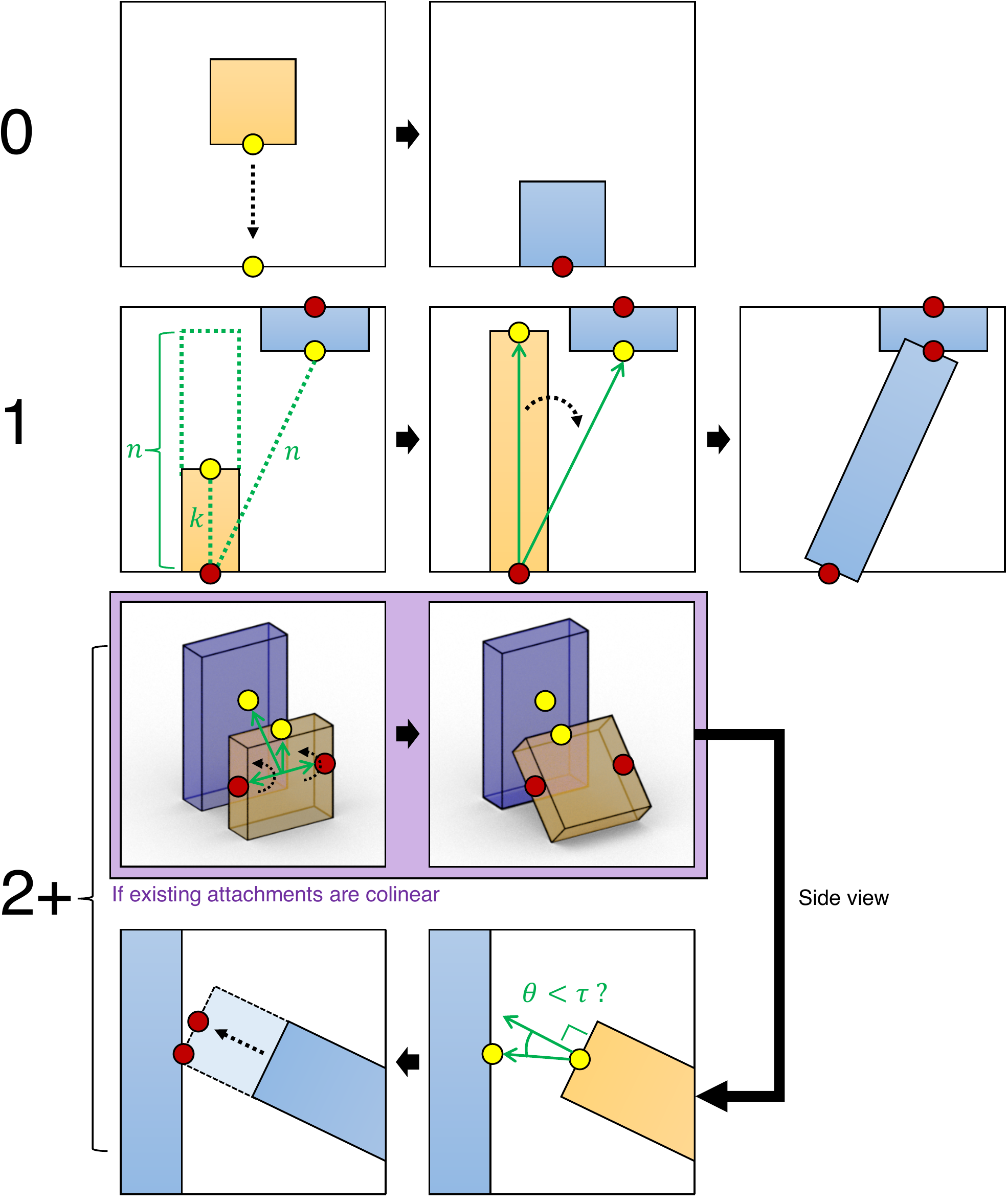}
\caption{
Illustrating how the \texttt{attach} command executes, depending on the number of existing attachments (left column) to the cuboid in question.
Cuboids with no existing attachments can simply be translated into place (top).
Cuboids with one existing attachment can be scaled along one axis and then rotated (middle).
Cuboids with two or more existing attachments are more complicated, and the attachment may not always be satisfiable.
Our interpreter attempts to rotate and scale the cuboid to get as close as possible to valid solution. 
}
\label{fig:attach_semantics}
\end{figure}

In designing the~\dslname~interpreter, our goal is to ensure that its internal operations stay limited to simple fixed-function, differentiable operations.
Thus, implementing the \texttt{attach} command, we opt not to use any constrained optimization routines which could resolve a globally-optimal configuration of cuboids given the attachment constraints.
Instead, the interpreter immediately executes each attachment as it is declared, i.e. it greedily solves for attachments.
To make the behavior of this procedure as predictable as possible, the greedy attachment procedure should induce the fewest changes possible to the current cuboid shapes.

With these desiderata in mind, we designed the following procedure for attaching cuboid $c_1$ to cuboid $c_2$ (see Figure~\ref{fig:attach_semantics}).
The logic that executes depends upon how many prior attachments $c_1$ has and the aligned flag of $c_1$:

\subsubsection*{No prior attachments}
In this case, cuboid $c_1$ can connect to cuboid $c_2$ by simply translating until the attach points are colocated.

\subsubsection*{One prior attachment}
Here, the interpreter scales cuboid $c_1$ along one of its axes and then rotates it such that the attachment is satisfied.
To choose the axis along which to scale $c_1$, the interpreter checks how quickly scaling each of its three dimensions would reduce the ratio $n/k$, where $n$ is the distance between $c_1$'s existing attachment point and the new target attachment point, and $k$ is the distance between $c_1$'s existing attachment point and the new source attachment point.
The interpreter then scales $c_1$ by $n/k$ along this dimension, which gives it the correct length.
Finally, $c_1$ is rotated such that the source and target attachment points are colinear (and thus colocated).

\subsubsection*{Two or more prior attachments}
In this case, it is not always possible to satisfy the attachment, as three point constraints on a cube may be overconstrained.
If a solution exists, however, our interpreter will find it.
And in the case where no solution exists, it attempts to approximately satisfy the attachment (which we decided to be more user-friendly behavior than throwing an error).

First, the interpreter checks if $c_1$'s existing attachment points are all colinear.
If they are, then it rotates $c_1$ about this axis of colinearity to make the source attachment point face the target attachment point.
The final step is to scale $c_1$ along the normal of the face containing the source attachment point.
If the existing attachment points were not colinear, and this face was not rotated to point toward the target attachment point, then this may not be a useful operation (i.e. it may introduce undesirable change to the cuboid shape while doing little to bring the source point closer to the target point).
Thus, the interpreter only executes this scale if the angle between the source face normal and the vector to the target point is smaller than a threshold $\tau$ (25 degrees in our implementation).

\subsubsection*{Aligned Cuboids}
Cuboids that are marked as aligned in ~\dslname~ programs cannot have their orientations changed through attachment. In fact, with correct cuboid dimension parameterization, a single attachment is enough to properly position and orient an aligned cuboid. However, in order to ensure that  aligned Cuboids remain connected through edits and predictions of our generative model, we minimally grow aligned cuboid dimensions to satisfy the part-to-part connectivity specified through attachments. That is, for aligned cuboids we do not guarantee attachment point colocation after the first attachment, as this is often impossible to exactly fulfill without changing a cuboid's orientation. Rather, we guarantee that aligned cuboids will fulfill attachment relationships with cuboids they are attached to at \textit{some} attachment point.

\section{Semantics of ~\dslname~ Macro Functions}
\label{sec:macro_expansion}

We provide an account of the logic for macro function expansion in ~\dslname~:

\subsubsection*{Squeeze.} The \texttt{squeeze} macro is parameterized by three cuboids ($c_{n1}$, $c_{n2}$, $c_{n3}$) a face $f$ and a $(u,v)$ position on $f$'s 2D coordinate system. 
A \texttt{squeeze} command expands into two \texttt{attach} functions.
The first attach function attaches the center of $c_{n1}$'s $f$ face to the $(u,v)$ position on the opposite face of $f$ on $c_{n2}$. 
The second attach function attaches the center of $c_{n1}$'s opposite face of $f$ to the $(u,v)$ position on the face of $f$ on $c_{n3}$. 
For example, the line \texttt{squeeze} ($c_{n1}, c_{n2}, c_{n3}$, left, .1, .4).
It expands into \texttt{attach}($c_{n1}, c_{n2}$, 0.0, .5, .5, 1.0, .1, .4) and \texttt{attach}($c_{n1}, c_{n3}$, 1.0, .5, .5, 0.0, .1, .4).

\subsubsection*{Reflect.} The \texttt{reflect} macro is parameterized by a cuboid $c_{n}$ and an axis $a$. 
A \texttt{reflect} command first expands into one \texttt{Cuboid} function, that creates a new cuboid $c_{n'}$ with the same parameters as $c_{n}$. Then for every previous attachment line pair that had moved $c_{n}$, of the form  
\texttt{attach}$(c_{n}, c_{m}, x_1, y_1, z_1, x_2, y_2, z_2)$, the reflect command creates a new attachment line:
\texttt{attach}$(c_{n'}, c_{m}$, $x_1, y_1, z_1$, 
$R(x_1, y_1, z_1, c_{n}, c_{m}, a))$. 
$R$ is a function that applies a reflection of the global point specified by $(x_1, y_1, z_1)$ in the local coordinate frame of $c_{n}$ about the axis $a$, and then returns the local coordinates of that point within $c_{m}$.

\subsubsection*{Translate.} The \texttt{translate} macro is parameterized by a cuboid $c_{n}$, an axis $a$, a number of members $m$, and a distance $d$. A \texttt{translate} command first expands into $m$ \texttt{Cuboid} functions, that each creates a new cuboid $c_{n_i}$ with the same parameters as $c_{n}$. Then for every previous attachment line pair that had moved $c_{n}$, of the form  \texttt{attach}($c_{n}, c_{m}$, $x_1, y_1, z_1$, $x_2, y_2, z_2$), the translate command creates a new attachment line \texttt{attach}($c_{n_i}, c_{m}$, $x_1, y_1, z_1$, 
$T(x_1, y_1, z_1, c_{n}, c_{m}, a, d))$. 
$T$ is a function that applies a translation of the global point specified by $(x_1, y_1, z_1)$ in the local coordinate frame of $c_{n}$ along the axis $a$ (of the bounding volume) for for a distance of $d$ (where d is normalized by the size of the bounding volume), and then returns the local coordinates of that point within $c_{m}$.

\section{Program Extraction Procedure}
\label{sec:detailed_prog_extract}

Here, we provide an account of our program extraction procedure in greater detail:

\subsubsection*{Part Shortening} Before any hierarchical processing, we first attempt to regularize any artifacts in the input data. Specifically, for each leaf cuboid part proxy, we check if any of its faces are completely contained within any other leaf cuboid. If we find that we can shorten a leaf cuboid without changing the visible, non-intersecting, geometry of the part graph, we do so. 

\subsubsection*{Semantic Hierarchy Arrangement} During our data preprocessing stage when converting PartNet part graphs into ~\dslname~ programs, we locally flatten part graph hierarchies based on semantic rules as depicted in Figure \ref{fig:progextract}. For chairs we flatten the following nodes: back, arm, base, seat, footrest and head. For tables we flatten the following nodes: top and base. For storage we flatten the following nodes: cabinet frame, cabinet base. For storage, we move the following nodes into the cabinet frame sub-program: countertop, shelf, drawer, cabinet door and mirror. We also perform a semantic collapsing step where the intermediate nodes containing detailed geometry are converted into leaf nodes and their children are discarded. For chairs we collapse the following nodes: caster and mechanical control. For tables we collapse the following nodes: caster, cabinet door, drawer, keyboard tray. For storage we collapse the following nodes: drawer, cabinet door, mirror and caster. Empirically we observed that this method of hierarchy re-arrangements produces cleaner and more regularized training data for our generative model. 

\subsubsection*{Attachment Point Detection} In order to identify which cuboids connect, and where they connect, we use a point cloud intersection procedure. We sample a uniform 20x20x20 point cloud within the volume defined by each cuboid. To check if two cuboids are attached, we find the set of points in the pairwise point cloud comparison that have a minimum distance to any point in the other point cloud within a distance threshold determined by the scale of the larger cuboid. For cuboids that attach (i.e. this intersection set is non-zero) we sample a denser 50x50x50 point cloud within the bounds of the detected intersection volume, forming a set of candidate attachment points. From this set we first filter all attachment points that are outside of either cuboid. If any remaining attachment points form face-to-face connections between cuboids we choose them, otherwise we define the attachment as taking place at the mean of the remaining attachment points. With the same procedure, we also record if cuboids connect to the top  or bottom of the bounding volume. Sampled points with bounding volume local y-coordinates in the ranges of $[0, 0.05]$ and $[.95, 1.0]$ are assigned to the bottom and top respectively.

\subsubsection*{Symmetry Detection} We enforce that all members of a symmetry group share the same connectivity structure in the input part graph. 
Cuboids are grouped together by symmetry if they: (i) connect to the same cuboids, (ii) share a reflectional or translational symmetry about the X, Y or Z axis of their parent bounding volume, and (iii) each attachment point involved in their outgoing connections also shares this same symmetrical relationship. 
Two cuboids, or two attachment points, are considered to share a symmetrical relationship if applying the symmetry transformation matrix to one member produces a parameterization close to that of the other member. 

Notice that this procedure can disqualify symmetry formation about groups of interconnected cuboids that share a symmetrical relationship. 
As such, before forming symmetry groups about individual cuboids, we attempt to form symmetry groups about connected components of multiple cuboids. 
Whenever such a component is found, we locally abstract its structure with a bounding volume, and create a symmetry group sub-program. 
In this manner, we capture additional spatial symmetries while continuing to enforce the relationship between symmetry and part connectivity.
The "H-leg" program (Program3) in Figure~\ref{fig:pipeline} shows an example of where such a symmetry sub-program was formed. \\

In total, our parsing procedure finds valid ~\dslname~ programs for 46\% of Chairs, 65\% of Tables and 58\% of Storage shapes in PartNet.

\section{Decoder Semantic Validity Checks}
\label{sec:validity_checks}

During the process of decoding a latent code, our generative network enforces the following semantic validity conditions on its outputs:
\begin{itemize}
    \item XYZ attachment coordinates are clamped between 0 and 1.0. Additionally,  attachments to the bounding box can only be at the top or bottom faces with an allowable error of .05. 
    \item Cuboid dimensions are clamped between 0.01 and the corresponding bounding box dimension
    \item Bounding box cuboids can have no sub-programs
    \item Cuboids only attach at a single location. As an exception, cuboids are allowed to attach to both the top and bottom faces of the bounding volume.
    \item The bounding box cannot be moved by an attach command
    \item Attachment orderings must be grounded. Upon terminating, any ungrounded cuboids instantiations are discarded. 
    \item Symmetries can only operate on grounded cuboids
    \item The ordering of \text{Cuboid}, \text{attach}, \text{squeeze}, \text{reflect}, and \text{translate} lines must be consistent with the~\dslname~ grammar.     
    \item Commands must keep cuboids within the bounds of the defined bounding volume with an allowable error of 10\%. 
\end{itemize}

During generation, if our model predicts a non-semantic program line, we attempt to back-track until we are able to find a semantically valid solution. For instance, if we predict a new line to be a \texttt{reflect} command, but no cuboids have been grounded, we pick a new command type for the line by zeroing out the logits for the \text{reflect} command index. 

In some cases, a combination of bad continuous parameters and program structure predictions produce a violating line that cannot be easily fixed. During unconditional generation, we reject the sample if we encounter this behavior (this happens for  10\% - 20\% of our random samples across the categories we consider). We run an ablation on this rejection sampling in Table \ref{tab:quality}. During interpolation, we never reject a sample. Instead, we simply do not add lines to the predicted program for which we could not find a fix.

\section{Shape Quality Metrics}

We provide additional details about the metrics used in Table~\ref{tab:quality}:

\label{sec:quant_metrics}

\begin{itemize}
    \item \textbf{Rootedness :} We check if a connected path exists between the ground and all parts in the shape. We judge two parts to be connected if they are separated by a distance no larger than 2\% of the overall shape's bounding box diagonal length.
    \item \textbf{Stability :} We convert generated 3D shape structures into rigid bodies and place them in a physical simulation with gravity. A vertical force is applied to each shape proportional to its mass, along with some other small random forces and torques. If the resting height of any connected component of the shape changes by more than 10\% after these perturbations we declare it unstable. Note that this is by definition less than or equal to the percentage of rooted shapes, as a shape must be rooted in order to be stable. 
    \item \textbf{Realism:} The percentage of test set shapes classified as ``generated'' by a binary PointNet classifier trained to distinguish between generated shapes and shapes from the training dataset. The classifier is trained on an equal amount of positive and negative examples for $300$ epochs. We hold out a portion of shapes from the test set, and measure the percentage of them incorrectly classified as ``fake". To reduce fluctuation, the percentage is averaged over the last $50$ epochs.
\end{itemize}

%% file: main.bbl

\begin{thebibliography}{51}

%%% the Web address.

\ifx \showCODEN    \undefined \def \showCODEN     #1{\unskip}     \fi
\ifx \showDOI      \undefined \def \showDOI       #1{#1}\fi
\ifx \showISBNx    \undefined \def \showISBNx     #1{\unskip}     \fi
\ifx \showISBNxiii \undefined \def \showISBNxiii  #1{\unskip}     \fi
\ifx \showISSN     \undefined \def \showISSN      #1{\unskip}     \fi
\ifx \showLCCN     \undefined \def \showLCCN      #1{\unskip}     \fi
\ifx \shownote     \undefined \def \shownote      #1{#1}          \fi
\ifx \showarticletitle \undefined \def \showarticletitle #1{#1}   \fi
\ifx \showURL      \undefined \def \showURL       {\relax}        \fi
% invisible to TeX
\providecommand\bibfield[2]{#2}
\providecommand\bibinfo[2]{#2}
\providecommand\natexlab[1]{#1}
\providecommand\showeprint[2][]{arXiv:#2}

\bibitem[\protect\citeauthoryear{Abbatematteo, Tellex, and
  Konidaris}{Abbatematteo et~al\mbox{.}}{2019}]%
        {BenKinematics}
\bibfield{author}{\bibinfo{person}{Ben Abbatematteo}, \bibinfo{person}{Stefanie
  Tellex}, {and} \bibinfo{person}{George Konidaris}.}
  \bibinfo{year}{2019}\natexlab{}.
\newblock \showarticletitle{Learning to Generalize Kinematic Models to Novel
  Objects}. In \bibinfo{booktitle}{\emph{Proceedings of the Third Conference on
  Robot Learning}}.
\newblock


\bibitem[\protect\citeauthoryear{Chen, Tagliasacchi, and Zhang}{Chen
  et~al\mbox{.}}{2019}]%
        {BSPNet}
\bibfield{author}{\bibinfo{person}{Zhiqin Chen}, \bibinfo{person}{Andrea
  Tagliasacchi}, {and} \bibinfo{person}{Hao Zhang}.}
  \bibinfo{year}{2019}\natexlab{}.
\newblock \bibinfo{title}{BSP-Net: Generating Compact Meshes via Binary Space
  Partitioning}.
\newblock   (\bibinfo{year}{2019}).
\newblock
\showeprint[arxiv]{cs.CV/1911.06971}


\bibitem[\protect\citeauthoryear{Chen and Zhang}{Chen and Zhang}{2019}]%
        {IMNet}
\bibfield{author}{\bibinfo{person}{Zhiqin Chen} {and} \bibinfo{person}{Hao
  Zhang}.} \bibinfo{year}{2019}\natexlab{}.
\newblock \showarticletitle{{Learning Implicit Fields for Generative Shape
  Modeling}}. In \bibinfo{booktitle}{\emph{IEEE Conference on Computer Vision
  and Pattern Recognition (CVPR)}}.
\newblock


\bibitem[\protect\citeauthoryear{{Demir}, {Aliaga}, and {Benes}}{{Demir}
  et~al\mbox{.}}{2016}]%
        {InverseProceduralArchitecture}
\bibfield{author}{\bibinfo{person}{İ. {Demir}}, \bibinfo{person}{D.~G.
  {Aliaga}}, {and} \bibinfo{person}{B. {Benes}}.}
  \bibinfo{year}{2016}\natexlab{}.
\newblock \showarticletitle{Proceduralization for Editing 3D Architectural
  Models}. In \bibinfo{booktitle}{\emph{2016 Fourth International Conference on
  3D Vision (3DV)}}.
\newblock


\bibitem[\protect\citeauthoryear{Du, Inala, Pu, Spielberg, Schulz, Rus,
  Solar-Lezama, and Matusik}{Du et~al\mbox{.}}{2018}]%
        {InverseCSG}
\bibfield{author}{\bibinfo{person}{Tao Du}, \bibinfo{person}{Jeevana~Priya
  Inala}, \bibinfo{person}{Yewen Pu}, \bibinfo{person}{Andrew Spielberg},
  \bibinfo{person}{Adriana Schulz}, \bibinfo{person}{Daniela Rus},
  \bibinfo{person}{Armando Solar-Lezama}, {and} \bibinfo{person}{Wojciech
  Matusik}.} \bibinfo{year}{2018}\natexlab{}.
\newblock \showarticletitle{InverseCSG: Automatic Conversion of {3D} Models to
  CSG Trees}.
\newblock \bibinfo{journal}{\emph{ACM Trans. Graph.}} \bibinfo{volume}{37},
  \bibinfo{number}{6} (\bibinfo{date}{Dec.} \bibinfo{year}{2018}).
\newblock


\bibitem[\protect\citeauthoryear{Ellis, Nye, Pu, Sosa, Tenenbaum, and
  Solar-Lezama}{Ellis et~al\mbox{.}}{2019}]%
        {ellis2019write}
\bibfield{author}{\bibinfo{person}{Kevin Ellis}, \bibinfo{person}{Maxwell Nye},
  \bibinfo{person}{Yewen Pu}, \bibinfo{person}{Felix Sosa},
  \bibinfo{person}{Josh Tenenbaum}, {and} \bibinfo{person}{Armando
  Solar-Lezama}.} \bibinfo{year}{2019}\natexlab{}.
\newblock \showarticletitle{Write, Execute, Assess: Program Synthesis with a
  REPL}. In \bibinfo{booktitle}{\emph{Advances in Neural Information Processing
  Systems (NeurIPS)}}.
\newblock


\bibitem[\protect\citeauthoryear{Ellis, Ritchie, Solar-Lezama, and
  Tenenbaum}{Ellis et~al\mbox{.}}{2018}]%
        {ellis2018learning}
\bibfield{author}{\bibinfo{person}{Kevin Ellis}, \bibinfo{person}{Daniel
  Ritchie}, \bibinfo{person}{Armando Solar-Lezama}, {and} \bibinfo{person}{Josh
  Tenenbaum}.} \bibinfo{year}{2018}\natexlab{}.
\newblock \showarticletitle{{Learning to Infer Graphics Programs from
  Hand-{Drawn} Images}}. In \bibinfo{booktitle}{\emph{Advances in Neural
  Information Processing Systems (NeurIPS)}}.
\newblock


\bibitem[\protect\citeauthoryear{Fan, Su, and Guibas}{Fan
  et~al\mbox{.}}{2017}]%
        {PointSetGeneration}
\bibfield{author}{\bibinfo{person}{Haoqiang Fan}, \bibinfo{person}{Hao Su},
  {and} \bibinfo{person}{Leonidas~J Guibas}.} \bibinfo{year}{2017}\natexlab{}.
\newblock \showarticletitle{A point set generation network for {3D} object
  reconstruction from a single image}. In \bibinfo{booktitle}{\emph{Proceedings
  of the IEEE conference on computer vision and pattern recognition}}.
  \bibinfo{pages}{605--613}.
\newblock


\bibitem[\protect\citeauthoryear{Gao, Yang, Wu, Yuan, Fu, Lai, and Zhang}{Gao
  et~al\mbox{.}}{2019}]%
        {gao2019sdmnet}
\bibfield{author}{\bibinfo{person}{Lin Gao}, \bibinfo{person}{Jie Yang},
  \bibinfo{person}{Tong Wu}, \bibinfo{person}{Yu-Jie Yuan},
  \bibinfo{person}{Hongbo Fu}, \bibinfo{person}{Yu-Kun Lai}, {and}
  \bibinfo{person}{Hao~(Richard) Zhang}.} \bibinfo{year}{2019}\natexlab{}.
\newblock \showarticletitle{SDM-NET: Deep Generative Network for Structured
  Deformable Mesh}. In \bibinfo{booktitle}{\emph{SIGGRAPH Asia}}.
\newblock


\bibitem[\protect\citeauthoryear{Groueix, Fisher, Kim, Russell, and
  Aubry}{Groueix et~al\mbox{.}}{2018}]%
        {AtlasNet}
\bibfield{author}{\bibinfo{person}{Thibault Groueix}, \bibinfo{person}{Matthew
  Fisher}, \bibinfo{person}{Vladimir~G. Kim}, \bibinfo{person}{Bryan~C.
  Russell}, {and} \bibinfo{person}{Mathieu Aubry}.}
  \bibinfo{year}{2018}\natexlab{}.
\newblock \showarticletitle{{{AtlasNet}: A {Papier-M{\^{a}}ch{\'{e}}} Approach
  to Learning {3D} Surface Generation}}. In \bibinfo{booktitle}{\emph{IEEE
  Conference on Computer Vision and Pattern Recognition (CVPR)}}.
\newblock


\bibitem[\protect\citeauthoryear{Heusel, Ramsauer, Unterthiner, Nessler, and
  Hochreiter}{Heusel et~al\mbox{.}}{2017}]%
        {FrechetInceptionDistance}
\bibfield{author}{\bibinfo{person}{Martin Heusel}, \bibinfo{person}{Hubert
  Ramsauer}, \bibinfo{person}{Thomas Unterthiner}, \bibinfo{person}{Bernhard
  Nessler}, {and} \bibinfo{person}{Sepp Hochreiter}.}
  \bibinfo{year}{2017}\natexlab{}.
\newblock \showarticletitle{GANs Trained by a Two Time-Scale Update Rule
  Converge to a Local Nash Equilibrium}. In
  \bibinfo{booktitle}{\emph{NeurIPS}}.
\newblock


\bibitem[\protect\citeauthoryear{Hwang, Stuhlm\"{u}ller, and Goodman}{Hwang
  et~al\mbox{.}}{2011}]%
        {BayesianProgramMerging}
\bibfield{author}{\bibinfo{person}{Irvin Hwang}, \bibinfo{person}{Andreas
  Stuhlm\"{u}ller}, {and} \bibinfo{person}{Noah~D. Goodman}.}
  \bibinfo{year}{2011}\natexlab{}.
\newblock \showarticletitle{{Inducing Probabilistic Programs by Bayesian
  Program Merging}}.
\newblock \bibinfo{journal}{\emph{CoRR}}  \bibinfo{volume}{arXiv:1110.5667}
  (\bibinfo{year}{2011}).
\newblock


\bibitem[\protect\citeauthoryear{Johnson, Hariharan, van~der Maaten, Hoffman,
  Fei-Fei, Zitnick, and Girshick}{Johnson et~al\mbox{.}}{2017}]%
        {VQAPrograms}
\bibfield{author}{\bibinfo{person}{Justin Johnson}, \bibinfo{person}{Bharath
  Hariharan}, \bibinfo{person}{Laurens van~der Maaten}, \bibinfo{person}{Judy
  Hoffman}, \bibinfo{person}{Li Fei-Fei}, \bibinfo{person}{C~Lawrence Zitnick},
  {and} \bibinfo{person}{Ross Girshick}.} \bibinfo{year}{2017}\natexlab{}.
\newblock \showarticletitle{Inferring and Executing Programs for Visual
  Reasoning}. In \bibinfo{booktitle}{\emph{ICCV}}.
\newblock


\bibitem[\protect\citeauthoryear{Kar, Prakash, Liu, Cameracci, Yuan, Rusiniak,
  Acuna, Torralba, and Fidler}{Kar et~al\mbox{.}}{2019}]%
        {MetaSim}
\bibfield{author}{\bibinfo{person}{Amlan Kar}, \bibinfo{person}{Aayush
  Prakash}, \bibinfo{person}{Ming-Yu Liu}, \bibinfo{person}{Eric Cameracci},
  \bibinfo{person}{Justin Yuan}, \bibinfo{person}{Matt Rusiniak},
  \bibinfo{person}{David Acuna}, \bibinfo{person}{Antonio Torralba}, {and}
  \bibinfo{person}{Sanja Fidler}.} \bibinfo{year}{2019}\natexlab{}.
\newblock \bibinfo{title}{Meta-Sim: Learning to Generate Synthetic Datasets}.
\newblock   (\bibinfo{year}{2019}).
\newblock
\showeprint[arxiv]{cs.CV/1904.11621}


\bibitem[\protect\citeauthoryear{Kingma and Ba}{Kingma and Ba}{2014}]%
        {Kingma2014AdamAM}
\bibfield{author}{\bibinfo{person}{Diederik~P. Kingma} {and}
  \bibinfo{person}{Jimmy Ba}.} \bibinfo{year}{2014}\natexlab{}.
\newblock \showarticletitle{Adam: A Method for Stochastic Optimization}.
\newblock \bibinfo{journal}{\emph{CoRR}}  \bibinfo{volume}{abs/1412.6980}
  (\bibinfo{year}{2014}).
\newblock


\bibitem[\protect\citeauthoryear{Kingma and Welling}{Kingma and
  Welling}{2014}]%
        {kingma2014auto}
\bibfield{author}{\bibinfo{person}{Diederik~P. Kingma} {and}
  \bibinfo{person}{Max Welling}.} \bibinfo{year}{2014}\natexlab{}.
\newblock \showarticletitle{{Auto-{Encoding} Variational Bayes}}. In
  \bibinfo{booktitle}{\emph{International Conference on Learning
  Representations (ICLR)}}.
\newblock


\bibitem[\protect\citeauthoryear{Knapitsch, Park, Zhou, and Koltun}{Knapitsch
  et~al\mbox{.}}{2017}]%
        {TanksAndTemples}
\bibfield{author}{\bibinfo{person}{Arno Knapitsch}, \bibinfo{person}{Jaesik
  Park}, \bibinfo{person}{Qian-Yi Zhou}, {and} \bibinfo{person}{Vladlen
  Koltun}.} \bibinfo{year}{2017}\natexlab{}.
\newblock \showarticletitle{Tanks and Temples: Benchmarking Large-Scale Scene
  Reconstruction}.
\newblock \bibinfo{journal}{\emph{ACM Transactions on Graphics}}
  \bibinfo{volume}{36}, \bibinfo{number}{4} (\bibinfo{year}{2017}).
\newblock


\bibitem[\protect\citeauthoryear{Kolve, Mottaghi, Gordon, Zhu, Gupta, and
  Farhadi}{Kolve et~al\mbox{.}}{2017}]%
        {AI2Thor}
\bibfield{author}{\bibinfo{person}{Eric Kolve}, \bibinfo{person}{Roozbeh
  Mottaghi}, \bibinfo{person}{Daniel Gordon}, \bibinfo{person}{Yuke Zhu},
  \bibinfo{person}{Abhinav Gupta}, {and} \bibinfo{person}{Ali Farhadi}.}
  \bibinfo{year}{2017}\natexlab{}.
\newblock \showarticletitle{{AI2-THOR:} An Interactive 3D Environment for
  Visual {AI}}.
\newblock \bibinfo{journal}{\emph{CoRR}}  \bibinfo{volume}{arXiv:1712.05474}
  (\bibinfo{year}{2017}).
\newblock


\bibitem[\protect\citeauthoryear{Kusner, Paige, and
  Hern\'{a}ndez-Lobato}{Kusner et~al\mbox{.}}{2017}]%
        {GrammarVAE}
\bibfield{author}{\bibinfo{person}{Matt~J. Kusner}, \bibinfo{person}{Brooks
  Paige}, {and} \bibinfo{person}{Jos\'{e}~Miguel Hern\'{a}ndez-Lobato}.}
  \bibinfo{year}{2017}\natexlab{}.
\newblock \showarticletitle{Grammar Variational Autoencoder}. In
  \bibinfo{booktitle}{\emph{Proceedings of the 34th International Conference on
  Machine Learning - Volume 70}} \emph{(\bibinfo{series}{ICML’17})}.
  \bibinfo{publisher}{JMLR.org}, \bibinfo{pages}{1945–1954}.
\newblock


\bibitem[\protect\citeauthoryear{Lau, Ohgawara, Mitani, and Igarashi}{Lau
  et~al\mbox{.}}{2011}]%
        {10.1145/2010324.1964980}
\bibfield{author}{\bibinfo{person}{Manfred Lau}, \bibinfo{person}{Akira
  Ohgawara}, \bibinfo{person}{Jun Mitani}, {and} \bibinfo{person}{Takeo
  Igarashi}.} \bibinfo{year}{2011}\natexlab{}.
\newblock \showarticletitle{Converting 3D Furniture Models to Fabricatable
  Parts and Connectors}.
\newblock \bibinfo{journal}{\emph{ACM Trans. Graph.}} \bibinfo{volume}{30},
  \bibinfo{number}{4}, Article \bibinfo{articleno}{85} (\bibinfo{date}{July}
  \bibinfo{year}{2011}), \bibinfo{numpages}{6}~pages.
\newblock
\showISSN{0730-0301}
\urldef\tempurl%
\url{https://doi.org/10.1145/2010324.1964980}
\showDOI{\tempurl}


\bibitem[\protect\citeauthoryear{Li, Xu, Chaudhuri, Yumer, Zhang, and
  Guibas}{Li et~al\mbox{.}}{2017}]%
        {GRASS}
\bibfield{author}{\bibinfo{person}{Jun Li}, \bibinfo{person}{Kai Xu},
  \bibinfo{person}{Siddhartha Chaudhuri}, \bibinfo{person}{Ersin Yumer},
  \bibinfo{person}{Hao Zhang}, {and} \bibinfo{person}{Leonidas Guibas}.}
  \bibinfo{year}{2017}\natexlab{}.
\newblock \showarticletitle{{GRASS}: Generative recursive autoencoders for
  shape structures}.
\newblock \bibinfo{journal}{\emph{ACM Transactions on Graphics (TOG)}}
  \bibinfo{volume}{36}, \bibinfo{number}{4} (\bibinfo{year}{2017}),
  \bibinfo{pages}{52}.
\newblock


\bibitem[\protect\citeauthoryear{Liu, Wu, Ritchie, Freeman, Tenenbaum, and
  Wu}{Liu et~al\mbox{.}}{2019}]%
        {liu2019learning}
\bibfield{author}{\bibinfo{person}{Yunchao Liu}, \bibinfo{person}{Zheng Wu},
  \bibinfo{person}{Daniel Ritchie}, \bibinfo{person}{William~T. Freeman},
  \bibinfo{person}{Joshua~B. Tenenbaum}, {and} \bibinfo{person}{Jiajun Wu}.}
  \bibinfo{year}{2019}\natexlab{}.
\newblock \showarticletitle{{Learning to Describe Scenes with Programs}}. In
  \bibinfo{booktitle}{\emph{International Conference on Learning
  Representations (ICLR)}}.
\newblock


\bibitem[\protect\citeauthoryear{Lu, Mao, Tenenbaum, and Wu}{Lu
  et~al\mbox{.}}{2019}]%
        {lu2019neurally}
\bibfield{author}{\bibinfo{person}{Sidi Lu}, \bibinfo{person}{Jiayuan Mao},
  \bibinfo{person}{Joshua~B. Tenenbaum}, {and} \bibinfo{person}{Jiajun Wu}.}
  \bibinfo{year}{2019}\natexlab{}.
\newblock \showarticletitle{{{Neurally-Guided} Structure Inference}}. In
  \bibinfo{booktitle}{\emph{International Conference on Machine Learning
  (ICML)}}.
\newblock


\bibitem[\protect\citeauthoryear{Maas}{Maas}{2013}]%
        {Maas2013RectifierNI}
\bibfield{author}{\bibinfo{person}{Andrew~L. Maas}.}
  \bibinfo{year}{2013}\natexlab{}.
\newblock \showarticletitle{Rectifier Nonlinearities Improve Neural Network
  Acoustic Models}.
\newblock


\bibitem[\protect\citeauthoryear{Martinovic and Van~Gool}{Martinovic and
  Van~Gool}{2013}]%
        {FacadeInduction}
\bibfield{author}{\bibinfo{person}{A. Martinovic} {and} \bibinfo{person}{L.
  Van~Gool}.} \bibinfo{year}{2013}\natexlab{}.
\newblock \showarticletitle{{Bayesian Grammar Learning for Inverse Procedural
  Modeling}}. In \bibinfo{booktitle}{\emph{CVPR}}.
\newblock


\bibitem[\protect\citeauthoryear{Michalkiewicz, Pontes, Jack, Baktashmotlagh,
  and Eriksson}{Michalkiewicz et~al\mbox{.}}{2019}]%
        {DeepLevelSets}
\bibfield{author}{\bibinfo{person}{Mateusz Michalkiewicz},
  \bibinfo{person}{Jhony~K. Pontes}, \bibinfo{person}{Dominic Jack},
  \bibinfo{person}{Mahsa Baktashmotlagh}, {and} \bibinfo{person}{Anders~P.
  Eriksson}.} \bibinfo{year}{2019}\natexlab{}.
\newblock \showarticletitle{Deep Level Sets: Implicit Surface Representations
  for {3D} Shape Inference}.
\newblock \bibinfo{journal}{\emph{CoRR}}  \bibinfo{volume}{abs/1901.06802}
  (\bibinfo{year}{2019}).
\newblock


\bibitem[\protect\citeauthoryear{Mo, Guerrero, Yi, Su, Wonka, Mitra, and
  Guibas}{Mo et~al\mbox{.}}{2019a}]%
        {StructureNet}
\bibfield{author}{\bibinfo{person}{Kaichun Mo}, \bibinfo{person}{Paul
  Guerrero}, \bibinfo{person}{Li Yi}, \bibinfo{person}{Hao Su},
  \bibinfo{person}{Peter Wonka}, \bibinfo{person}{Niloy Mitra}, {and}
  \bibinfo{person}{Leonidas Guibas}.} \bibinfo{year}{2019}\natexlab{a}.
\newblock \showarticletitle{{StructureNet}: Hierarchical Graph Networks for
  {3D} Shape Generation}. In \bibinfo{booktitle}{\emph{SIGGRAPH Asia}}.
\newblock


\bibitem[\protect\citeauthoryear{Mo, Zhu, Chang, Yi, Tripathi, Guibas, and
  Su}{Mo et~al\mbox{.}}{2019b}]%
        {PartNet}
\bibfield{author}{\bibinfo{person}{Kaichun Mo}, \bibinfo{person}{Shilin Zhu},
  \bibinfo{person}{Angel~X. Chang}, \bibinfo{person}{Li Yi},
  \bibinfo{person}{Subarna Tripathi}, \bibinfo{person}{Leonidas~J. Guibas},
  {and} \bibinfo{person}{Hao Su}.} \bibinfo{year}{2019}\natexlab{b}.
\newblock \showarticletitle{{PartNet}: A Large-Scale Benchmark for Fine-Grained
  and Hierarchical Part-Level {3D} Object Understanding}. In
  \bibinfo{booktitle}{\emph{The IEEE Conference on Computer Vision and Pattern
  Recognition (CVPR)}}.
\newblock


\bibitem[\protect\citeauthoryear{M\"{u}ller, Wonka, Haegler, Ulmer, and
  Van~Gool}{M\"{u}ller et~al\mbox{.}}{2006}]%
        {CGAShape}
\bibfield{author}{\bibinfo{person}{Pascal M\"{u}ller}, \bibinfo{person}{Peter
  Wonka}, \bibinfo{person}{Simon Haegler}, \bibinfo{person}{Andreas Ulmer},
  {and} \bibinfo{person}{Luc Van~Gool}.} \bibinfo{year}{2006}\natexlab{}.
\newblock \showarticletitle{Procedural Modeling of Buildings}. In
  \bibinfo{booktitle}{\emph{SIGGRAPH}}.
\newblock


\bibitem[\protect\citeauthoryear{Nishida, Bousseau, and Aliaga}{Nishida
  et~al\mbox{.}}{2018}]%
        {Nishida2018}
\bibfield{author}{\bibinfo{person}{Gen Nishida}, \bibinfo{person}{Adrien
  Bousseau}, {and} \bibinfo{person}{Daniel~G. Aliaga}.}
  \bibinfo{year}{2018}\natexlab{}.
\newblock \showarticletitle{Procedural Modeling of a Building from a Single
  Image}.
\newblock \bibinfo{journal}{\emph{Computer Graphics Forum (Eurographics)}}
  \bibinfo{volume}{37}, \bibinfo{number}{2} (\bibinfo{year}{2018}).
\newblock


\bibitem[\protect\citeauthoryear{Nishida, Garcia-Dorado, Aliaga, Benes, and
  Bousseau}{Nishida et~al\mbox{.}}{2016}]%
        {Nishida2016}
\bibfield{author}{\bibinfo{person}{Gen Nishida}, \bibinfo{person}{Ignacio
  Garcia-Dorado}, \bibinfo{person}{Daniel~G Aliaga}, \bibinfo{person}{Bedrich
  Benes}, {and} \bibinfo{person}{Adrien Bousseau}.}
  \bibinfo{year}{2016}\natexlab{}.
\newblock \showarticletitle{{Interactive Sketching of Urban Procedural
  Models}}.
\newblock \bibinfo{journal}{\emph{ACM Transactions on Graphics (TOG)}}
  \bibinfo{volume}{35}, \bibinfo{number}{4} (\bibinfo{year}{2016}),
  \bibinfo{pages}{130}.
\newblock


\bibitem[\protect\citeauthoryear{Parish and M\"{u}ller}{Parish and
  M\"{u}ller}{2001}]%
        {CityEngine}
\bibfield{author}{\bibinfo{person}{Yoav I.~H. Parish} {and}
  \bibinfo{person}{Pascal M\"{u}ller}.} \bibinfo{year}{2001}\natexlab{}.
\newblock \showarticletitle{Procedural Modeling of Cities}. In
  \bibinfo{booktitle}{\emph{SIGGRAPH}}.
\newblock


\bibitem[\protect\citeauthoryear{Park, Florence, Straub, Newcombe, and
  Lovegrove}{Park et~al\mbox{.}}{2019}]%
        {DeepSDF}
\bibfield{author}{\bibinfo{person}{Jeong~Joon Park}, \bibinfo{person}{Peter
  Florence}, \bibinfo{person}{Julian Straub}, \bibinfo{person}{Richard
  Newcombe}, {and} \bibinfo{person}{Steven Lovegrove}.}
  \bibinfo{year}{2019}\natexlab{}.
\newblock \showarticletitle{DeepSDF: Learning Continuous Signed Distance
  Functions for Shape Representation}. In \bibinfo{booktitle}{\emph{The IEEE
  Conference on Computer Vision and Pattern Recognition (CVPR)}}.
\newblock


\bibitem[\protect\citeauthoryear{Paszke, Gross, Chintala, Chanan, Yang, DeVito,
  Lin, Desmaison, Antiga, and Lerer}{Paszke et~al\mbox{.}}{2017}]%
        {paszke2017automatic}
\bibfield{author}{\bibinfo{person}{Adam Paszke}, \bibinfo{person}{Sam Gross},
  \bibinfo{person}{Soumith Chintala}, \bibinfo{person}{Gregory Chanan},
  \bibinfo{person}{Edward Yang}, \bibinfo{person}{Zachary DeVito},
  \bibinfo{person}{Zeming Lin}, \bibinfo{person}{Alban Desmaison},
  \bibinfo{person}{Luca Antiga}, {and} \bibinfo{person}{Adam Lerer}.}
  \bibinfo{year}{2017}\natexlab{}.
\newblock \showarticletitle{Automatic differentiation in PyTorch}.
\newblock  (\bibinfo{year}{2017}).
\newblock


\bibitem[\protect\citeauthoryear{Prusinkiewicz and Lindenmayer}{Prusinkiewicz
  and Lindenmayer}{1996}]%
        {LSystemsBook}
\bibfield{author}{\bibinfo{person}{Przemyslaw Prusinkiewicz} {and}
  \bibinfo{person}{Aristid Lindenmayer}.} \bibinfo{year}{1996}\natexlab{}.
\newblock \bibinfo{booktitle}{\emph{The Algorithmic Beauty of Plants}}.
\newblock \bibinfo{publisher}{Springer-Verlag}, \bibinfo{address}{Berlin,
  Heidelberg}.
\newblock
\showISBNx{0-387-94676-4}


\bibitem[\protect\citeauthoryear{Qi, Yi, Su, and Guibas}{Qi
  et~al\mbox{.}}{2017}]%
        {qi2017pointnet++}
\bibfield{author}{\bibinfo{person}{Charles~Ruizhongtai Qi}, \bibinfo{person}{Li
  Yi}, \bibinfo{person}{Hao Su}, {and} \bibinfo{person}{Leonidas~J Guibas}.}
  \bibinfo{year}{2017}\natexlab{}.
\newblock \showarticletitle{Pointnet++: Deep hierarchical feature learning on
  point sets in a metric space}. In \bibinfo{booktitle}{\emph{Advances in
  neural information processing systems}}. \bibinfo{pages}{5099--5108}.
\newblock


\bibitem[\protect\citeauthoryear{Richter, Vineet, Roth, and Koltun}{Richter
  et~al\mbox{.}}{2016}]%
        {PlayingForData}
\bibfield{author}{\bibinfo{person}{Stephan~R Richter}, \bibinfo{person}{Vibhav
  Vineet}, \bibinfo{person}{Stefan Roth}, {and} \bibinfo{person}{Vladlen
  Koltun}.} \bibinfo{year}{2016}\natexlab{}.
\newblock \showarticletitle{Playing for data: Ground truth from computer
  games}. In \bibinfo{booktitle}{\emph{European conference on computer
  vision}}. Springer, \bibinfo{pages}{102--118}.
\newblock


\bibitem[\protect\citeauthoryear{Ritchie, Jobalia, and Thomas}{Ritchie
  et~al\mbox{.}}{2018}]%
        {ProcmodLearn}
\bibfield{author}{\bibinfo{person}{Daniel Ritchie}, \bibinfo{person}{Sarah
  Jobalia}, {and} \bibinfo{person}{Anna Thomas}.}
  \bibinfo{year}{2018}\natexlab{}.
\newblock \showarticletitle{Example-based Authoring of Procedural Modeling
  Programs with Structural and Continuous Variability}. In
  \bibinfo{booktitle}{\emph{EUROGRAPHICS}}.
\newblock


\bibitem[\protect\citeauthoryear{Savva, Kadian, Maksymets, Zhao, Wijmans, Jain,
  Straub, Liu, Koltun, Malik, Parikh, and Batra}{Savva et~al\mbox{.}}{2019}]%
        {Habitat}
\bibfield{author}{\bibinfo{person}{Manolis Savva}, \bibinfo{person}{Abhishek
  Kadian}, \bibinfo{person}{Oleksandr Maksymets}, \bibinfo{person}{Yili Zhao},
  \bibinfo{person}{Erik Wijmans}, \bibinfo{person}{Bhavana Jain},
  \bibinfo{person}{Julian Straub}, \bibinfo{person}{Jia Liu},
  \bibinfo{person}{Vladlen Koltun}, \bibinfo{person}{Jitendra Malik},
  \bibinfo{person}{Devi Parikh}, {and} \bibinfo{person}{Dhruv Batra}.}
  \bibinfo{year}{2019}\natexlab{}.
\newblock \showarticletitle{{H}abitat: {A} {P}latform for {E}mbodied {AI}
  {R}esearch}. In \bibinfo{booktitle}{\emph{The IEEE International Conference
  on Computer Vision (ICCV)}}.
\newblock


\bibitem[\protect\citeauthoryear{Sharma, Goyal, Liu, Kalogerakis, and
  Maji}{Sharma et~al\mbox{.}}{2018}]%
        {sharma2018csgnet}
\bibfield{author}{\bibinfo{person}{Gopal Sharma}, \bibinfo{person}{Rishabh
  Goyal}, \bibinfo{person}{Difan Liu}, \bibinfo{person}{Evangelos Kalogerakis},
  {and} \bibinfo{person}{Subhransu Maji}.} \bibinfo{year}{2018}\natexlab{}.
\newblock \showarticletitle{{{CSGNet}: Neural Shape Parser for Constructive
  Solid Geometry}}. In \bibinfo{booktitle}{\emph{IEEE Conference on Computer
  Vision and Pattern Recognition (CVPR)}}.
\newblock


\bibitem[\protect\citeauthoryear{Stava, Benes, Mech, Aliaga, and Kristof}{Stava
  et~al\mbox{.}}{2010}]%
        {InverseLSystems}
\bibfield{author}{\bibinfo{person}{Ondrej Stava}, \bibinfo{person}{Bedrich
  Benes}, \bibinfo{person}{Radom{\'i}r Mech}, \bibinfo{person}{Daniel~G.
  Aliaga}, {and} \bibinfo{person}{Peter Kristof}.}
  \bibinfo{year}{2010}\natexlab{}.
\newblock \showarticletitle{Inverse Procedural Modeling by Automatic Generation
  of L-systems}.
\newblock \bibinfo{journal}{\emph{Comput. Graph. Forum}}  \bibinfo{volume}{29}
  (\bibinfo{year}{2010}), \bibinfo{pages}{665--674}.
\newblock


\bibitem[\protect\citeauthoryear{Sung, Su, Kim, Chaudhuri, and Guibas}{Sung
  et~al\mbox{.}}{2017}]%
        {ComplementMe}
\bibfield{author}{\bibinfo{person}{Minhyuk Sung}, \bibinfo{person}{Hao Su},
  \bibinfo{person}{Vladimir~G. Kim}, \bibinfo{person}{Siddhartha Chaudhuri},
  {and} \bibinfo{person}{Leonidas Guibas}.} \bibinfo{year}{2017}\natexlab{}.
\newblock \showarticletitle{Complement{Me}: Weakly-Supervised Component
  Suggestions for {3D} Modeling}.
\newblock \bibinfo{journal}{\emph{ACM Transactions on Graphics (Proc. of
  SIGGRAPH Asia)}} (\bibinfo{year}{2017}).
\newblock


\bibitem[\protect\citeauthoryear{Talton, Yang, Kumar, Lim, Goodman, and
  Mech}{Talton et~al\mbox{.}}{2012}]%
        {BayesianGrammarInduction}
\bibfield{author}{\bibinfo{person}{Jerry~O. Talton}, \bibinfo{person}{Lingfeng
  Yang}, \bibinfo{person}{Ranjitha Kumar}, \bibinfo{person}{Maxine Lim},
  \bibinfo{person}{Noah~D. Goodman}, {and} \bibinfo{person}{Radom{\'i}r Mech}.}
  \bibinfo{year}{2012}\natexlab{}.
\newblock \showarticletitle{Learning design patterns with {Bayesian} grammar
  induction}. In \bibinfo{booktitle}{\emph{UIST}}.
\newblock


\bibitem[\protect\citeauthoryear{Tian, Luo, Sun, Ellis, Freeman, Tenenbaum, and
  Wu}{Tian et~al\mbox{.}}{2019}]%
        {tian2019learning}
\bibfield{author}{\bibinfo{person}{Yonglong Tian}, \bibinfo{person}{Andrew
  Luo}, \bibinfo{person}{Xingyuan Sun}, \bibinfo{person}{Kevin Ellis},
  \bibinfo{person}{William~T. Freeman}, \bibinfo{person}{Joshua~B. Tenenbaum},
  {and} \bibinfo{person}{Jiajun Wu}.} \bibinfo{year}{2019}\natexlab{}.
\newblock \showarticletitle{{Learning to Infer and Execute {3D} Shape
  Programs}}. In \bibinfo{booktitle}{\emph{International Conference on Learning
  Representations (ICLR)}}.
\newblock


\bibitem[\protect\citeauthoryear{Wu, Zhang, Xue, Freeman, and Tenenbaum}{Wu
  et~al\mbox{.}}{2016}]%
        {3DGAN}
\bibfield{author}{\bibinfo{person}{Jiajun Wu}, \bibinfo{person}{Chengkai
  Zhang}, \bibinfo{person}{Tianfan Xue}, \bibinfo{person}{William~T. Freeman},
  {and} \bibinfo{person}{Joshua~B. Tenenbaum}.}
  \bibinfo{year}{2016}\natexlab{}.
\newblock \showarticletitle{{Learning a Probabilistic Latent Space of Object
  Shapes via {3D} {Generative-Adversarial} Modeling}}. In
  \bibinfo{booktitle}{\emph{Advances in Neural Information Processing Systems
  (NeurIPS)}}.
\newblock


\bibitem[\protect\citeauthoryear{Xia, R.~Zamir, He, Sax, Malik, and
  Savarese}{Xia et~al\mbox{.}}{2018}]%
        {GibsonEnv}
\bibfield{author}{\bibinfo{person}{Fei Xia}, \bibinfo{person}{Amir R.~Zamir},
  \bibinfo{person}{Zhi-Yang He}, \bibinfo{person}{Alexander Sax},
  \bibinfo{person}{Jitendra Malik}, {and} \bibinfo{person}{Silvio Savarese}.}
  \bibinfo{year}{2018}\natexlab{}.
\newblock \showarticletitle{Gibson env: real-world perception for embodied
  agents}. In \bibinfo{booktitle}{\emph{CVPR}}.
\newblock


\bibitem[\protect\citeauthoryear{Z.~Wu}{Z.~Wu}{2015}]%
        {3DShapeNets}
\bibfield{author}{\bibinfo{person}{A.~Khosla F. Yu L. Zhang X. Tang J.~Xiao
  Z.~Wu, S.~Song}.} \bibinfo{year}{2015}\natexlab{}.
\newblock \showarticletitle{3D ShapeNets: A Deep Representation for Volumetric
  Shapes}. In \bibinfo{booktitle}{\emph{Computer Vision and Pattern
  Recognition}}.
\newblock


\bibitem[\protect\citeauthoryear{Zhang, Song, Yumer, Savva, Lee, Jin, and
  Funkhouser}{Zhang et~al\mbox{.}}{2017}]%
        {RenderingSUNCG}
\bibfield{author}{\bibinfo{person}{Yinda Zhang}, \bibinfo{person}{Shuran Song},
  \bibinfo{person}{Ersin Yumer}, \bibinfo{person}{Manolis Savva},
  \bibinfo{person}{Joon-Young Lee}, \bibinfo{person}{Hailin Jin}, {and}
  \bibinfo{person}{Thomas Funkhouser}.} \bibinfo{year}{2017}\natexlab{}.
\newblock \showarticletitle{Physically-Based Rendering for Indoor Scene
  Understanding Using Convolutional Neural Networks}.
\newblock \bibinfo{journal}{\emph{The IEEE Conference on Computer Vision and
  Pattern Recognition (CVPR)}} (\bibinfo{year}{2017}).
\newblock


\bibitem[\protect\citeauthoryear{Zhou, Li, and Poczos}{Zhou
  et~al\mbox{.}}{2019}]%
        {zhou2019treeLSTM}
\bibfield{author}{\bibinfo{person}{Chenghui Zhou}, \bibinfo{person}{Chun-liang
  Li}, {and} \bibinfo{person}{Barnabas Poczos}.}
  \bibinfo{year}{2019}\natexlab{}.
\newblock \showarticletitle{Program Synthesis for Images using Tree-Structured
  LSTM}. In \bibinfo{booktitle}{\emph{PGR Workshop at NeurIPS}}.
\newblock


\bibitem[\protect\citeauthoryear{Zhu, Xu, Chaudhuri, Yi, and Zhang}{Zhu
  et~al\mbox{.}}{2018}]%
        {SCORES}
\bibfield{author}{\bibinfo{person}{Chenyang Zhu}, \bibinfo{person}{Kai Xu},
  \bibinfo{person}{Siddhartha Chaudhuri}, \bibinfo{person}{Renjiao Yi}, {and}
  \bibinfo{person}{Hao Zhang}.} \bibinfo{year}{2018}\natexlab{}.
\newblock \showarticletitle{{SCORES}: Shape Composition with Recursive
  Substructure Priors}.
\newblock \bibinfo{journal}{\emph{ACM Transactions on Graphics (TOG)}}
  \bibinfo{volume}{37}, \bibinfo{number}{6} (\bibinfo{year}{2018}),
  \bibinfo{pages}{211:1--211:14}.
\newblock


\bibitem[\protect\citeauthoryear{Zou, Yumer, Yang, Ceylan, and Hoiem}{Zou
  et~al\mbox{.}}{2017}]%
        {zou20173d}
\bibfield{author}{\bibinfo{person}{Chuhang Zou}, \bibinfo{person}{Ersin Yumer},
  \bibinfo{person}{Jimei Yang}, \bibinfo{person}{Duygu Ceylan}, {and}
  \bibinfo{person}{Derek Hoiem}.} \bibinfo{year}{2017}\natexlab{}.
\newblock \showarticletitle{{3D-PRNN: Generating Shape Primitives with
  Recurrent Neural Networks}}. In \bibinfo{booktitle}{\emph{IEEE International
  Conference on Computer Vision (ICCV)}}.
\newblock


\end{thebibliography}
